\documentclass[a4paper,11pt]{article}
\pdfoutput=1
\usepackage{jcappub}
\usepackage{mathtools}
\usepackage[normalem]{ulem}
\graphicspath{{plots/}}

\newcommand{\gs}{g_\star}
\newcommand{\gss}{g_{\star s}}
\newcommand{\gdm}{g}
\newcommand{\Trh}{T_\text{rh}}
\newcommand{\Tmax}{T_\text{max}}
\newcommand{\Tfo}{T_\text{fo}}
\newcommand{\Tpfo}{T'_\text{fo}}
\newcommand{\Tk}{T_\text{k}}
\newcommand{\xrh}{x_\text{rh}}
\newcommand{\xfo}{x_\text{fo}}
\newcommand{\xfoc}{x_\text{fo}^\text{c}}
\newcommand{\xpfo}{{x'_\text{fo}}}
\newcommand{\xpfom}{{x^{\prime \text{min}}_\text{fo}}}
\newcommand{\xk}{x_\text{k}}
\newcommand{\xkm}{x_\text{k}^\text{min}}
\newcommand{\arh}{a_\text{rh}}

\newcommand{\afo}{a_\text{fo}}
\newcommand{\rp}{\rho_\phi}
\newcommand{\rR}{\rho_R}
\newcommand{\svel}{\langle\sigma v\rangle_\text{el}}
\newcommand{\svann}{\langle\sigma v\rangle_{2\to 2}}
\newcommand{\svcan}{\langle\sigma v^2\rangle_{3\to 2}}
\newcommand{\Gel}{\Gamma_\text{el}}
\newcommand{\Gann}{\Gamma_{2\to 2}}
\newcommand{\Gcan}{\Gamma_{3\to 2}}
\newcommand{\yef}{y_\text{eff}}
\newcommand{\eef}{\epsilon_\text{eff}}

\title{Thermal Dark Matter with Low-Temperature Reheating}

\author{Nicolás Bernal,}
\author{Kuldeep Deka,}
\author{and Marta Losada}

\affiliation{New York University Abu Dhabi\\
PO Box 129188, Saadiyat Island, Abu Dhabi, United Arab Emirates}

\emailAdd{nicolas.bernal@nyu.edu}
\emailAdd{kuldeep.deka@nyu.edu}
\emailAdd{marta.losada@nyu.edu}

\abstract{We explore the production of thermal dark matter (DM) candidates (WIMPs, SIMPs, ELDERs and Cannibals) during cosmic reheating. Assuming a general parametrization for the scaling of the inflaton energy density and the standard model (SM) temperature, we study the requirements for kinetic and chemical DM freeze-out in a model-independent way. For each of the mechanisms, up to two solutions that fit the entire observed DM relic density exist, for a given reheating scenario and DM mass.  As an example, we assume a simple particle physics model in which DM interacts with itself and with SM through contact interactions. We find that low-temperature reheating can accommodate a wider range of couplings and larger masses than those permitted in the usual instantaneous high-temperature reheating. This results in DM solutions for WIMPs reaching masses as high as $10^{14}$~GeV, whereas for SIMPs and ELDERs, we can reach masses of $10^{13}$~GeV. Interestingly, current experimental data already constrain the enlarged parameter space of these models with low-reheating temperatures. Next-generation experiments could further probe these scenarios.}

\begin{document}
\begin{flushright}
\end{flushright}
\maketitle

\section{Introduction}
The evidence for the presence of nonbaryonic dark matter (DM) in the Universe is strongly supported by both astrophysical and cosmological data. For a DM candidate to be considered viable, it must meet several criteria: It must be electromagnetically neutral, cosmologically stable, and nonrelativistic at the time of Big Bang nucleosynthesis (BBN). Additionally, it should exhibit a relic density of $\Omega h^2 \simeq 0.12$, which accounts for 27\% of the total energy content of the Universe~\cite{Planck:2018vyg}; for a recent review see Ref.~\cite{Cirelli:2024ssz}.

The most popular mechanism for DM production in the early Universe is the weakly interacting massive particle (WIMP) model. In this case, DM possesses a mass at the electroweak scale and couples significantly with the standard model (SM) thermal plasma, as is common in electroweak interactions. WIMPs achieve thermal equilibrium with the SM thermal plasma and subsequently undergo chemical freeze-out, leading to the observed DM relic abundance. It is generally assumed that this freeze-out occurs well after reheating has ended, at a time when the Universe's energy density is dominated by SM radiation. Observational data typically require a thermally averaged annihilation cross-section $\langle\sigma v\rangle \simeq \text{few} \times 10^{-26}$~cm$^3$/s~\cite{Steigman:2012nb}. The WIMP paradigm can have different realizations: one can imagine 2-to-2 annihilation of DM particles into SM states~\cite{Lee:1977ua}, the co-annihilation of a pair of states of the dark sector with only one being the DM~\cite{Griest:1990kh}, or even the semi-annihilation of DM particles into a DM and a SM states~\cite{Hambye:2008bq, Hambye:2009fg, DEramo:2010keq, Belanger:2012zr, Belanger:2014bga}. The WIMP mechanism is notably intriguing, as it can be explored through various complementary techniques such as direct, indirect, and collider probes. However, the absence of positive experimental results and strict constraints on its expected parameter space urge investigations beyond the conventional WIMP framework~\cite{Arcadi:2017kky, Roszkowski:2017nbc, Arcadi:2024ukq}.

Alternatively, instead of WIMPs, one can consider the Strongly Interacting Massive Particle (SIMP) paradigm~\cite{Hochberg:2014dra} where the freeze-out proceeds through $N$-to-$N'$ number-changing processes, where $N$ DM particles annihilate into $N'$ of them (with $N > N' \ge 2$). The most studied cases of the $N$-to-$N'$ processes correspond to 3-to-2, because it is typically dominant. However, 3-to-2 annihilations, necessarily induced by interaction vertices with an odd number of DM particles, are forbidden in the most common models where the stability of the DM is guaranteed by a $\mathbb{Z}_2$ symmetry. To allow 3-to-2 annihilations, one has to assume that DM is protected by a different symmetry such as $\mathbb{Z}_3$~\cite{Choi:2015bya, Bernal:2015bla, Bernal:2015lbl, Ko:2014nha, Choi:2017mkk, Chu:2017msm}, or consider models where DM stability emerges as a result of DM dynamics~\cite{Bernal:2015ova, Yamanaka:2014pva, Hochberg:2014kqa, Lee:2015gsa, Hansen:2015yaa}. If DM is stabilized by a $\mathbb{Z}_2$ symmetry, the 4-to-2 reactions would be those that give rise to the DM relic abundance, while the 3-to-2 annihilations are forbidden~\cite{Bernal:2015xba, Heikinheimo:2016yds, Bernal:2017mqb, Heikinheimo:2017ofk, Bernal:2018hjm}. One potential problem of this mechanism is that when $N$-to-$N'$ annihilations are effective, the DM reheats itself, significantly modifying the formation of structures~\cite{deLaix:1995vi}. However, this can be avoided by imposing kinetic equilibrium between the DM and the visible sector~\cite{Hochberg:2014dra}.\footnote{Other solutions exist. For example, one can consider an enlarged dark sector containing new particles that are relativistic at the moment of DM freeze-out, so that the heating of the dark sector is only due to the (small) change of relativistic degrees of freedom~\cite{Bernal:2015bla}, or abandon the kinetic equilibrium and start with a colder dark sector~\cite{Bernal:2015ova, Bernal:2015xba}.} In the SIMP paradigm, kinetic equilibrium is broken {\it after} chemical equilibrium, so that during DM production, the visible and dark sectors share the same temperature.

However, the kinetic equilibrium could be broken {\it before} chemical equilibrium. On the one hand, if this happens when DM is {\it non-relativistic}, it corresponds to an ELastically DEcoupling Relic (ELDER)~\cite{Kuflik:2015isi, Kuflik:2017iqs}. In this case, at the moment of chemical freeze-out, the DM is warmer than the SM because of efficient $N$-to-$N'$ annihilations that tend to keep the DM at a constant temperature as the Universe expands. Interestingly, the near-constant temperature (and therefore density) of DM implies that its present relic abundance is mainly determined by the cross section of its {\it elastic} scattering on SM particles (instead of the inelastic processes as in the case of WIMPs or SIMPs).

On the other hand, the scenario where DM goes out of kinetic equilibrium at a very high temperature when it is still {\it relativistic}, {\it before} breaking the chemical equilibrium corresponds to cannibal or self-interacting DM~\cite{Carlson:1992fn, Pappadopulo:2016pkp, Farina:2016llk}. In this case, the increase in temperature of the dark sector with respect to the visible sector (occurring between the moment when DM becomes nonrelativistic and its chemical freeze-out) is maximal. In contrast to ELDERs, for cannibal DM, the present relic abundance becomes independent of the detail of kinetic decoupling and is only determined by the chemical freeze-out.

Interestingly, all previously presented {\it thermal} production mechanisms can be organized as a function of the temperatures at which DM departs from chemical and kinetic equilibrium.\footnote{DM could also have been produced nonthermally, as in the case of the feebly interacting massive particle (FIMP) paradigm~\cite{McDonald:2001vt, Choi:2005vq, Kusenko:2006rh, Petraki:2007gq, Hall:2009bx, Elahi:2014fsa, Bernal:2017kxu}.} Calling $\Tk$ the temperature of SM at kinetic decoupling and $\Tpfo$ the temperature of DM at chemical decoupling, the WIMP and SIMP scenarios correspond to $m > \Tpfo > \Tk$, ELDER to $m > \Tk > \Tpfo$ and cannibal DM to $\Tk > m > \Tpfo$, where $m$ stands for the mass of the DM particle. We note that since we are interested in {\it cold} DM, the hierarchy $m \gg \Tpfo$ is required.

It is important to note that the present DM abundance is determined not only by the particle-physics dynamics but also by the cosmological history of the Universe. As the evolution of the early Universe is largely unknown, the standard (i.e., the simplest!) assumption corresponds to a Universe dominated by SM radiation from the end of cosmological inflation until matter-radiation equality at a SM temperature $T \simeq 0.8$~eV~\cite{Planck:2018vyg}. Additionally, the transition from an inflaton-dominated to a radiation-dominated Universe, that is, the reheating era, is typically assumed to be instantaneous and to occur at a very high temperature, well before the production of DM. However, this picture cannot be taken for granted. In fact, many well-motivated non-standard deviations could have been realized by nature and therefore should be studied~\cite{Allahverdi:2020bys}. The reheating temperature $\Trh$ (that is, the SM temperature from which the Universe begins to be dominated by SM radiation) must satisfy $\Trh > T_\text{bbn} \simeq 4$~MeV~\cite{Sarkar:1995dd, Kawasaki:2000en, Hannestad:2004px, DeBernardis:2008zz, deSalas:2015glj}, in order not to spoil the success of BBN. The production of DM in scenarios with a nonstandard expansion phase has recently gained increasing interest, particularly for WIMPs~\cite{Barrow:1982ei, Kamionkowski:1990ni, McDonald:1989jd, Salati:2002md, Fornengo:2002db, Comelli:2003cv, Rosati:2003yw, Pallis:2004yy, Gelmini:2006pw, Drees:2006vh, Gelmini:2006pq, Arbey:2008kv, Cohen:2008nb, Arbey:2009gt, Roszkowski:2014lga, Berlin:2016vnh, Berlin:2016gtr, DEramo:2017gpl, Hamdan:2017psw, Visinelli:2017qga, Drees:2017iod, Hardy:2018bph, Bernal:2018kcw, Drees:2018dsj, Betancur:2018xtj, Maldonado:2019qmp, Poulin:2019omz, Arias:2019uol, Han:2019vxi, Chanda:2019xyl, Arcadi:2020aot, Bhatia:2020itt, Barman:2021ifu, Cheek:2021cfe, Arcadi:2021doo, Bernal:2020bjf, Bernal:2022wck, Haque:2023yra, Silva-Malpartida:2023yks, Bernal:2023ura, Arcadi:2024jzv}, but also for SIMPs~\cite{Bernal:2018ins, Bernal:2020kse, Bernal:2023ura}.

During reheating the Universe is dominated by the inflaton. Its energy density is typically assumed to scale as non-relativistic matter or as radiation, corresponding to cases where the inflaton oscillates on a quadratic or quartic potential, respectively. However, it can also scale faster than radiation, as in the case of kination~\cite{Spokoiny:1993kt, Ferreira:1997hj}, or even faster, as in the context of ekpyrotic~\cite{Khoury:2001wf, Khoury:2003rt} or cyclic scenarios~\cite{Gasperini:2002bn, Erickson:2003zm, Barrow:2010rx, Ijjas:2019pyf}; see also Ref.~\cite{Scherrer:2022nnz}. Additionally, the inflaton could decay or annihilate into different kinds of SM particles, and therefore the dependence of the temperature of the SM bath could feature different behaviors with the cosmic scale factor.

Here, we focus on thermal DM production that occurs {\it during} low-temperature reheating.\footnote{Alternatively, this could also correspond to the end of a non-standard cosmological scenario produced by the decay of a long-lived heavy particle (e.g. a moduli field) different from the inflaton, or to a Universe dominated by primordial black holes that evaporate by emitting Hawking radiation.} In particular, we study the phenomenology of the thermal production mechanisms mentioned above (WIMP, SIMP, ELDER, and cannibal DM) in the case where kinetic and/or chemical decoupling does not occur in the SM radiation-dominated era but during reheating. For that, in Section~\ref{sec:LTR} we present the general parameterization used to describe the low-temperature reheating scenario. In Section~\ref{sec:after}, the different thermal DM production mechanisms are presented, highlighting how thermal decoupling occurs, for the usual case where the reheating temperature is very high, much higher than the scales at which the DM is produced. The case in which DM is thermally produced during reheating is studied in Section~\ref{sec:during}. In Section~\ref{sec:model}, a simplified particle physics model is presented as an example to materialize our findings. Finally, in Section~\ref{sec:concl} we summarize and conclude.

\section{Low-Temperature Reheating} \label{sec:LTR}
Cosmic reheating is the era in which the Universe transits from being dominated by inflaton to SM radiation energy density. During reheating, the inflaton $\phi$ has an effective equation of state $\omega$, which implies that its energy density $\rp$ scales as
\begin{equation}
    \rp(a) \propto a^{-3\, (1 + \omega)},
\end{equation}
where $a$ corresponds to the cosmic scale factor of the Universe. Particularly interesting cases correspond to $\omega = 0$, $\omega = 1/3$, and $\omega = 1$, corresponding to an inflaton oscillating in a quadratic or quartic potential, and to kination, respectively. More generally, if the inflaton oscillates at the bottom of a monomial potential $V(\phi) \propto \phi^n$, its equation of state tends to be $\omega = (n - 2)/(n + 2)$~\cite{Turner:1983he}.\footnote{Preheating effects due to possible self-interaction of the inflaton could result in a transition into a radiation-dominated regime within $\mathcal{O}(1)$ $e$-folds~\cite{Lozanov:2016hid, Lozanov:2017hjm}.} In the case where $\omega > 1/3$, $\rp(a)$ gets diluted faster than free radiation and, therefore, eventually the SM energy density dominates without the decay or annihilation of the inflatons. However, if $\omega \leq 1/3$, $\phi$ has to decay or annihilate, injecting entropy into SM particles. We assume that the inflaton only decays or annihilates into SM states, producing a scaling of the SM temperature
\begin{equation} \label{eq:Ta}
    T(a) = \Trh \left(\frac{\arh}{a}\right)^\alpha,
\end{equation}
where $\Trh$ is the bath temperature at the end of reheating, and the corresponding scale factor $\arh \equiv a(\Trh)$. Furthermore, $\alpha$ depends on the properties of the inflaton during reheating. For an inflaton oscillating in a quadratic potential ($\omega = 0$) and a constant decay width, $\alpha = 3/8$. However, even in the case of $\omega = 0$, other scalings are possible, for example, in the case of a nontrivial dependence of the decay width of the inflaton with the scale factor~\cite{Bodeker:2006ij, Mukaida:2012qn, Daido:2017wwb, Co:2020xaf, Garcia:2020wiy, Ahmed:2021fvt, Barman:2022tzk, Banerjee:2022fiw, Arias:2022qjt, Chowdhury:2023jft}. Interestingly, in a quadratic potential, one can also have an era with constant temperature $\alpha =0$, as elaborated in Refs.~\cite{Co:2017pyf, Co:2020xaf,Cosme:2024ndc}. Alternatively, if the inflaton oscillates in a quartic potential ($\omega = 1/3$), the decay into scalar or fermionic states gives rise to $\alpha = 1/4$ or 3/4, respectively. In general, in the case of a monomial potential $\alpha = \frac32\, \frac{1}{2 + n}$ for a scalar decay, and $\alpha = \frac32\, \frac{n - 1}{2 + n}$ if $n \leq 7$ or $\alpha = 1$ if $n \geq 7$ for a fermionic decay~\cite{Shtanov:1994ce, Ichikawa:2008ne, Bernal:2022wck}. If instead of decaying, reheating occurs through inflaton annihilations into bosons, one gets $\alpha = \frac{9}{2\, n + 4}$ for $n \geq 3$ or $\alpha = \frac{3(7-2n)}{2\, n + 4}$, for the case of a heavy or light mediator, respectively~\cite{Bernal:2023wus, Barman:2024mqo}. The case of inflaton annihilations into fermions gives $\alpha = 1$ or $\alpha = \frac{3(5-n)}{2\, n + 4}$, for a heavy or light mediator, respectively~\cite{Barman:2024mqo}. Finally, if the inflaton energy density is diluted faster than free radiation, that is, if $\omega > 1/3$, it is not necessary for the inflaton to decay or annihilate away, and then one can have $\alpha = 1$, as in the case of kination~\cite{Spokoiny:1993kt, Ferreira:1997hj}. We note that, in cases where $\alpha > 0$, during reheating the SM bath temperature rises to a maximal value $\Tmax$ that could exceed $\Trh$ by several orders of magnitude~\cite{Giudice:2000ex}.

Taking into account that the SM radiation energy density $\rR$ is given by
\begin{equation}
    \rR(T) = \frac{\pi^2}{30}\, \gs(T)\, T^4,
\end{equation}
with $\gs(T)$ being the number of relativistic degrees of freedom contributing to $\rR$~\cite{Drees:2015exa}, it is interesting to note that a viable reheating, that is, an eventual onset of the SM radiation domination, requires that at some point $\rR(a) > \rp(a)$, which in turn implies
\begin{equation} \label{eq:noRH}
    \alpha \leq \frac34\, (1 + \omega)\,.
\end{equation}

\begin{figure}[t!]
    \def\sepf{0.55}
    \centering
    \includegraphics[scale=\sepf]{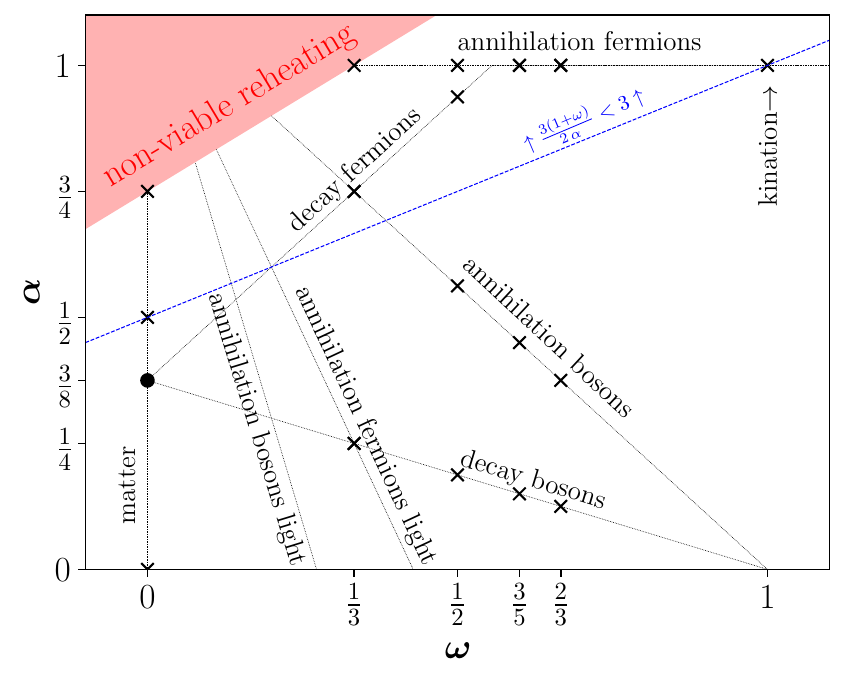}
    \caption{Summary of the different reheating scenarios. The black dot corresponds to the standard case where the inflaton scales as non-relativistic matter and decays into SM particles with a constant decay width, while the black crosses correspond to the alternatives described in the text. The red area in the upper left corner does not give rise to viable reheating. Above the blue dotted line $\frac{3(1+\omega)}{2\, \alpha} < 3$.}
    \label{fig:cosmo}
\end{figure}

In the plane $[\omega,\, \alpha]$, Fig.~\ref{fig:cosmo} shows with crosses the different reheating options previously discussed. The black dot corresponds to the standard case with $\omega = 0$ and $\alpha = 3/8$. The thin black dotted lines correspond to the previously described scenarios: the vertical one to $\omega = 0$ (dust-like inflaton), the horizontal one to a kination-like faster-than-radiation expansion, and the five diagonals to decay and annihilations into scalars and fermions. Importantly, the red area in the upper left corner does not give rise to a viable reheating, cf. Eq.~\eqref{eq:noRH}, and is therefore ignored.

The Hubble expansion rate $H$ depends on the total energy density of the Universe and is given by the Friedmann equation
\begin{equation} \label{eq:H}
    H^2 = \frac{\rp + \rR}{3\, M_P^2}\,,
\end{equation}
where $M_P \simeq 2.4 \times 10^{18}$~GeV is the reduced Planck mass. It can be approximately written as
\begin{equation}
    H(a) \simeq H(\arh) \times
    \begin{dcases}
        \left(\frac{\arh}{a}\right)^\frac{3\, (1 + \omega)}{2} &\text{ for } a \leq \arh,\\
        \left(\frac{\arh}{a}\right)^2 &\text{ for } \arh \leq a,
    \end{dcases}
\end{equation}
depending on whether we are during ($a \leq \arh$) or after ($a \geq \arh$) reheating, which corresponds to
\begin{equation}
    H(T) \simeq H(\Trh) \times
    \begin{dcases}
        \left(\frac{T}{\Trh}\right)^\frac{3\, (1 + \omega)}{2\, \alpha} &\text{ for } T \geq \Trh,\\
        \left(\frac{T}{\Trh}\right)^2 &\text{ for } \Trh \geq T,
    \end{dcases}
\end{equation}
where the reheating temperature $\Trh$ is implicitly defined by the equality $\rp(\Trh) = \rR(\Trh)$. We note that if $\frac{3 (1+\omega)}{2\alpha} = 2$, $H$ has the same scaling with temperature during and after reheating.

\section{Dark Matter Production \textit{\textbf{After}} Reheating} \label{sec:after}
In this section, different thermal production mechanisms for DM in the early Universe are studied, assuming the standard scenario where the reheating temperature is much higher than the relevant scales for DM genesis. The two cases $\Tfo \gg \Tk$ (corresponding to WIMPs and SIMPs) and $\Tk \gg \Tfo$ (corresponding to ELDERs and cannibals) are studied separately.

\subsection[WIMPs and SIMPs: $\Trh \gg \Tfo \gg \Tk$]{\boldmath WIMPs and SIMPs: $\Trh \gg \Tfo \gg \Tk$}
Both in the WIMP and the SIMP paradigms, chemical equilibrium is broken before kinetic equilibrium, $\Tfo \gg \Tk$, which guarantees that at chemical freeze-out the two sectors share the same temperature $T$.

The equilibrium number density $n_\text{eq}$ for DM particles of mass $m$ and $\gdm$ internal degrees of freedom is given by
\begin{equation} \label{eq:neq}
    n_\text{eq}(T) = \frac{\gdm}{2 \pi^2}\, m^2\, T\, K_2\left(\frac{m}{T}\right),
\end{equation}
where $K_i$ is the modified Bessel function, and where the Maxwell-Boltzmann statistic was assumed. For simplicity, in all numerical evaluations we fix $g = 1$. After chemical freeze-out, the DM yield defined as $Y(T) \equiv n(T)/s(T)$ is conserved, where $n(T)$ is the DM number density,
\begin{equation} \label{eq:sSM}
    s(T) \equiv \frac{2 \pi^2}{45}\, \gss(T)\, T^3
\end{equation}
is the SM entropy density, and $\gss(T)$ corresponds to the effective number of degrees of freedom contributing to the SM entropy~\cite{Drees:2015exa}. At present, the DM yield $Y_0$ can be estimated as
\begin{equation} \label{eq:WIMPSIMP-MD}
    Y_0 \simeq Y_\text{fo} = \frac{n_\text{eq}(\Tfo)}{s(\Tfo)} = \frac{45}{4 \pi^4}\, \frac{\gdm}{\gss(\Tfo)}\, \xfo^2\, K_2(\xfo) \simeq \frac{45}{2^{5/2}\, \pi^{7/2}}\, \frac{\gdm}{\gss(\Tfo)}\, \xfo^{3/2}\, e^{-\xfo},
\end{equation}
with $\xfo \equiv m/\Tfo$. We emphasize that at this level, WIMPs and SIMPs are indistinguishable, as the reaction fixing the DM freeze-out has not been specified.

\begin{figure}[t!]
    \def\sepf{0.55}
    \centering
    \includegraphics[scale=\sepf]{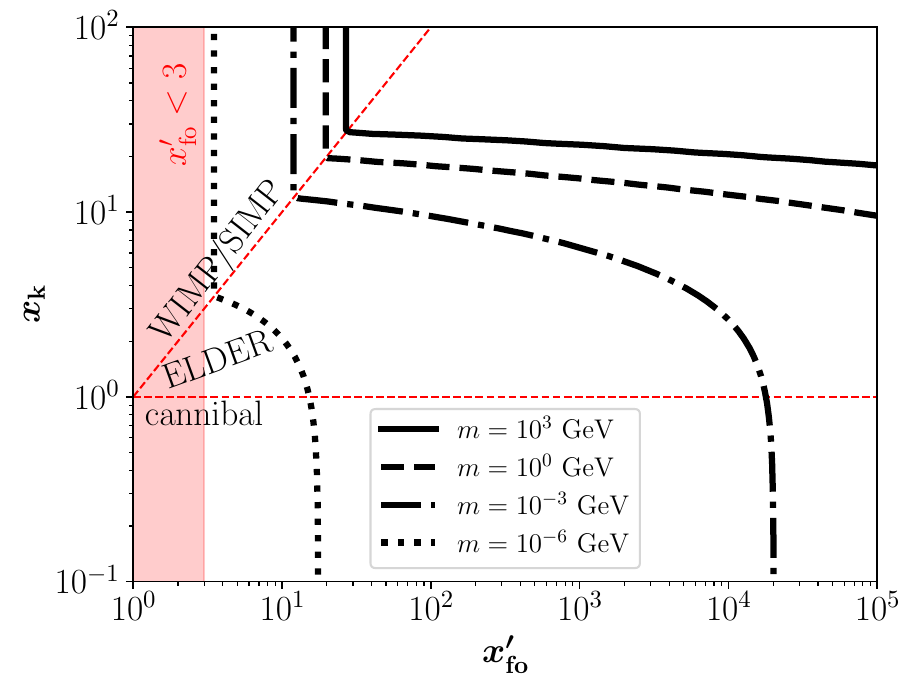}
    \caption{Parameter space that fits the observed DM relic abundance for different DM masses, assuming a production {\it after} reheating (that is, in the radiation-dominated era). In the red band ($\xpfo < 3$), DM freezes-out when relativistic, and is therefore not a cold relic.}
    \label{fig:RD}
\end{figure}

To match the entire observed DM relic density, it is required that
\begin{equation} \label{relic-density-no}
    m\, Y_0 = \frac{\Omega h^2\, \rho_c}{s_0\, h^2} \simeq 4.3 \times 10^{-10}~\text{GeV},
\end{equation}
where $Y_0$ is the asymptotic value of the DM yield at low temperatures, $s_0 \simeq 2.69 \times 10^3$~cm$^{-3}$ is the present entropy density~\cite{ParticleDataGroup:2022pth}, $\rho_c \simeq 1.05 \times 10^{-5}~h^2$~GeV/cm$^3$ is the critical energy density of the Universe, and $\Omega h^2 \simeq 0.12$ is the observed DM relic abundance~\cite{Planck:2018vyg}. Figure~\ref{fig:RD} shows, for different DM masses, the parameter space that fits the observed DM relic abundance, in an extravagantly large plane $[\xpfo,\, \xk]$.\footnote{We are aware that a 1~keV is in tension with observations of the Lyman-$\alpha$ forest, which set a lower limit on the DM mass of around 5.3~keV~\cite{Viel:2013fqw, Irsic:2017ixq}, and with BBN and CMB observations, set a lower bound on the thermal DM mass of 0.4~MeV~\cite{Sabti:2019mhn, Sabti:2021reh}. We used, however, 1~keV for reference.} We recall that in this case $\xpfo = \xfo$, given that for WIMPs and SIMPs $\Tfo \gg \Tk$. As expected from Eq.~\eqref{eq:WIMPSIMP-MD}, the DM relic abundance is independent of the kinetic decoupling (i.e. the lines are vertical) and has a small logarithmic dependence on the DM mass. In the red band, corresponding to $\xpfo < 3$, DM freezes-out when still relativistic, and therefore it is not a cold relic.

\subsection[Cannibals and ELDERs: $\Trh \gg \Tk \gg \Tfo$]{\boldmath Cannibals and ELDERs: $\Trh \gg \Tk \gg \Tfo$}
Opposite to the WIMP and the SIMP cases, for ELDERs and cannibals kinetic equilibrium is broken before chemical equilibrium, $\Tk \gg \Tfo$. It is important to note that kinetic equilibrium guarantees that the two sectors have a common temperature and therefore if it is broken, the two sectors will be characterized by different temperatures: $T$ for the SM and $T'$ for the dark sector.\footnote{Here we assume that self-interactions within the dark sector are strong enough to create a thermal dark plasma. We further neglect possible asymmetries in the dark sector, and therefore chemical potentials.}

The DM yield at chemical freeze-out is given by
\begin{equation} \label{eq:elder0}
    Y_\text{fo} = \frac{n_\text{eq}(\Tpfo)}{s(\Tfo)} = \frac{n_\text{eq}(\Tpfo)}{s(\Tk)}\, \frac{s'(\Tk)}{s'(\Tpfo)}\,,
\end{equation}
where in the first equality we have used the fact that at freeze out the DM has a temperature $\Tpfo$ while the SM temperature is $\Tfo$, while in the second, the separate conservation of the entropies of the visible and dark sectors ($s$ and $s'$ respectively) was taken into account.
Additionally, given that the energy density $\rho'$ and the pressure $p'$ of the dark sector are
\begin{align}
    \rho'(T') &= \frac{\gdm}{2 \pi^2}\, m^3\, T' \left[K_1\left(\frac{m}{T'}\right) + 3\, \frac{T'}{m}\, K_2\left(\frac{m}{T'}\right)\right],\\
    p'(T') &= \frac{\gdm}{2 \pi^2}\, m^2\, {T'}^2\, K_2\left(\frac{m}{T'}\right),
\end{align}
for a Maxwell-Boltzmann distribution without chemical potential, the entropy density of the dark sector can be written as
\begin{equation}
    s'(T') = \frac{\rho'(T') + p'(T')}{T'} = \frac{\gdm}{2 \pi^2}\, m^3\, K_3\left(\frac{m}{T'}\right).
\end{equation}
Therefore, the DM yield at present can be conveniently expressed as
\begin{equation} \label{eq:elder1}
    Y_0 \simeq Y_\text{fo} = \frac{45}{4 \pi^4}\, \frac{\gdm}{\gss(\Tk)}\, \frac{\xk^3}{\xpfo}\, \frac{K_2(\xpfo)\, K_3(\xk)}{K_3(\xpfo)}\,,
\end{equation}
with $\xk \equiv m/\Tk$ and $\xpfo \equiv m/\Tpfo$. As expected, Eqs.~\eqref{eq:elder1} and~\eqref{eq:WIMPSIMP-MD} coincide in the limit $\xk = \xfo = \xpfo$.  Figure ~\ref{fig:RD} also displays the parameter space that fits the observed abundance of DM, for the case $\xpfo > \xk$.

Equation~\eqref{eq:elder1} has two interesting limits extensively discussed in the literature. The case where chemical decoupling occurs non-relativistically, but kinetic decoupling when DM is still relativistic ($\xk \ll 1 \ll \xpfo$) corresponds to the cannibal scenario, and therefore Eq.~\eqref{eq:elder1} reduces to
\begin{equation}
    Y_0 \simeq \frac{90}{\pi^4}\, \frac{\gdm}{\gss(\Tk)}\, \frac{1}{\xpfo}\,,
\end{equation}
and is independent of the kinetic decoupling~\cite{Carlson:1992fn}. Alternatively, if both kinetic and chemical decoupling happen when DM is non-relativistic ($1 \ll \xk \ll \xpfo$)
\begin{equation}
    Y_0 \simeq \frac{45}{2^{5/2}\, \pi^{7/2}}\, \frac{\gdm}{\gss(\Tk)}\, \frac{\xk^{5/2}\, e^{-\xk}}{\xpfo}\,,
\end{equation}
corresponding to the ELDER scenario~\cite{Kuflik:2015isi, Kuflik:2017iqs}.\footnote{We found a difference of $\pi^2$ with respect to the original result reported in Ref.~\cite{Kuflik:2015isi}.} Even if it also depends on $\xpfo$, the ELDER limit has a strong exponential dependence on the kinetic decoupling. The two regimes can be recognized in Fig.~\ref{fig:RD}.

Before closing this section, we note that the separate conservation of entropies allows us to compute the SM temperature at freeze-out $\Tfo$
\begin{equation}
    \xfo \equiv \frac{m}{\Tfo} = \xk \left[\frac{\gss(\Tfo)}{\gss(\Tk)}\, \frac{K_3(\xk)}{K_3(\xpfo)}\right]^{1/3},
\end{equation}
as a function of $\xk$ and $\xpfo$.

\section{Dark Matter Production \textit{\textbf{During}} Reheating} \label{sec:during}
 Although it is commonly assumed that cosmic reheating finishes at a very high temperature, it may not be the case. In this section, we study different thermal mechanisms in the case where they are effective during reheating. We emphasize that, by construction, the inflaton transmits its energy to the SM bath and not to the DM. This is typically a good assumption as long as its branching ratio to the dark sector is smaller than $\mathcal{O}(10^{-4}) \times m/(100~\text{GeV})$~\cite{Drees:2017iod, Arias:2019uol}.

\subsection[WIMPs and SIMPs: $\Tfo \gg \Trh$]{\boldmath WIMPs and SIMPs: $\Tfo \gg \Trh$}
The impact on the final DM relic abundance produced through the WIMP and the SIMP mechanisms occurring during reheating is two-fold. On the one hand, chemical decoupling occurs earlier, increasing the DM yield; on the other hand, injection of entropy into the SM bath because of inflaton decays dilutes the DM abundance.

The dilution corresponds to the change in SM entropy $S$ from a given moment until the end of the reheating and is given by
\begin{equation} \label{eq:entropyinj}
    \frac{S(T)}{S(\Trh)} = \frac{s(T)}{s(\Trh)} \left(\frac{a(T)}{a(\Trh)}\right)^3  = \frac{\gss(T)}{\gss(\Trh)} \left(\frac{\Trh}{T}\right)^\frac{3\, (1 - \alpha)}{\alpha},
\end{equation}
where Eqs.~\eqref{eq:Ta} and~\eqref{eq:sSM} were used. Given that the SM entropy is conserved after the end of reheating, the DM yield at present can be estimated by
\begin{equation} \label{eq:WIMPSIMP-RH}
    Y_0 \simeq Y_\text{rh} = Y_\text{fo}\, \frac{S(\Tfo)}{S(\Trh)} = \frac{45}{4 \pi^4}\, \frac{\gdm}{\gss(\Trh)}\, \xfo^2\, K_2(\xfo) \left(\frac{\xfo}{\xrh}\right)^\frac{3\, (1 - \alpha)}{\alpha},
\end{equation}
where $Y_\text{fo}$ is the same as the one computed in Eq.~\eqref{eq:WIMPSIMP-MD}. Interestingly, Eq.~\eqref{eq:WIMPSIMP-RH} is valid as long as $\Tfo > \Trh$, in the two cases $\Tfo > \Trh > \Tk$ and $\Tfo > \Tk > \Trh$. Additionally, it can be approximated in the two limits
\begin{equation} \label{eq:wimprhlimits}
    Y_0 \simeq
    \begin{dcases}
        \frac{45}{2 \pi^4}\, \frac{\gdm}{\gss(\Trh)} \left(\frac{\xfo}{\xrh}\right)^\frac{3\, (1 - \alpha)}{\alpha} &\text{ for } \xfo \ll \xfoc\,,\\
        \frac{45}{4 \pi^4} \sqrt{\frac{\pi}{2}}\, \frac{\gdm}{\gss(\Trh)}\, \xfo^\frac32\, e^{-\xfo} \left(\frac{\xfo}{\xrh}\right)^\frac{3\, (1 - \alpha)}{\alpha} &\text{ for } \xfo \gg \xfoc\,,
    \end{dcases}
\end{equation}
where the critical value $\xfoc$ (after extremizing Eq.~\eqref{eq:WIMPSIMP-RH} w.r.t. $\xfo$) is implicitly defined by
\begin{equation} \label{eq:xfoc}
    \alpha\, \xfoc\, K_1(\xfoc) + 3\, (\alpha - 1)\, K_2(\xfoc) =0\,.
\end{equation}

Several comments are in order with respect to Eq.~\eqref{eq:wimprhlimits}.
$i)$ To be realized, the first expression (corresponding to $\xfo \ll \xfoc$) requires interaction rates with a higher temperature dependence than the Hubble expansion rate, as will be shown in Section~\ref{sec:model}. $ii)$ For the second expression (corresponding to $\xfo \gg \xfoc$) one only has to require $\xfo < \xrh$. And $iii)$, with respect to $\xfoc$, for a given $\Trh$, a critical value of the DM mass can be defined as the mass required to fit the entire abundance of DM if $\xfo = \xfoc$. DM with a mass larger than that value overshoots the observations if it is produced during reheating.

\begin{figure}[t!]
    \def\sepf{0.50}
    \centering
    \includegraphics[scale=\sepf]{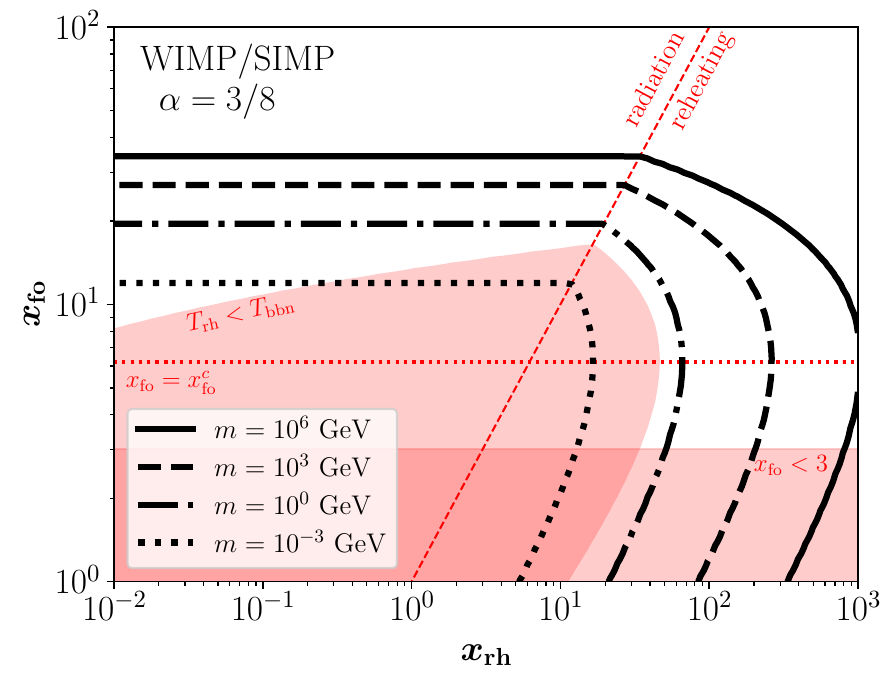}
    \includegraphics[scale=\sepf]{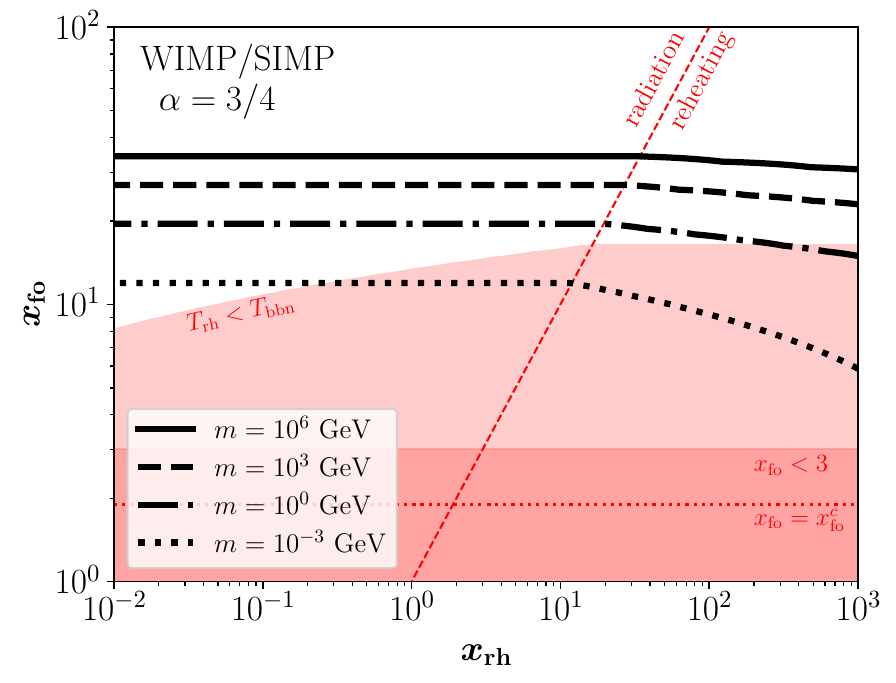}
    \includegraphics[scale=0.52]{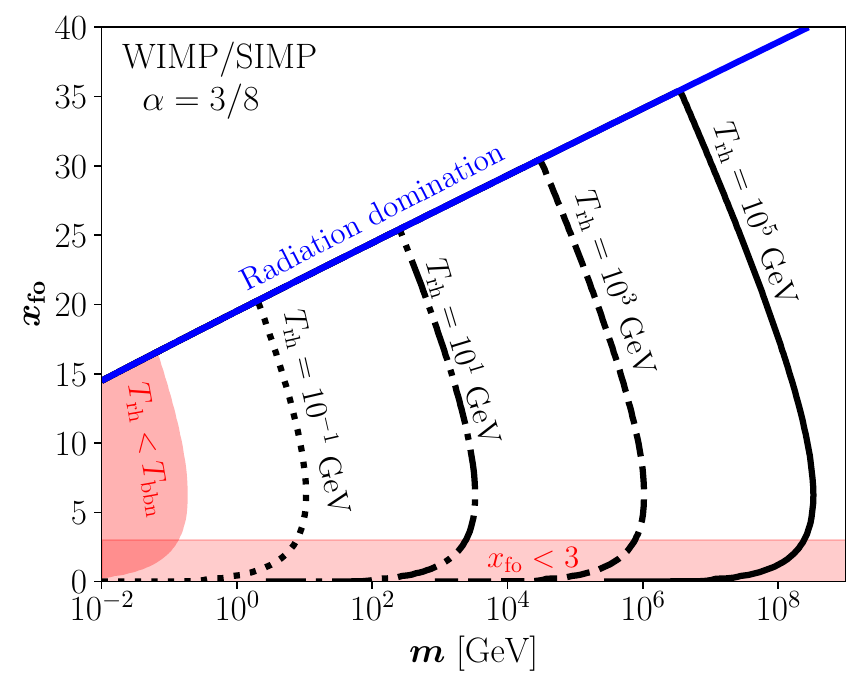}
    \includegraphics[scale=0.52]{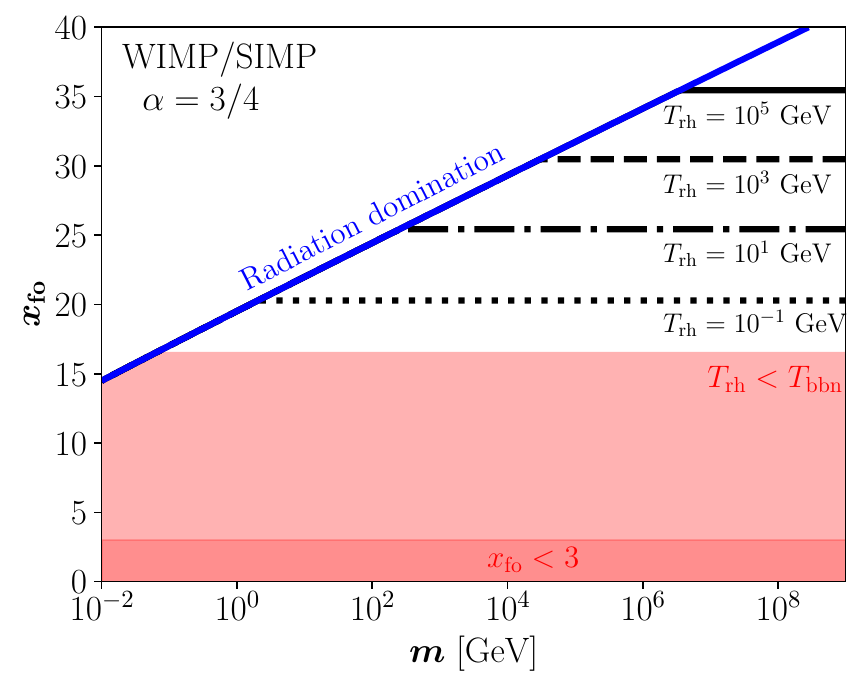}
    \caption{Value of $\xfo$ required to fit the entire DM relic abundance for $\alpha = 3/8$ (left) or $\alpha=3/4$ (right), as a function of $\xrh$ (top) or $m$ (bottom), in the WIMP and SIMP scenarios. The red bands are in tension with BBN or the non-relativistic freeze-out.}
    \label{fig:wimp1}
\end{figure} 
The parameter space that fits the observed DM relic abundance for WIMPs and SIMPs, taking $\alpha = 3/8$ (left) or $\alpha = 3/4$ (right), as a function of $\xrh$ (top) or $m$ (bottom), is shown in Fig.~\ref{fig:wimp1}. The red dotted horizontal lines correspond to the critical value $\xfoc \simeq 6.2$ (1.9) for $\alpha = 3/8$ (3/4), while the red dashed diagonal lines in the upper panel depict the border between freeze-out during radiation domination and reheating (that is, when $\xfo=\xrh$). The red bands are in tension with BBN or the non-relativistic freeze-out. In the upper panels of Fig.~\ref{fig:wimp1}, if $\xfo > \xrh$, DM freezes out in the radiation-dominated era and therefore the abundance of DM is independent of $\xrh$. In contrast, if $\xfo < \xrh$, DM is produced during reheating, at a higher temperature (that is, smaller $\xfo$) compared to the standard case. For a fixed $\xrh$, the two solutions during reheating described in Eq.~\eqref{eq:wimprhlimits} are visible. Furthermore, the corresponding maximum mass (for a given $\Trh$) clearly appears in the lower left panel of Fig.~\ref{fig:wimp1}, at $\xfo = \xfoc$. This novel feature of a double solution during reheating is further investigated in Appendix~\ref{sec:numerics}, where a complete numerical calculation was performed, solving the system of Boltzmann equations for the inflaton and SM radiation energy densities and the DM number density. Finally, the minimal viable DM mass that can be produced during reheating is also visible, and corresponds to point where $\Tfo = \Trh = T_\text{bbn}$.

\subsection[Cannibals and ELDERs: $\Tk \gg \Tfo$]{\boldmath Cannibals and ELDERs: $\Tk \gg \Tfo$}
This section is conveniently divided into two cases, $\Trh \gg \Tfo$ and $\Tfo \gg \Trh$, depending on whether reheating finishes after or before the DM chemical freeze-out.

\subsubsection[$\Tk \gg \Trh \gg \Tfo$]{\boldmath $\Tk \gg \Trh \gg \Tfo$}
In this case where $\Tk \gg \Trh \gg \Tfo$, the DM yield at present is given by
\begin{align} \label{eq:elder2}
    Y_0 &\simeq Y_\text{fo} = \frac{n_\text{eq}(\Tpfo)}{s(\Tfo)} = \frac{n_\text{eq}(\Tpfo)}{s(\Trh)}\, \frac{s'(\Trh')}{s'(\Tpfo)} = \frac{n_\text{eq}(\Tpfo)}{s(\Trh)}\, \frac{s'(\Tk)}{s'(\Tpfo)} \left(\frac{\Trh}{\Tk}\right)^\frac{3}{\alpha}\nonumber\\
    &= \frac{45}{4 \pi^4}\, \frac{\gdm}{\gss(\Trh)}\, \frac{\xrh^3}{\xpfo} \left(\frac{\xk}{\xrh}\right)^\frac{3}{\alpha} \frac{K_2(\xpfo)\, K_3(\xk)}{K_3(\xpfo)},
\end{align}
where the separate conservation of entropies between the end of reheating and the DM freeze-out, and the scaling of the DM temperature during reheating were used.

In this case where $\Tk \gg \Tfo$, at the time of chemical freeze out the two sectors are not expected to share the same temperature. Using the conservation of the dark sector entropy between $a(\Tk)$ and $\afo$, the conservation of the SM entropy between $\arh$ and $\afo$, and the entropy injection in Eq.~\eqref{eq:entropyinj}, one gets that
\begin{equation} \label{eq:xfoelder2}
    \xfo = \xrh \left(\frac{\xk}{\xrh}\right)^\frac{1}{\alpha} \left[\frac{\gss(\Tfo)}{\gss(\Trh)}\, \frac{K_3(\xk)}{K_3(\xpfo)}\right]^\frac13.
\end{equation}

\subsubsection[$\Tk \gg \Tfo \gg \Trh$]{\boldmath $\Tk \gg \Tfo \gg \Trh$}
Alternatively, in the case where $\Tk \gg \Tfo \gg \Trh$, the DM yield at present is given by
\begin{align} \label{eq:elder3}
    Y_0 &\simeq Y_\text{rh} = \frac{n_\text{eq}(\Trh')}{s(\Trh)} = \frac{n_\text{eq}(\Tpfo)}{s(\Trh)}\, \frac{s'(\Trh')}{s'(\Tpfo)}= \frac{n_\text{eq}(\Tpfo)}{s(\Trh)}\, \frac{s'(\Tk)}{s'(\Tpfo)} \left(\frac{\Trh}{\Tk}\right)^\frac{3}{\alpha}\nonumber\\
    &= \frac{45}{4 \pi^4}\, \frac{\gdm}{\gss(\Trh)}\, \frac{\xrh^3}{\xpfo} \left(\frac{\xk}{\xrh}\right)^\frac{3}{\alpha} \frac{K_2(\xpfo)\, K_3(\xk)}{K_3(\xpfo)}\,,
\end{align}
which, as expected, is equal to the expression in Eq.~\eqref{eq:elder2}, and simply corresponds to Eq.~\eqref{eq:elder1} times the dilution factor between $\Tk$ and $\Trh$. However, even if the expressions for the DM yield are the same, the SM temperature at which the chemical freeze-out occurs is different. Using the conservation of the dark sector entropy between $a(\Tk)$ and $\afo$, and the entropy injection to the visible sector in Eq.~\eqref{eq:entropyinj}, one gets that the SM temperature at chemical freeze-out is
\begin{equation}
    \xfo = \xk \left[\frac{K_3(\xk)}{K_3(\xpfo)}\right]^\frac{\alpha}{3},
\end{equation}
independently of the reheating temperature, and which only coincides with Eq.~\eqref{eq:xfoelder2} in the limit $\xk = \xfo$.

\begin{figure}[t!]
    \def\sepf{0.49}
    \centering
    \includegraphics[scale=\sepf]{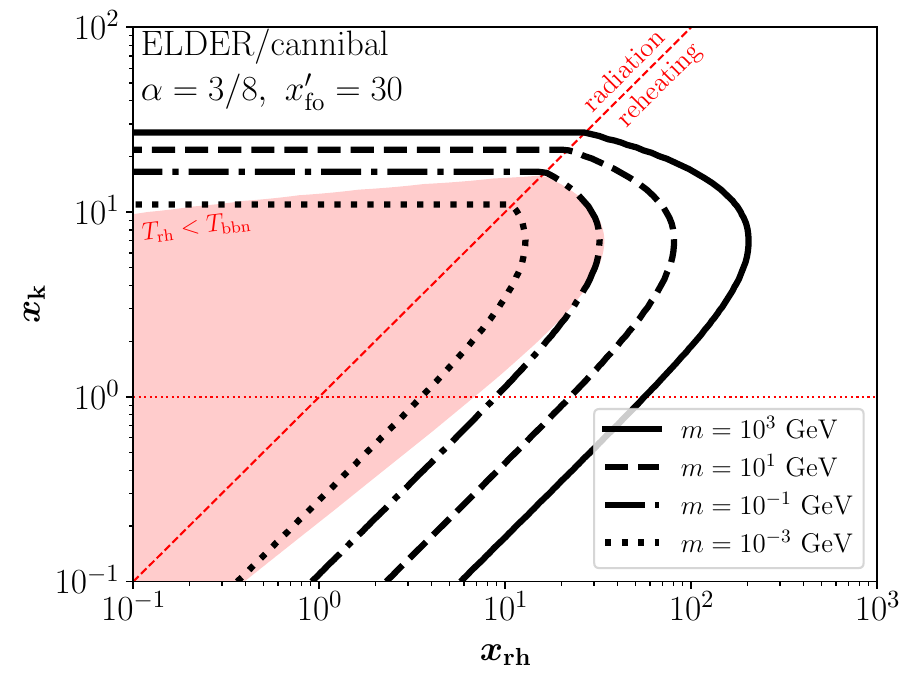}
    \includegraphics[scale=\sepf]{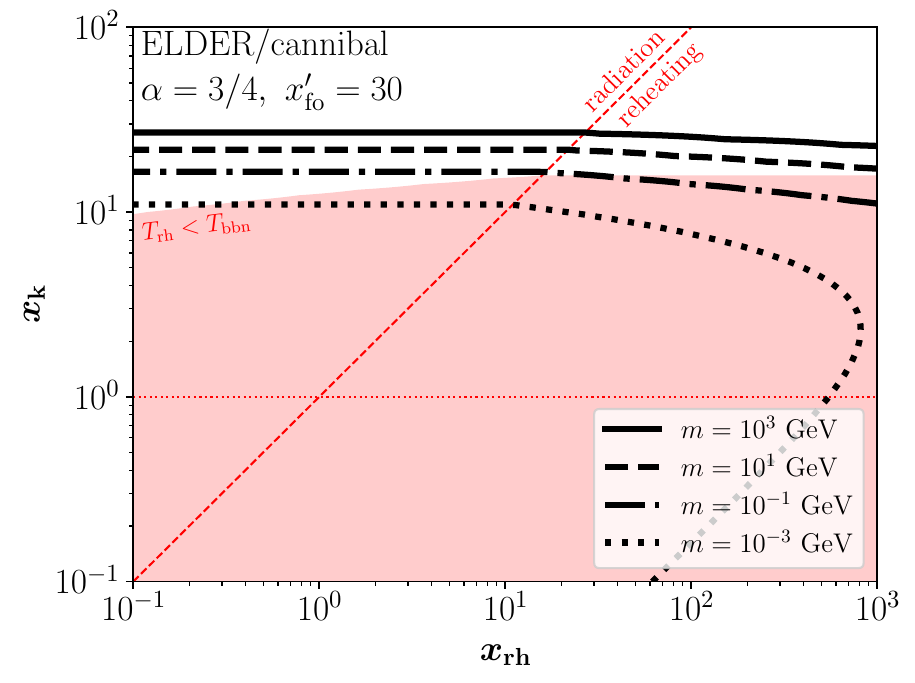}
    \caption{Parameter space that fits the observed DM relic abundance, for ELDERs and cannibals, for $\alpha = 3/8$ (left) or $\alpha=3/4$ (right), $\xpfo = 30$ and different DM masses. The red bands are in tension with BBN.}
    \label{fig:elder1}
\end{figure} 
Considering Eq.~\eqref{eq:elder2} (or equivalently Eq.~\eqref{eq:elder3}), the ELDER limit occurs if both kinetic and chemical decoupling happen when DM is non-relativistic ($1 \ll \xk \ll \xpfo$)
\begin{equation}
    Y_0 \simeq \frac{45}{2^{5/2}\, \pi^{7/2}}\, \frac{\gdm}{\gss(\Trh)}\, \frac{\xk^{5/2}\, e^{-\xk}}{\xpfo} \left(\frac{\xk}{\xrh}\right)^\frac{3(1 - \alpha)}{\alpha},
\end{equation}
while in the case where chemical decoupling occurs nonrelativistically, but kinetic decoupling when DM is still relativistic ($\xk \ll 1 \ll \xpfo$) corresponds to the cannibal scenario and therefore the DM yield reduces to
\begin{equation}
    Y_0 \simeq \frac{90}{\pi^4}\, \frac{\gdm}{\gss(\Trh)}\, \frac{1}{\xpfo} \left(\frac{\xk}{\xrh}\right)^\frac{3(1-\alpha)}{\alpha},
\end{equation}
which, contrary to the case in radiation domination, shows a dependence on the kinetic decoupling coming from the dilution factor. The parameter space that fits the observed DM relic abundance for ELDERs and cannibals, taking $\alpha = 3/8$ (left) or $\alpha = 3/4$ (right), $\xpfo = 30$ and different masses of DM, is shown in Fig.~\ref{fig:elder1}. In a way equivalent to WIMPs/SIMPs, if $\xk < \xrh$ DM kinetically decouples during reheating, at a higher temperature (that is, smaller $\xk$) compared to the standard case. Also, for a fixed $\xrh$, two solutions during reheating are visible.\\

\begin{figure}[t!]
    \def\sepf{0.49}
    \centering
    \includegraphics[scale=\sepf]{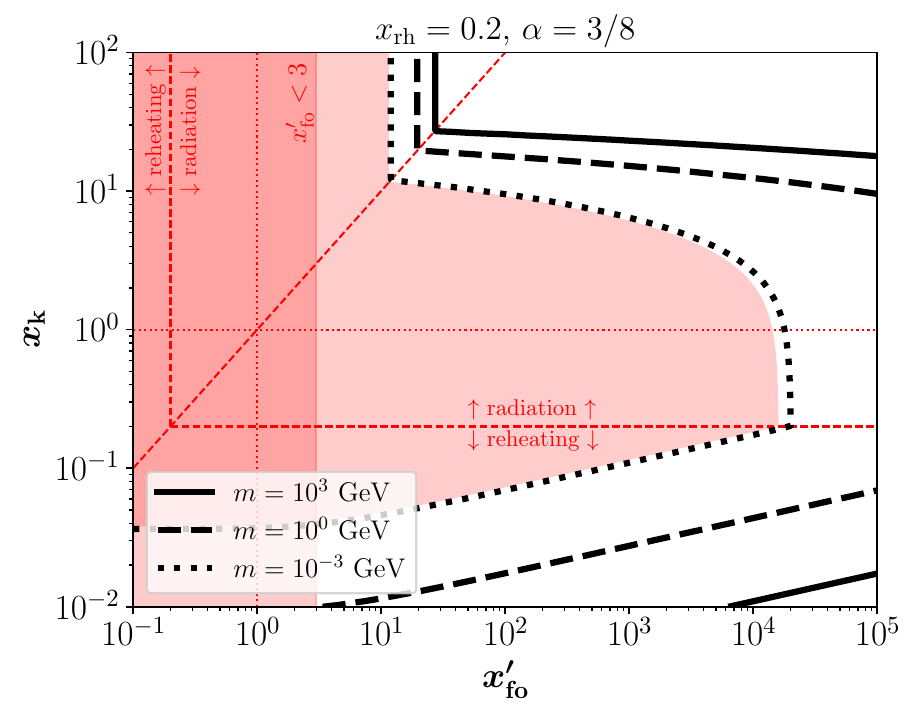}
    \includegraphics[scale=\sepf]{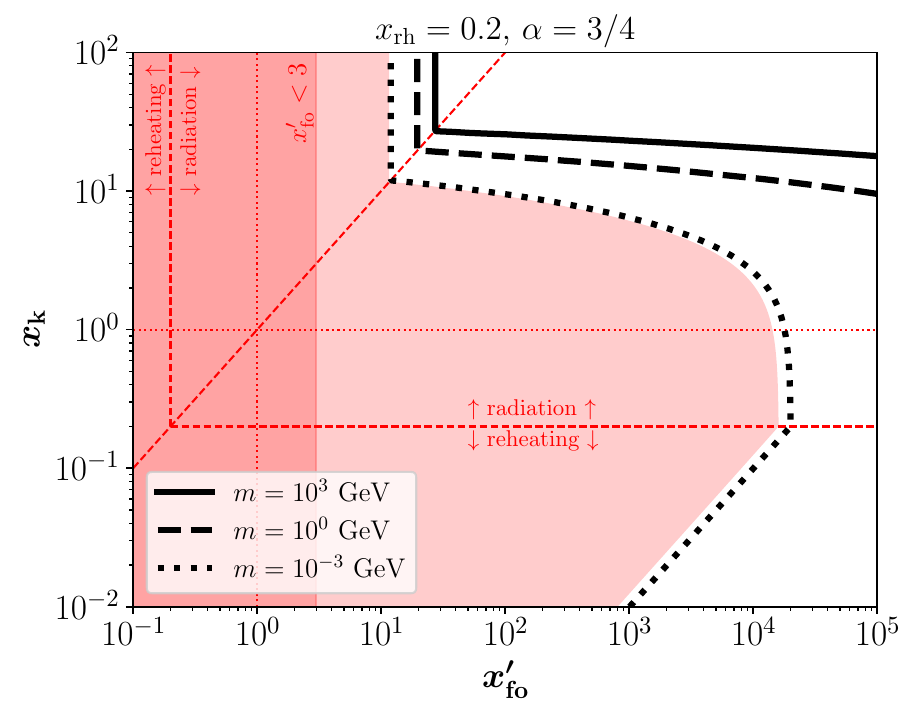}
    \includegraphics[scale=\sepf]{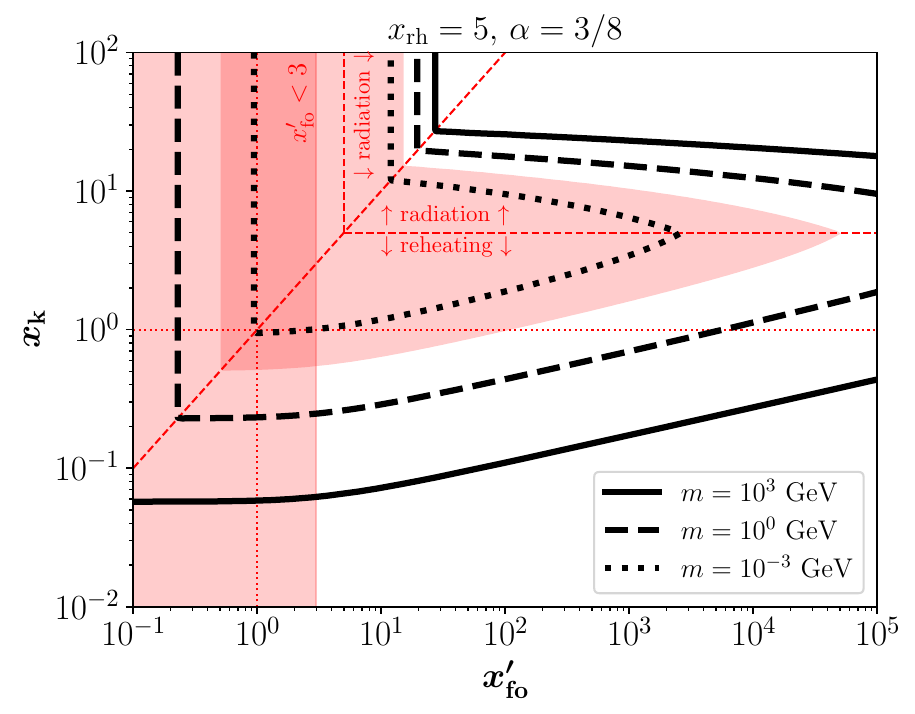}
    \includegraphics[scale=\sepf]{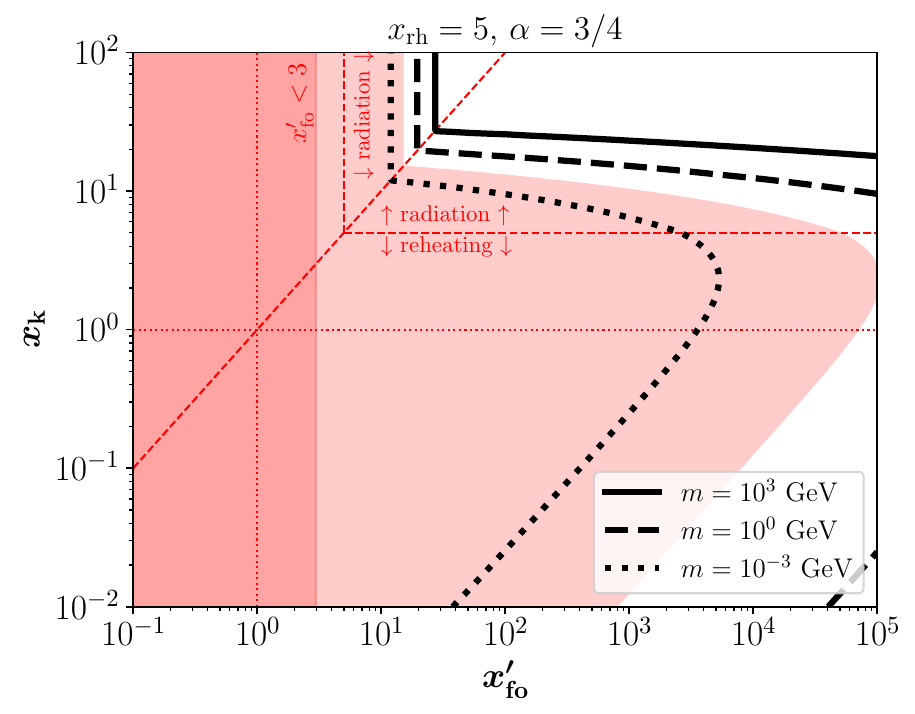}
    \includegraphics[scale=\sepf]{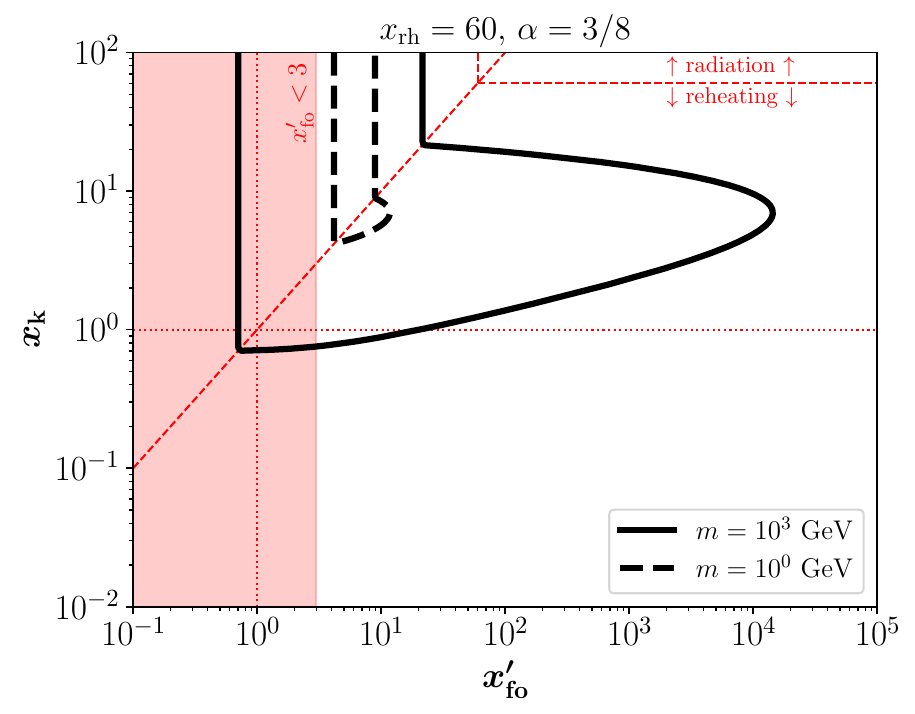}
    \includegraphics[scale=\sepf]{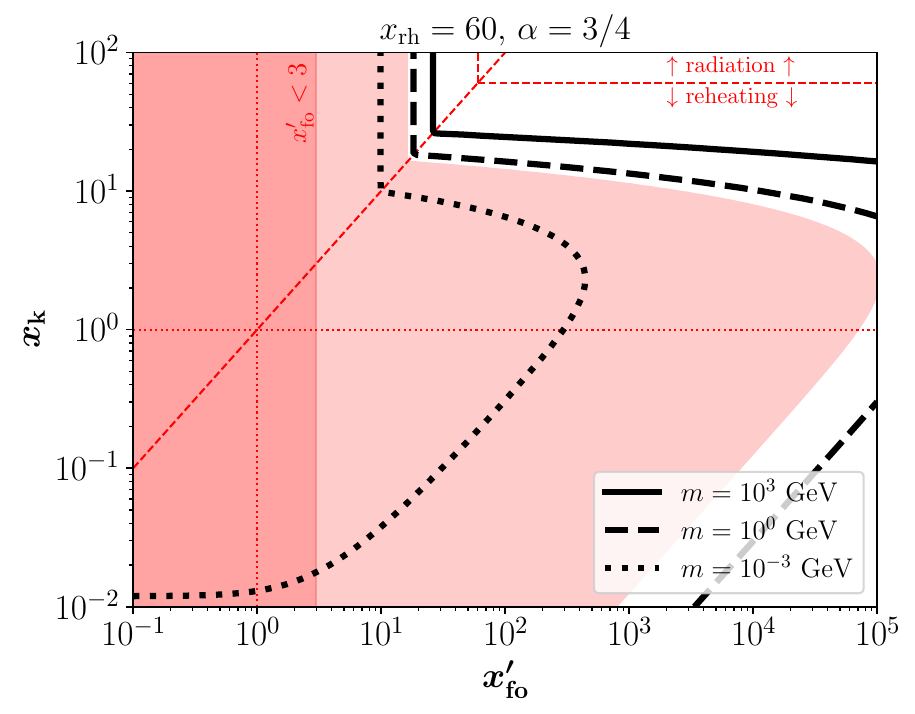}
    \caption{Parameter space that fits the observed DM relic abundance, for different DM masses. For reheating, we assume $\alpha = 3/8$ (left) or $\alpha=3/4$ (right), and $\xrh = 0.5$ (top), 5 (middle) or 50 (bottom). The thin red dotted lines show $\xk = 1$ and $\xpfo = 1$, while the dashed red lines correspond to $\xk = \xpfo$, $\xk = \xrh$, and $\xpfo = \xrh$. The red bands are in tension with BBN or the non-relativistic freeze-out.}
    \label{fig:RH}
\end{figure} 
All in all, the previously discussed cases are summarized in Fig.~\ref{fig:RH}, for $\alpha = 3/8$ (left) or $\alpha=3/4$ (right) and three reheating temperatures $\xrh = 0.2$ (top), $\xrh = 5$ (middle), and $\xrh = 60$ (bottom); it is equivalent to Fig.~\ref{fig:RD} but for low-reheating scenarios. The thin red dotted lines show $\xk = 1$ and $\xpfo = 1$, while the dashed red lines correspond to $\xk = \xpfo$, $\xk = \xrh$, and $\xpfo = \xrh$. The red bands are in tension with BBN or the non-relativistic freeze-out. We notice that while in cases $\xrh = 0.2$ and $\xrh = 5$ there is a single solution during reheating, for $\xrh = 60$ and $\alpha=3/8$ the two solutions are reachable. A fully numerical solution of a case where the two solutions occur is presented in Appendix~\ref{sec:numerics}.

\section{Particle Physics Realization} \label{sec:model}
In this section we give a simple example of a specific particle physics model that realizes the previous DM production mechanisms. In the case of contact interactions, the cross sections for the 2-to-2 annihilation of DM into SM particles $\svann$ and the elastic scattering of a DM particle with a SM particle $\svel$ are given by
\begin{equation} \label{eq:sv22}
    \svel \sim \svann \sim \left[\frac{\eef}{m}\, \frac{K_1(x)}{K_2(x)}\right]^2 \simeq
    \begin{dcases}
        \frac{\eef^2}{4\, T^2} &\text{ for } x \ll 1\,,\\
        \frac{\eef^2}{m^2} &\text{ for } x \gg 1\,,
    \end{dcases}
\end{equation}
as a function of the dimensionless effective coupling $\eef$. Additionally, 3-to-2 DM annihilations are characterized by the thermally averaged cross section
\begin{equation}
    \svcan \sim \frac{\yef^3}{m^5}
\end{equation}
for $x \gg 1$, and depends on the the dimensionless effective coupling $\yef$. Therefore, the corresponding interaction rates are
\begin{align}
    \Gel &= n^\text{sm}_\text{eq}\, \svel\,,\\
    \Gann &= n_\text{eq}\, \svann\,,\\ 
    \Gcan &= n_\text{eq}^2\, \svcan\,, 
\end{align}
where $n_\text{eq}$ is the DM number density in equilibrium given in Eq.~\eqref{eq:neq}, and
\begin{equation}
    n^\text{sm}_\text{eq}(T) = \frac{\zeta(3)}{\pi^2}\, g_\text{sm}\, T^3
\end{equation}
is the equilibrium number density for SM particles, where $g_\text{sm}(T)$ is the number of relativistic SM degrees of freedom that interact with DM (each fermionic degree of freedom counts as 3/4 due to statistics), and $\zeta(3) \simeq 1.2$ is the Riemann zeta function. Here we fix $g_\text{sm} = 2$.

As 3-to-2 DM annihilations are effective, it is expected that elastic 2-to-2 DM self-scattering, with a cross section $\sigma_\text{dm}$ in the non-relativistic limit given by
\begin{equation}
    \sigma_\text{dm} \sim \frac{\yef^2}{m^2},
\end{equation}
is efficient as well. Here, for simplicity, we have assumed the same coupling $\yef$ for the 3-to-2 and 2-to-2 scatterings. The non-observation of an offset between the mass distributions of DM and hot baryonic gas in the Bullet cluster constrains the ratio of the self-scattering cross section over DM mass to be $\sigma_\text{dm}/m < 1.25$~cm$^2$/g~\cite{Clowe:2003tk, Markevitch:2003at, Randall:2008ppe}. Further observations of other cluster collisions reinforce this bound to $\sigma_\text{dm}/m < 0.2$ - $0.5$~cm$^2$/g~\cite{Harvey:2015hha, Bondarenko:2017rfu, Harvey:2018uwf}.

\subsection{After Reheating}
\begin{figure}[t!]
    \def\sepf{0.55}
    \centering
    \includegraphics[scale=\sepf]{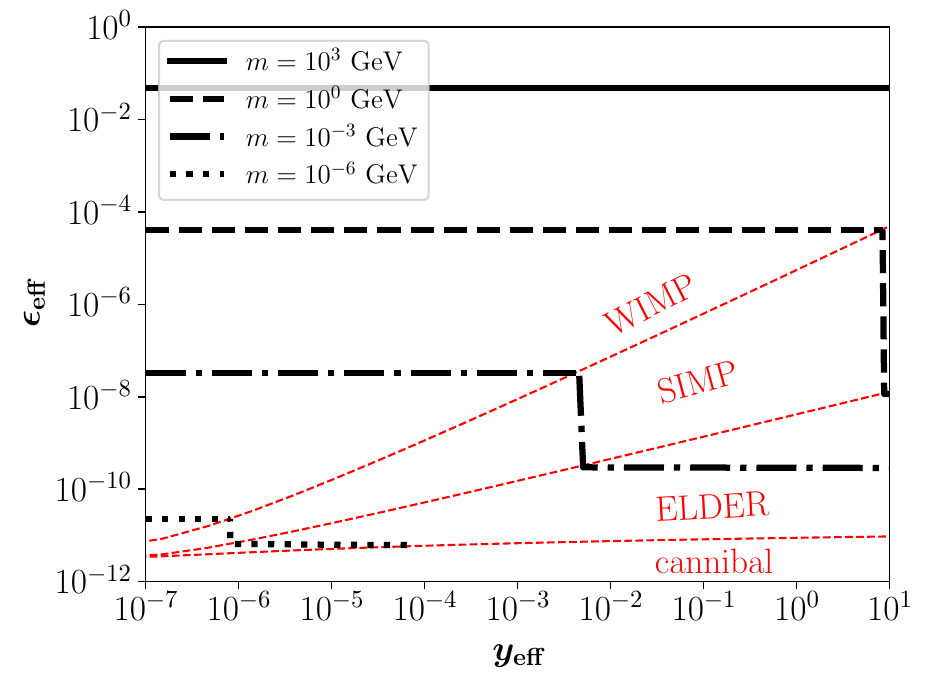}
    \caption{Parameter space that fits the observed DM relic abundance, for a production {\it after} reheating (that is, during the radiation-dominated era) for different DM masses. The same as in Fig.~\ref{fig:RD} but projected in the parameter space $[\yef,\, \eef]$.}
    \label{fig:RD2}
\end{figure}

For the case in which DM is produced well after the end of reheating, Fig.~\ref{fig:RD2} shows with thick black lines the parameter space that fits the entire observed abundance of DM in the case of thermal production after reheating (that is, during radiation domination), for different DM masses. It corresponds to the information in Fig.~\ref{fig:RD} projected in the plane $[\yef,\, \eef]$. The dashed red lines correspond to the boundaries between the different thermal production mechanisms. Several comments are in order: $i)$ The WIMP case depends only on $\eef$, and can be realized for a wide range of masses: from few keV (Lyman-$\alpha$ bound~\cite{Irsic:2017ixq}) to $\sim 130$~TeV (unitarity bound~\cite{Griest:1989wd}). However, there is an upper limit for $\yef$ from which DM becomes a SIMP. $ii)$ Due to the consideration of the perturbativity of $\yef$, the SIMP and ELDER mechanisms can occur only in the sub-GeV range. $iii)$ SIMPs only depend on $\yef$. The upper and lower bounds on $\eef$ appear to avoid the WIMP and ELDER solutions, respectively. $iv)$ ELDERs depend mainly on $\eef$, having only a logarithmic dependence on $\yef$. $v)$ As a result of the choice of the interaction between the visible and dark sectors (i.e., Eq.~\eqref{eq:sv22}), and in particular of the suppression of the cross section at large temperatures, the cannibal mechanism cannot be realized. In fact, kinetic equilibrium, given by the equality $\Gel(\xk) = H(\xk)$, can only occur if $\xk \geq \xkm \simeq 1.8$, implicitly defined by
\begin{equation} \label{eq:xkm-soln-eq-sc}
    2\, {\xkm}^2\, K_0^2(\xkm) + 3\, \xkm\, K_0(\xkm)\, K_1(\xkm) = 2 \left[1 + {\xkm}^2\right] K_1^2(\xkm)\,,
\end{equation}
where $\xkm$ corresponds to the minimal value of $\xk$ required to guarantee kinetic equilibrium between the dark and visible sectors. We remind the unsuspecting reader that $\xk < 1$ is required for the cannibal solution to take place.

\subsection{During Reheating} \label{sec:model-nsc}
\begin{figure}[t!]
    \def\sepf{0.40}
    \centering
    \includegraphics[scale=\sepf]{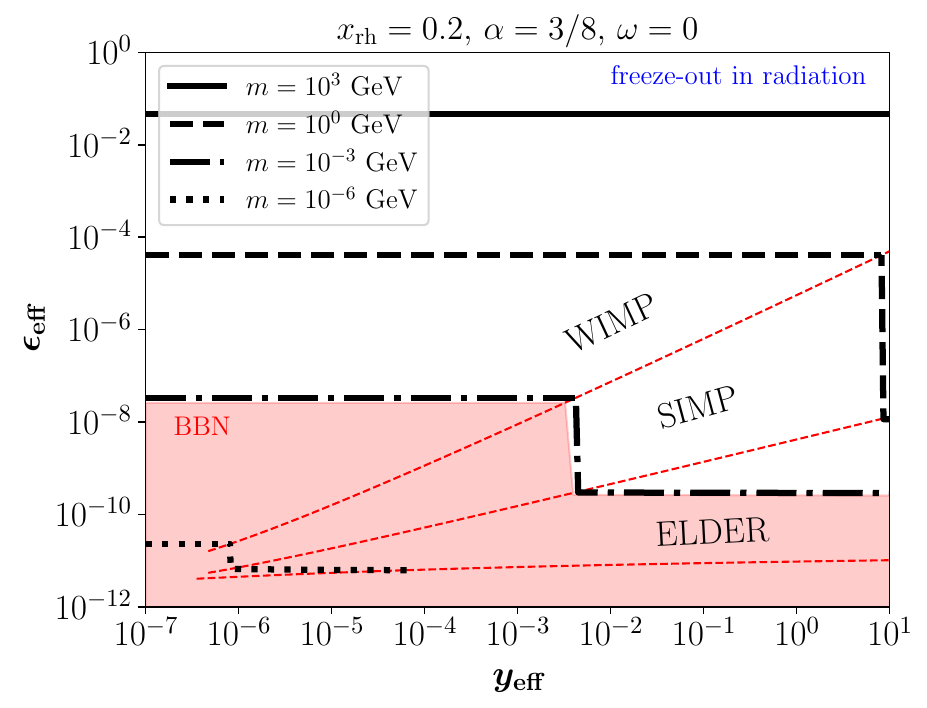}
    \includegraphics[scale=\sepf]{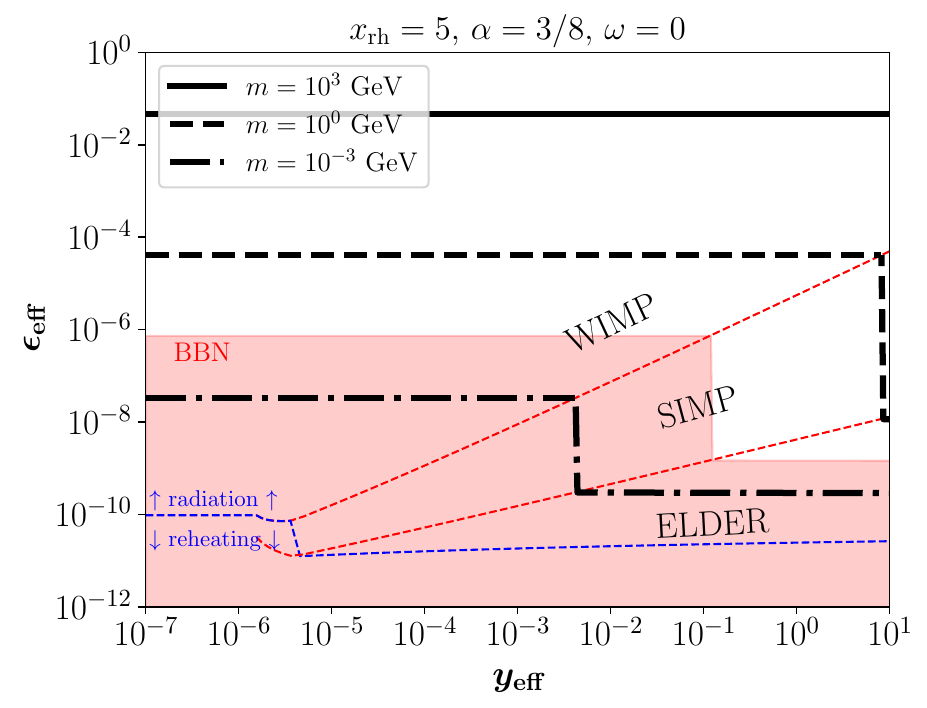}
    \includegraphics[scale=\sepf]{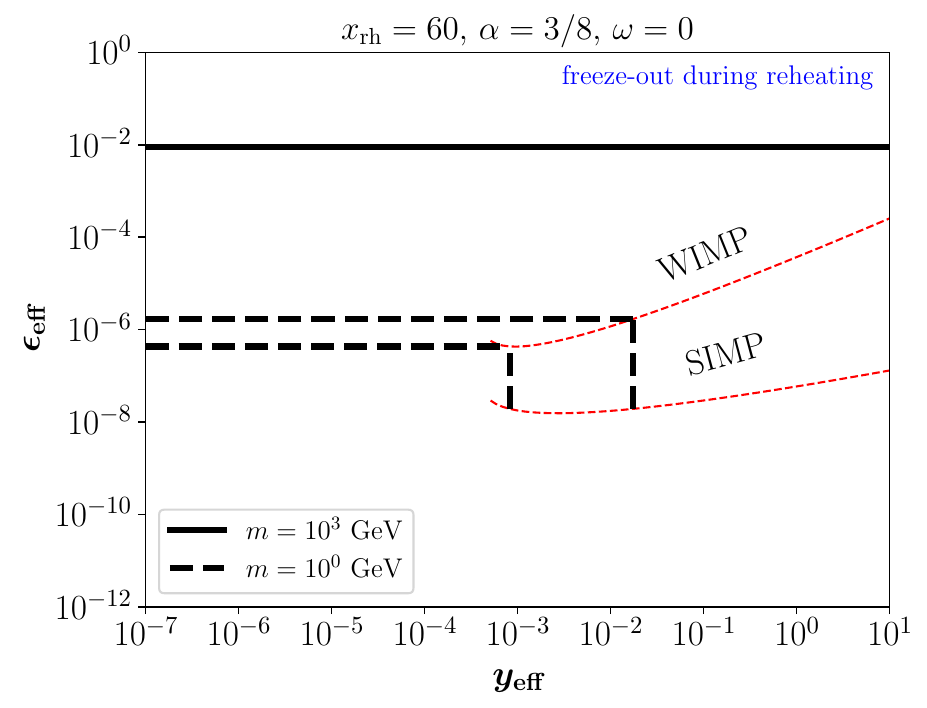}
    \caption{Parameter space that fits the observed DM relic abundance, for $\omega = 0$ and $\alpha = 3/8$. The different panels correspond to $\xrh = 0.2$, 5 and 60. The same as in the left panels of Fig.~\ref{fig:RH} but projected in the parameter space $[\yef,\, \eef]$. In the top-left panel the DM production always occurs during the radiation-dominated era, while in the lower panel, during reheating. However, in the top right panel, it can happen in the two regimes, depending on the DM mass. The transition between the two regimes is shown with a dashed blue line. The red bands are in tension with BBN.}
    \label{fig:EMD}
\end{figure}

Alternatively, for a DM decoupling during reheating, Fig.~\ref{fig:EMD} shows the parameter space that fits the entire observed abundance of DM assuming $\omega = 0$ and $\alpha = 3/8$, with $\xrh = 0.2$ (top left), $\xrh = 5$ (top right), or $\xrh = 60$ (bottom). It compares to Fig.~\ref{fig:RD2} but for different low-temperature reheating scenarios, and corresponds to the information in the left panels of Fig.~\ref{fig:RH} projected in the plane $[\yef,\, \eef]$.

As a first comment, we note that, as expected from Fig.~\ref{fig:RD2}, here the cannibal solution cannot be realized either. In fact, following the same procedure as in the previous section, the minimal value $\xkm$ of $\xk$ required to have kinetic equilibrium between the two sectors is implicitly defined by
\begin{align} \label{eq:xkmin}
    &2 \xkm\, K_1^2(\xkm) + \left[1 + \frac{3(1+\omega)}{2 \alpha}\right] K_1(\xkm)\, K_2(\xkm) \nonumber\\
    &\qquad\qquad\qquad = \xkm\, K_2(\xkm) \left[K_0(\xkm) + K_2(\xkm)\right],
\end{align}
during reheating. We note that if $\xkm$ is higher than $\xrh$, Eq.~\eqref{eq:xkmin} ceases to be valid and therefore $\xkm = \xrh$. Furthermore, the lower bound $\frac{3(1+\omega)}{2 \alpha} \geq 2$, coming from a viable reheating (cf. Eq.~\eqref{eq:noRH}) implies that $\xkm \gtrsim 1.8$ as in the case of radiation domination; cf. Eq.~\eqref{eq:xkm-soln-eq-sc}. We emphasize that, as $\xk > 1$, the cannibal solution cannot be realized in this particle-physics scenario. Moreover, in addition to cannibals, the minimal value of $\xk$ also limits ELDERs. In fact, the ELDER solution during reheating requires that $\xk < \xrh$, and that is only possible in the range $2 \leq \frac{3(1+\omega)}{2 \alpha} < 3$.

\begin{figure}[t!]
    \def\sepf{0.51}
    \centering
    \includegraphics[scale=\sepf]{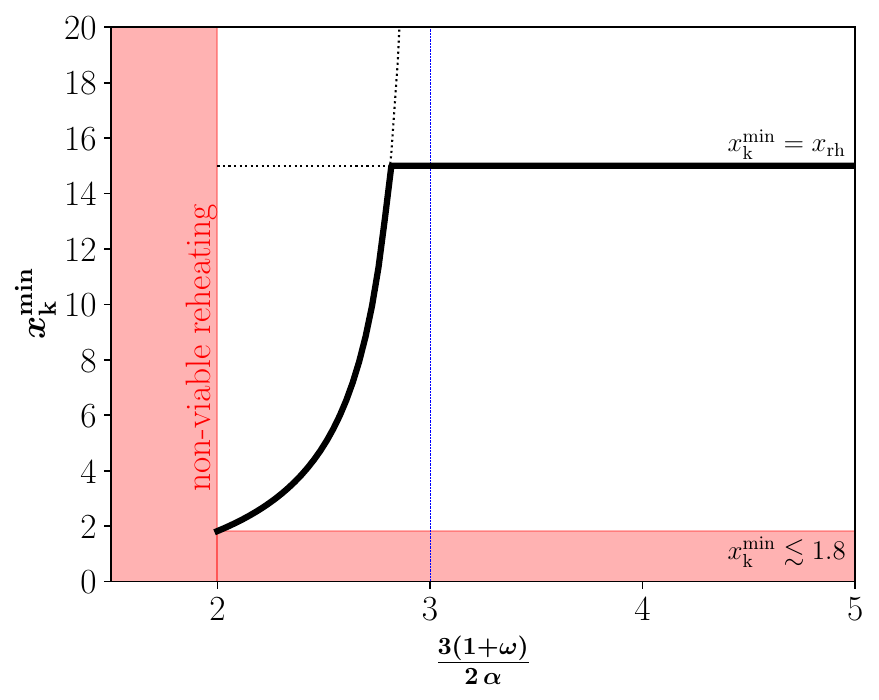}
    \includegraphics[scale=\sepf]{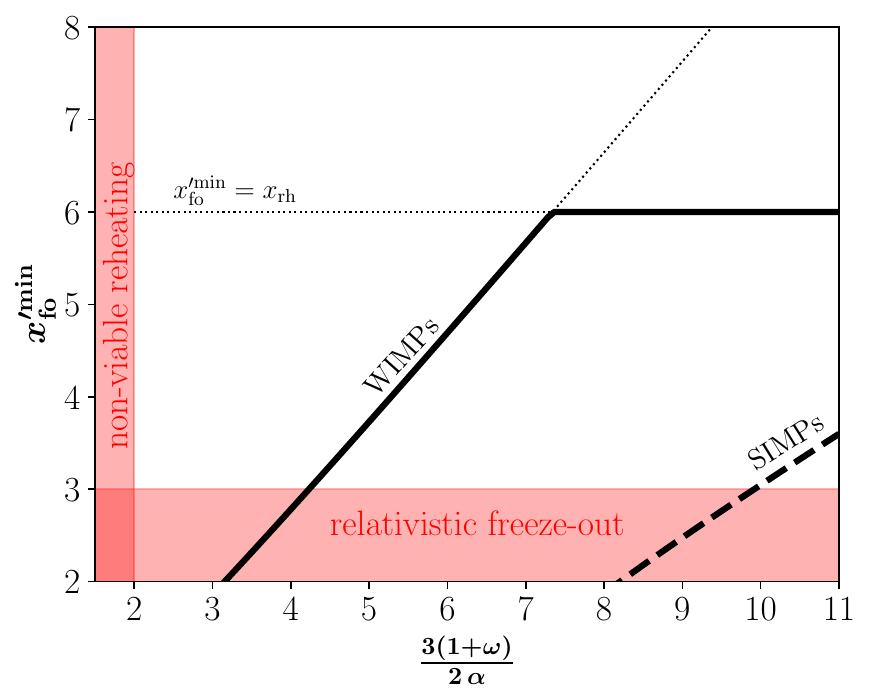}
    \caption{Left: Example of the minimal value $\xkm$ of $\xk$ to have kinetic equilibrium between the dark and visible sectors during reheating (black thick line) for $\xrh=15$. Right: Example of the minimal value $\xpfom$ of $\xpfo$ to have a chemical equilibrium (black thick lines) for $\xrh = 6$. The solid line corresponds to WIMPs, while the dashed line corresponds to SIMPs. For $\frac{3(1+\omega)}{2 \alpha} < 2$ the reheating is not viable, while for $\xpfo < 3$ freeze out is relativistic (red bands).}
    \label{fig:xkmin}
\end{figure}

The left panel of Fig.~\ref{fig:xkmin} shows an example of the value of $\xkm$ (black thick line) as a function of $\frac{3(1+\omega)}{2 \alpha}$ for $\xrh=15$. The black dotted lines correspond to the solution of Eq.~\eqref{eq:xkmin} and to $\xkm = \xrh$. Here again, as a result of the choice of the interaction between the visible and dark sectors, the cannibal mechanism cannot be realized, even during reheating, as $\xk$ can never be smaller than 1. We emphasize that the previous conclusion is valid for {\it any} reheating scenario (any value of $\omega$, $\alpha$ and $\Trh$). Also, it is interesting to note that for this reheating scenario the ELDER solution cannot occur during reheating; it can happen, however, in the radiation-dominated era, as shown in the top panels of Fig.~\ref{fig:EMD}. In the top-left panel the DM freeze-out always occurs during the radiation-dominated era, whereas in the lower panel the DM freeze-out occurs during reheating. However, in the top right panel, it can happen in the two regimes, depending on the DM mass. The transition between the two regimes is shown with a dashed blue line.

In the top-left panel of Fig.~\ref{fig:EMD} ($\xrh = 0.2$), the DM freeze-out occurs in the radiation-dominated era, and therefore the curves coincide with those of Fig.~\ref{fig:RD2}. However, $\xrh$ could correspond to a sufficiently small value of $\Trh$ that rules out masses smaller than $m \lesssim 8 \times 10^{-4}$~GeV due to the BBN constraint (red area). Even higher DM masses are in tension with BBN for higher values of $\xrh$, as seen in the top right panel ($\xrh = 5$), where $m \gtrsim 2 \times 10^{-2}$~GeV. Furthermore, the solution for $m = 10^{-6}$~GeV is not viable, as it is smaller than the minimum mass $m \gtrsim 3\times 10^{-6}$~GeV for which chemical equilibrium is granted.

It is important to note that, in the two top panels of Fig.~\ref{fig:EMD}, only one of the two branches appearing in Fig.~\ref{fig:RH} (the one with higher values of $\xpfo$) can be realized. Interestingly, for even larger values of $\xrh$ (that is, small reheating temperatures), the two branches can occur, as seen in the lower panel of Fig.~\ref{fig:EMD} ($\xrh = 60$). For a given mass, two disconnected solutions for WIMPs and two for SIMPs are realized for different values of the corresponding couplings. In addition, masses $m = 10^{-6}$~GeV and $m = 10^{-3}$~GeV cannot be generated, and the BBN bound disappears because large masses correspond to large reheating temperatures.

Concerning the possibility of the two solutions, recall that the first branch corresponds to values of $\xpfo$ smaller than $\xfoc$, as defined in Eq.~\eqref{eq:xfoc}. However, the existence of the first branch is challenged by the requirement of chemical equilibrium. Taking into account that $\xpfo$ is defined by the equality $H(\xpfo) = \Gann(\xpfo)$ for WIMPs and $H(\xpfo) = \Gcan(\xpfo)$ for SIMPs, the minimal value $\xpfom$ of $\xpfo$ is implicitly defined as
\begin{equation}
    \xpfom\, K_1(\xpfom) + \left[3 - \beta\, \frac{3(1+\omega)}{2 \alpha}\right] K_2(\xpfom) = 0\,,
\end{equation}
with $\beta =1$ for WIMPs and $\beta =1/2$ for SIMPs. The right panel of Fig.~\ref{fig:xkmin} shows an example of $\xpfom$ for WIMPs (solid black line) and SIMPs (dashed black line), as a function of $\frac{3(1+\omega)}{2 \alpha}$ assuming $\xrh = 6$. For choice made in Fig.~\ref{fig:EMD} ($\omega = 0$ and $\alpha = 3/8$), $\frac{3(1+\omega)}{2 \alpha} = 4$ and therefore $\xpfom \simeq 2.8$. We emphasize that even if the first branch is, in general, difficult to realize, it is not impossible to have it. As a sanity cross check, in Appendix~\ref{sec:numerics} a complete numerical calculation was performed, solving the system of Boltzmann equations for the inflaton and SM radiation energy densities and the DM number density. Oscillations at the bottom of a quadratic potential and a constant decay width for the inflaton were assumed, which corresponds to $\omega = 0$ and $\alpha = 3/8$. An example of a point for which the two branches is explicitly shown.

\begin{figure}[t!]
    \def\sepf{0.50}
    \centering
    \includegraphics[scale=\sepf]{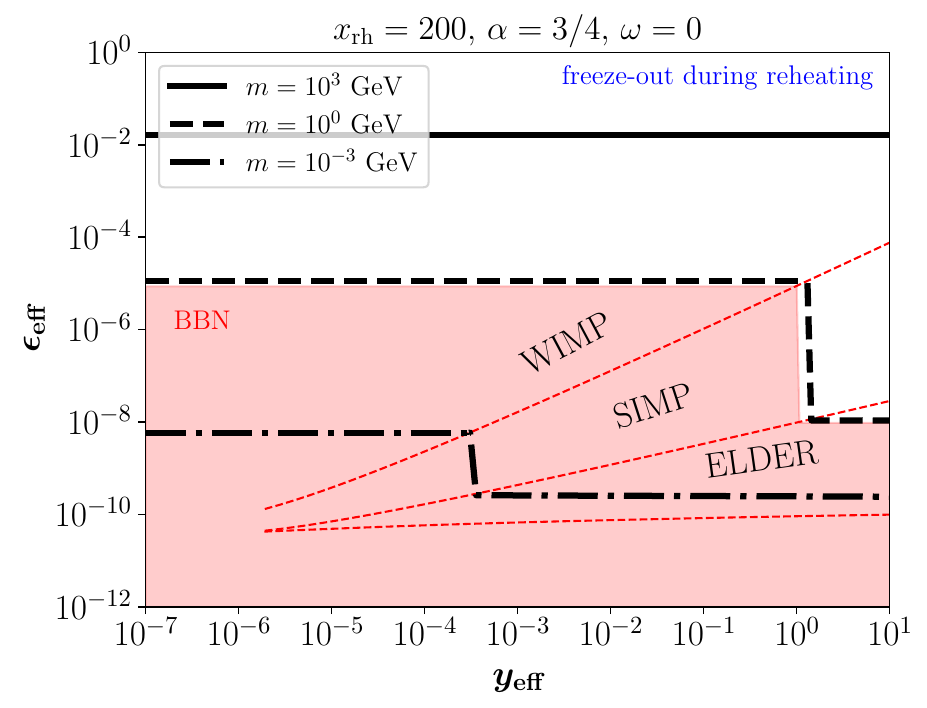}
    \caption{Parameter space that fits the observed DM relic abundance, for $\omega = 0$, $\alpha = 3/4$, and $\xrh = 200$. The red band is in tension with BBN. Here, DM production always occurs during reheating.}
    \label{fig:al3o5}
\end{figure}

In Fig.~\ref{fig:EMD}, corresponding to $\omega=0$ and $\alpha=3/8$, the ELDER solution cannot be realized during reheating, as $\frac{3(1+\omega)}{2\alpha} > 3$. However, it can occur in the range $2 \leq \frac{3(1+\omega)}{2\alpha} < 3$. For example, Fig.~\ref{fig:al3o5} shows the parameter space that fits the observed DM relic abundance for $\omega = 0$, $\alpha = 3/4$ and $\xrh = 200$. Here, the ELDER solution is clearly realized (and compatible with BBN) for DM masses above the GeV scale.

In the following subsections, the three viable DM production mechanisms for this given particle-physics scenario will be analyzed in detail.

\subsubsection{WIMPs}
\begin{figure}[t!]
    \def\sepf{0.49}
    \centering
    \includegraphics[scale=\sepf]{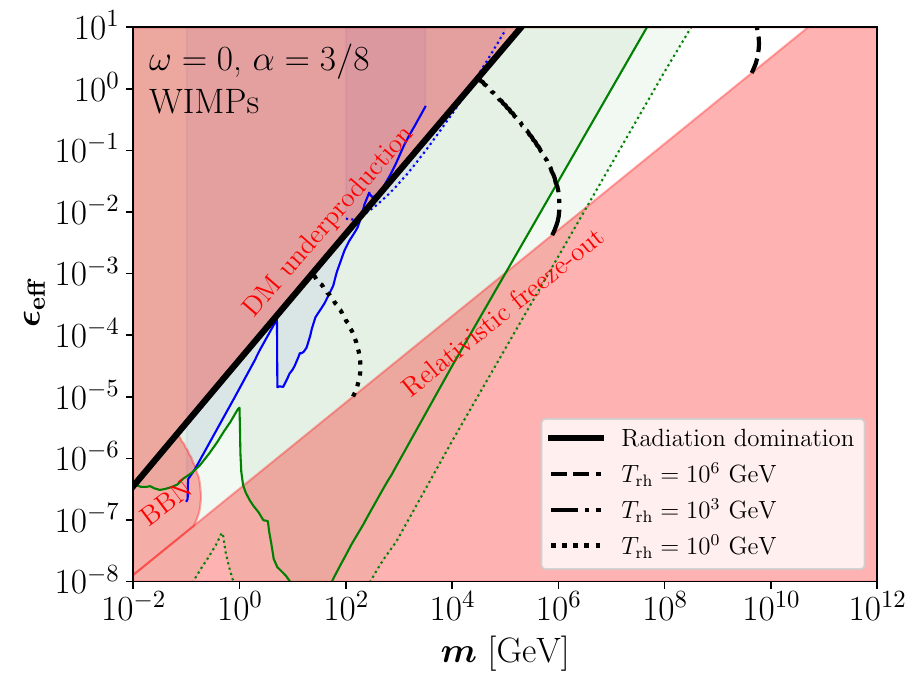}
    \includegraphics[scale=\sepf]{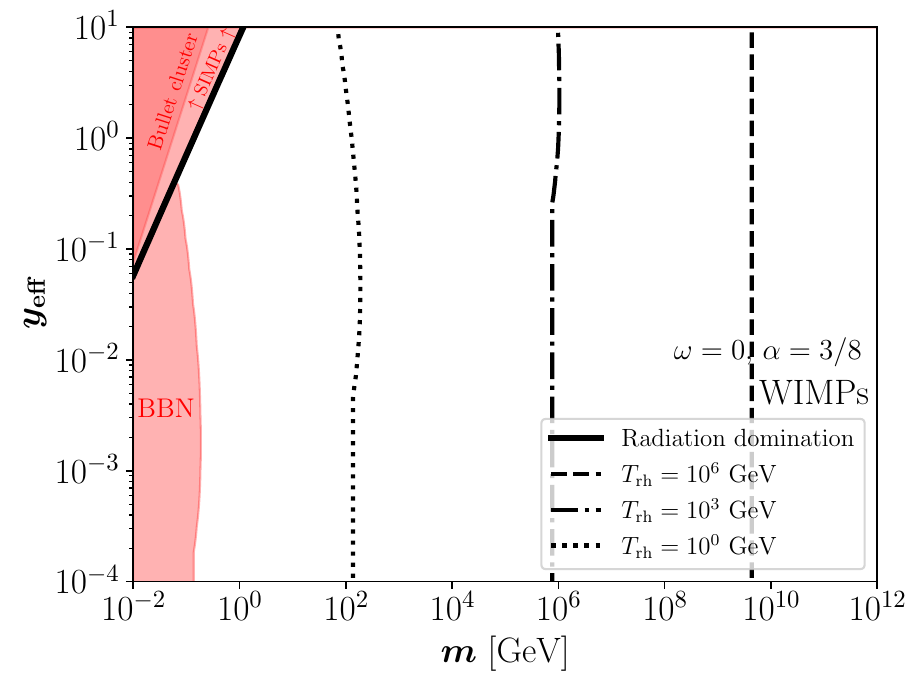}
    \includegraphics[scale=\sepf]{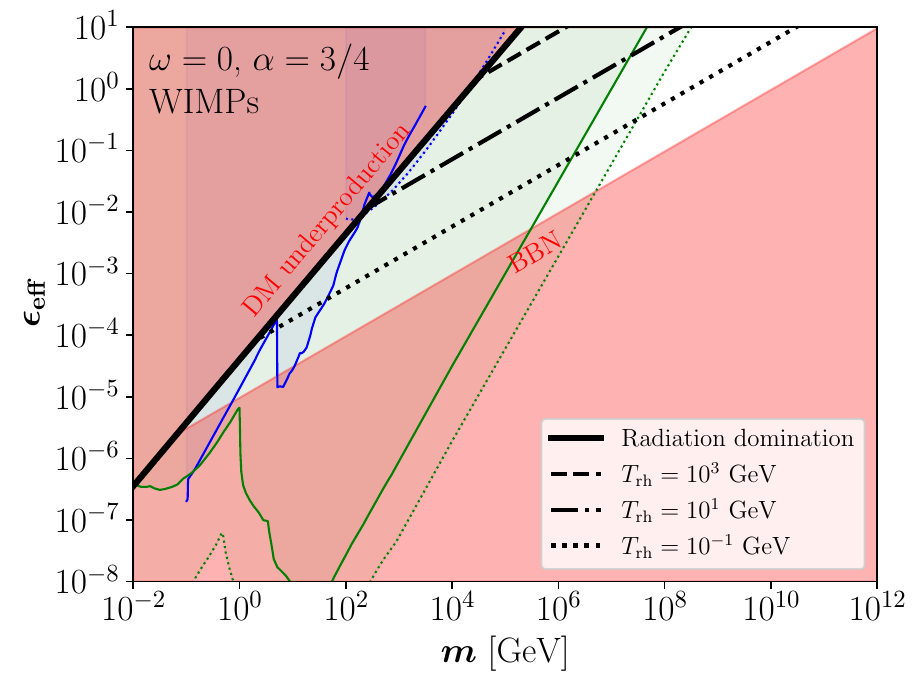}
    \includegraphics[scale=\sepf]{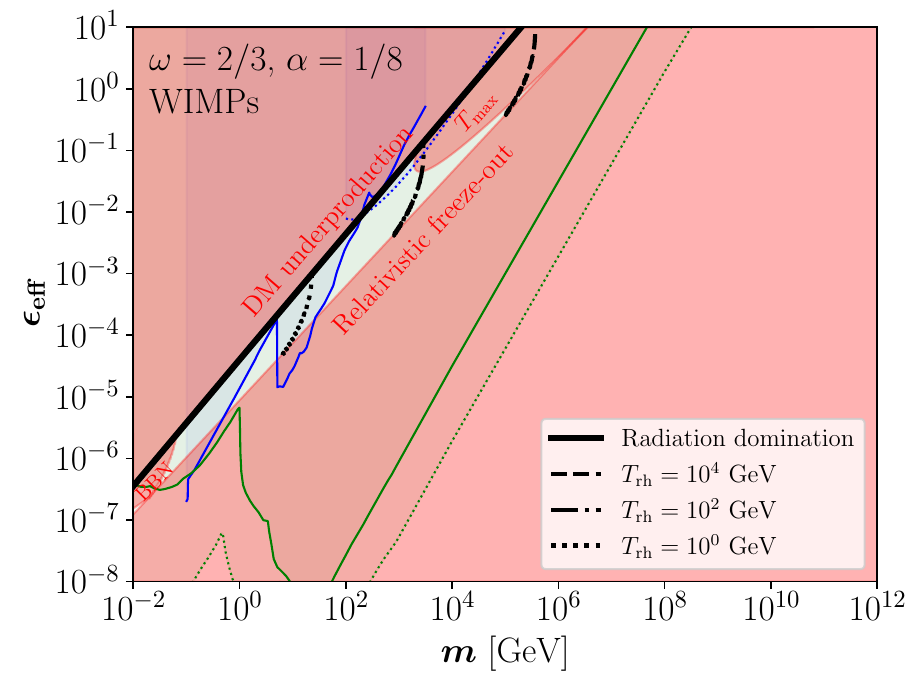}
    \caption{WIMPs. Top left and bottom: Parameter space $[m,\, \eef]$ required to fit the observed DM relic abundance for $\omega = 0$ and $\alpha = 3/8$ (top left), $\omega = 0$ and $\alpha = 3/4$ (bottom left), and $\omega = 2/3$ and $\alpha = 1/8$ (bottom right), with different reheating temperatures. Top right: Allowed range of $\yef$ values required to guarantee a WIMP solution, for $\omega = 0$ and $\alpha = 3/8$. The red bands are excluded because of DM underproduction, BBN, relativistic DM freeze-out, the maximum temperature of the SM bath, Bullet cluster, or the SIMP solution. The blue and green regions correspond to bounds (solid borders) or projections (dotted borders) from DM indirect and direct detection, respectively.}
    \label{fig:wimpsimp-nsc-nb}
\end{figure} 
In addition to the broad parameter-space scans shown in Fig.~\ref{fig:EMD}, one can also perform a specific analysis for the different production mechanisms. The top left and the lower panels of Fig.~\ref{fig:wimpsimp-nsc-nb} show the values of $\eef$ required to fit the entire observed abundance of WIMP DM for different reheating temperatures. The thick black lines correspond to the usual high-temperature reheating scenario, where DM freezes out in the radiation-dominated era. As expected, the perturbativity bound on $\eef$ restricts the WIMPs to be lighter than $\sim 130$~TeV~\cite{Lee:1977ua}. Higher couplings produce a later chemical decoupling and, therefore, a DM underabundance. However, smaller couplings and larger masses can be explored in low-reheating scenarios because of the injection of entropy. In the case of $\omega = 0$ and $\alpha = 3/8$ (top left panel), the maximum WIMP mass is $m \sim \mathcal{O}(10^{11})$~GeV~\cite{Berlin:2016vnh, Bernal:2022wck, Bernal:2023ura, Coy:2024itg}, and is limited by perturbativity ($\eef \lesssim 10$) and the condition of having a non-relativistic freeze-out ($\xpfo > 3$). In other cosmological scenarios, for example, if $\omega = 0$ and $\alpha = 3/4$ (bottom left panel), DM can reach up to $m \sim \mathcal{O}(10^{12})$~GeV and this time it is bounded by perturbativity and the BBN limit on $\Trh$. A different scenario occurs for small values of $\alpha$. We remind unsuspecting readers that the parameter $\alpha$ controls the slope to the SM temperature during reheating (cf. Eq.~\eqref{eq:Ta}), and therefore small values of $\alpha$ imply a small separation between $\Trh$ and $\Tmax$. This is a potential issue, as DM to be a {\it thermal} relic must have a mass $m < \Tmax$. In Appendix~\ref{sec:Tmax} a detailed derivation of the maximal temperature $\Tmax$ reached by the thermal bath is presented. That bound is shown in the bottom right panel of Fig.~\ref{fig:wimpsimp-nsc-nb}, for $\omega = 2/3$ and $\alpha = 1/8$. In this particular case, the upper bound on the DM mass is $m \sim \mathcal{O}(10^6)$~GeV.

WIMP solutions require the dominance of DM annihilation into SM particles over DM self-annihilation. This can be guaranteed if the self-coupling interaction rate $\Gcan$ is sufficiently suppressed. The top right panel of Fig.~\ref{fig:wimpsimp-nsc-nb} shows the maximal allowed value for $\yef$ to have a WIMP solution. For higher values of $\yef$, the self-annihilation interaction rates of DM dominate, rendering DM a SIMP relic. Interestingly, beyond $\mathcal{O}$(1)~MeV, $\yef$ could be arbitrarily large (within the perturbative range) and still have a subdominant interaction rate. This is due to the strong Boltzmann suppression factor in the equilibrium DM number density.

We comment on some present and future possibilities of testing the current scenario using the different channels offered by DM indirect detection experiments. In Fig.~\ref{fig:wimpsimp-nsc-nb} we overlay in blue (solid lines) present constraints coming from: $i)$ CMB spectral distortions from WIMP annihilation into charged particles~\cite{Leane:2018kjk}, $ii)$ the combined analysis of $\gamma$-ray data using Fermi-LAT, HAWC, H.E.S.S., MAGIC, and VERITAS~\cite{Hess:2021cdp}, and $iii)$ AMS-02 measurements of antimatter in cosmic rays, in the positron~\cite{Bergstrom:2013jra} and the antiproton~\cite{Calore:2022stf} channels.\footnote{It is interesting to recall that the bounds presented depend on the DM annihilation channels, and suffer from large uncertainties arising from e.g. assumptions on the DM density profile, the local DM density, and the propagation of charged particles in the interstellar medium~\cite{Zemp:2008gw, Pato:2010yq, Bernal:2014mmt, Boudaud:2014dta, Bernal:2015oyn, Bernal:2016guq, Benito:2016kyp}.} Additionally, Fig.~\ref{fig:wimpsimp-nsc-nb} also shows with blue dotted lines the projected sensitivity of the ground-based CTA experiment, which could be capable of testing WIMP DM at the TeV scale and above~\cite{CTA:2020qlo}. Assuming that the DM is coupled to all SM particles with the same contact interaction defined by Eq.~\eqref{eq:sv22}, direct detection bounds from the elastic scattering of the DM with both electrons and nucleons become important and are shown by the shaded light green region (solid lines). The sub-GeV DM masses are constrained by the DM-electron elastic scattering, where the most stringent bounds come from XENON10, XENON100~\cite{Essig:2017kqs} and DarkSide-50~\cite{DarkSide:2022knj}. Higher DM masses are constrained by bounds from DM-nucleon elastic scattering probes; the most stringent being DarkSide-50~\cite{DarkSide-50:2022qzh}, XENON1T~\cite{XENON:2017vdw} and LZ~\cite{LZ:2022lsv}. The future projected sensitivities are indicated by the dotted light-green line, where the maximum reach for DM-electron scattering can be achieved by DAMIC and SuperCDMS~\cite{Essig:2015cda}, while DARWIN~\cite{DARWIN:2016hyl}, ARGO and DarkSide-20k~\cite{Billard:2021uyg} can attain the maximum reach for DM-nucleon scattering. It is interesting to note that actual direct and indirect DM detection experiments are already probing regions of parameter space where DM is produced deep in the reheating era. In particular, the scenario presented in the bottom right plot of Fig.~\ref{fig:wimpsimp-nsc-nb} is completely ruled out. Finally, we also note that, as expected, all direct and indirect bounds and projections are the same in all panels of Fig.~\ref{fig:wimpsimp-nsc-nb}, because they depend only on the local details of the DM and not on its cosmological evolution.

\begin{figure}[t!]
    \def\sepf{0.50}
    \centering
    \includegraphics[scale=\sepf]{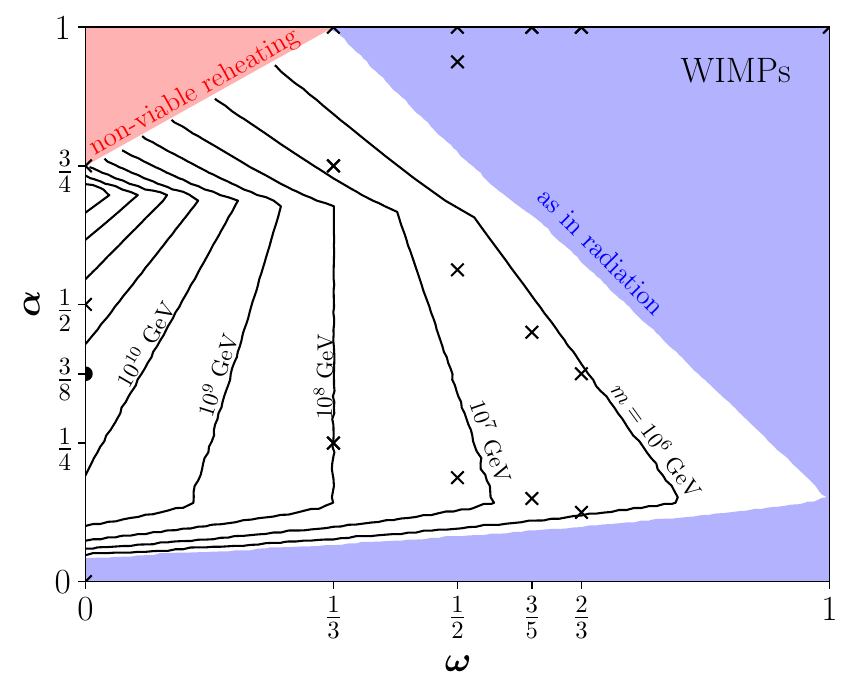}\\
    \includegraphics[scale=\sepf]{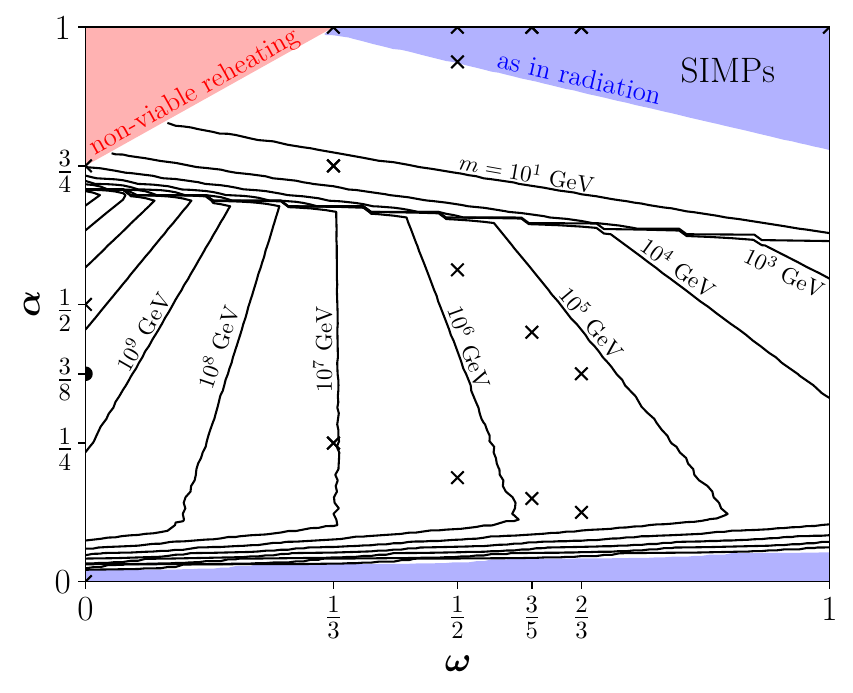}
    \includegraphics[scale=\sepf]{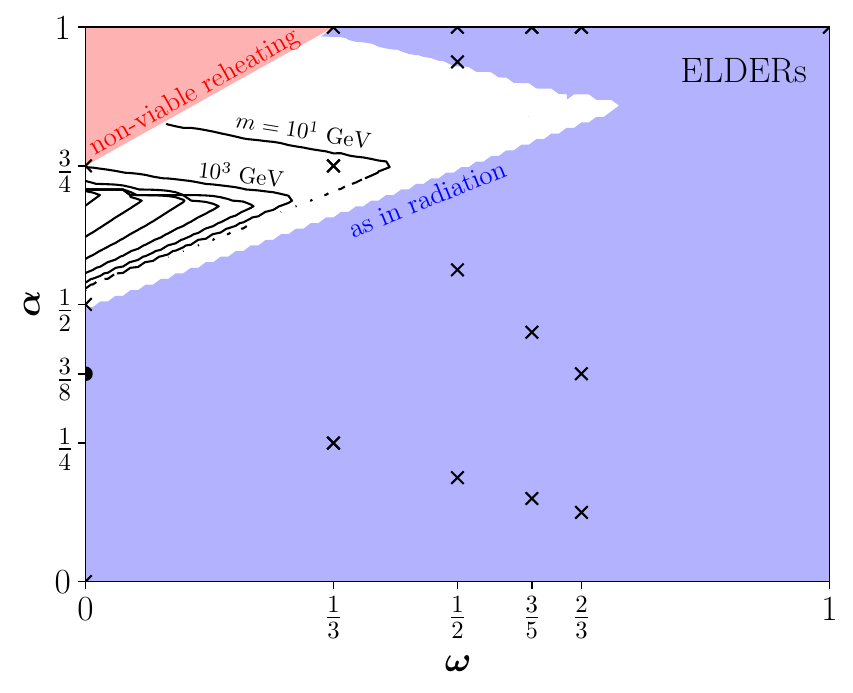}
    \caption{Contours for the maximal DM mass that can be produced by the different mechanisms: WIMP from $10^6$~GeV until $10^{14}$~GeV (top), SIMP from $10^1$~GeV until $10^{13}$~GeV (bottom left) and ELDER from $10^1$~GeV until $10^{13}$~GeV (bottom right). In the blue areas the maximal DM mass is the same as in radiation era, while in the red bands reheating is not viable.}
    \label{fig:maxwimp}
\end{figure} 
A complete analysis of the maximal mass that can be produced by the WIMP mechanism compatible with all constraints is shown in the top panel of Fig.~\ref{fig:maxwimp}, using the same parameter space presented in Fig.~\ref{fig:cosmo}. In particular, we ensure that $\Trh > T_\text{bbn}$, $\xpfo > 3$, $m < \Tmax$, $\eef < 10$, and $\xk > \xpfo$. In the figure, the contours show three behaviors: $i)$ For $0.7 \lesssim \alpha < 1$, the maximal mass is set by the combination of perturbativity and BBN (cf. left bottom panel of Fig.~\ref{fig:wimpsimp-nsc-nb}), $ii)$ for $0.15 \lesssim \alpha \lesssim 0.7$, it comes from perturbativity and $\xpfo > 3$ (cf. top panel of Fig.~\ref{fig:wimpsimp-nsc-nb}), while $iii)$ for $0 < \alpha \lesssim 0.15$ from $\Tmax$ (cf. right bottom panel of Fig.~\ref{fig:wimpsimp-nsc-nb}). The maximum WIMP mass is $m \simeq 5 \times 10^{14}$~GeV and occurs near $\omega \simeq 0$ and $\alpha \simeq 0.7$. Finally, we note that in the blue area the maximum DM mass is the same as in radiation domination.

\subsubsection{SIMPs}
\begin{figure}[t!]
    \def\sepf{0.49}
    \centering
    \includegraphics[scale=\sepf]{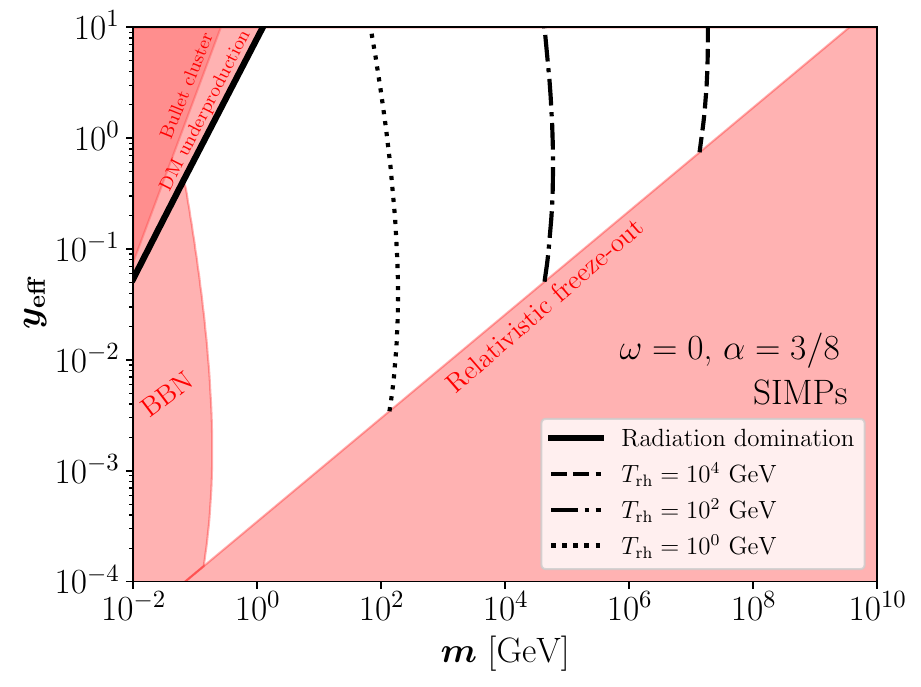}
    \includegraphics[scale=\sepf]{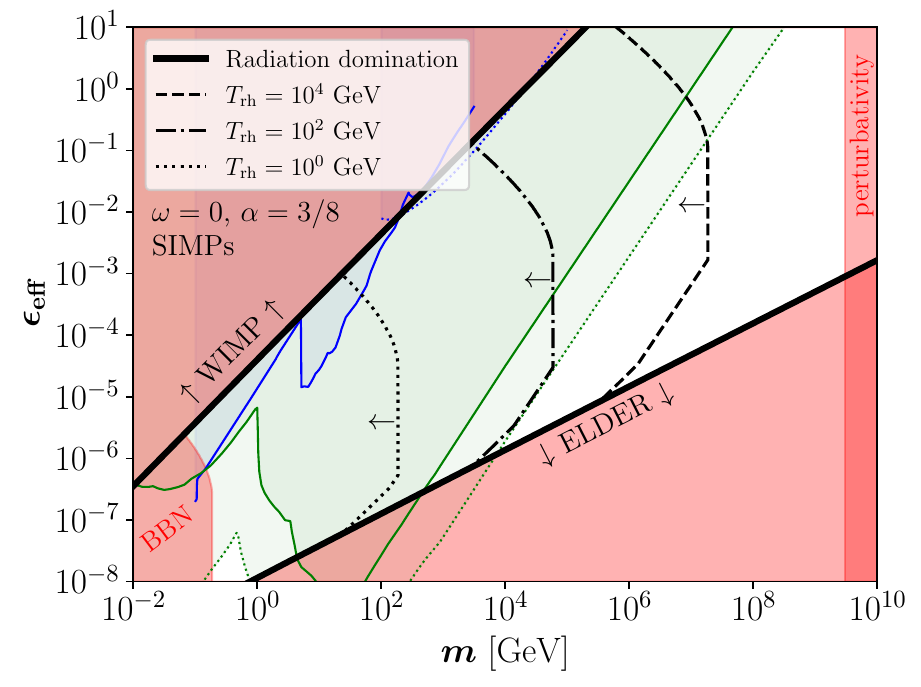}
    \caption{SIMPs. Left: Parameter space $[m,\, \yef]$ that fits the observed DM relic abundance for $\omega = 0$ and $\alpha = 3/8$ and different reheating temperatures. Right: Maximal and minimal values of $\eef$ required to have a SIMP solution. The red bands are excluded because of DM underproduction, BBN, relativistic DM freeze-out, or data from the Bullet cluster. The blue and green regions correspond to bounds (solid borders) or projections (dotted borders) from DM indirect and direct detection, respectively.}
    \label{fig:wimpsimp-nsc}
\end{figure}

In the left panel of Fig.~\ref{fig:wimpsimp-nsc}, the parameter space that accounts for the entire observed abundance of DM through the SIMP mechanism is depicted with thick black lines for various reheating temperatures, assuming $\omega = 0$ and $\alpha=3/8$. It compares to Fig.~\ref{fig:wimpsimp-nsc-nb} but for the SIMP scenario. The solid black line represents the conventional high-temperature reheating scenario, where DM undergoes a chemical freeze-out through 3-to-2 annihilations during the radiation-dominated epoch. As anticipated, the perturbativity constraint on the coupling limits the SIMP DM to masses below approximately $\sim \mathcal{O}(1)$~GeV. Larger couplings result in delayed chemical decoupling, leading to a DM underabundance. Conversely, smaller couplings and larger masses can be investigated in low-reheating scenarios because of the entropy injection. It is interesting to note that in this case, SIMP DM can be as heavy as $\sim \mathcal{O}(10^9)$~GeV with perturbative couplings, being limited by the requirement of a non-relativistic freeze-out. Additionally, light DM particles below the GeV ballpark, with sizable self-interactions, are in conflict with Bullet cluster data. 

Even if the abundance of SIMP DM is fixed by the self-scattering coupling $\yef$, there are consistency bounds on the interaction of the DM with the SM from above and below: too large values for $\eef$ could make the DM a WIMP, while too small values for $\eef$ could make it an ELDER (cf. Fig.~\ref{fig:EMD}). The right panel of Fig.~\ref{fig:wimpsimp-nsc} shows the interval in which $\eef$ can vary while being compatible with a SIMP solution. Interestingly, for low-temperature reheating, such a range shrinks. Furthermore, the red vertical band on the right panel corresponds to the maximal mass allowed by perturbativity on the $\yef$ coupling. Finally, we also overlay direct and indirect bounds and projections in the parameter space $[m,\, \eef]$. It can be seen that a large fraction of the parameter space favored by low-temperature reheating is already in tension, especially with DM direct detection data, and even a larger fraction could be probed in next-generation experiments.

It is important to warn the reader that the apparent overlap of the parameter space for WIMPs and SIMPs in Figs.~\ref{fig:wimpsimp-nsc-nb} and~\ref{fig:wimpsimp-nsc} is just an artifact of the projection of the full 4-dimensional parameter space ($m$, $\eef$, $\yef$ and $\Trh$) in the 2-dimensional planes $[m,\, \yef]$ and $[m,\, \eef]$.

The lower left panel of Fig.~\ref{fig:maxwimp} shows the contours of the maximal SIMP DM mass that can be produced in a low-temperature reheating scenario. The behavior of these curves resembles the one in the WIMP case (cf. the top panel of Fig.~\ref{fig:maxwimp}). The contours show again three behaviors: $i)$ For $0.7 \lesssim \alpha < 1$, the maximal mass is set by the combination of perturbativity and BBN, $ii)$ for $0.15 \lesssim \alpha \lesssim 0.7$, it comes from perturbativity and $\xpfo > 3$, while $iii)$ for $0 < \alpha \lesssim 0.15$ from $\Tmax$. The maximum SIMP mass is $m \simeq 3 \times 10^{13}$~GeV and occurs near $\omega \simeq 0$ and $\alpha \simeq 0.7$. Finally, we note that in the blue area the maximum DM mass is the same as in radiation domination.

\subsubsection{ELDERs}
\begin{figure}[t!]
    \def\sepf{0.49}
    \centering
    \includegraphics[scale=\sepf]{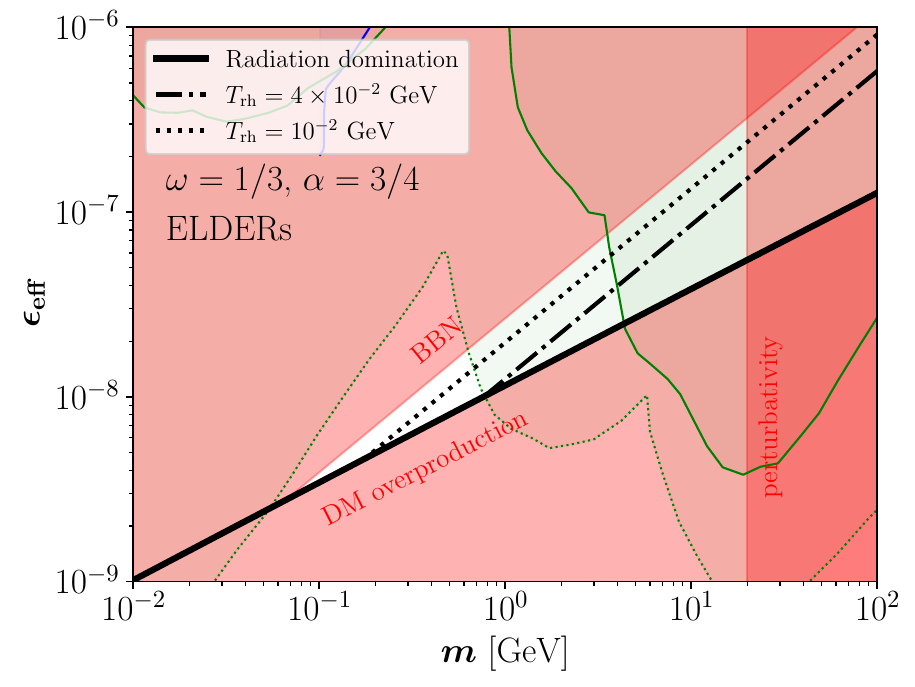}
    \includegraphics[scale=\sepf]{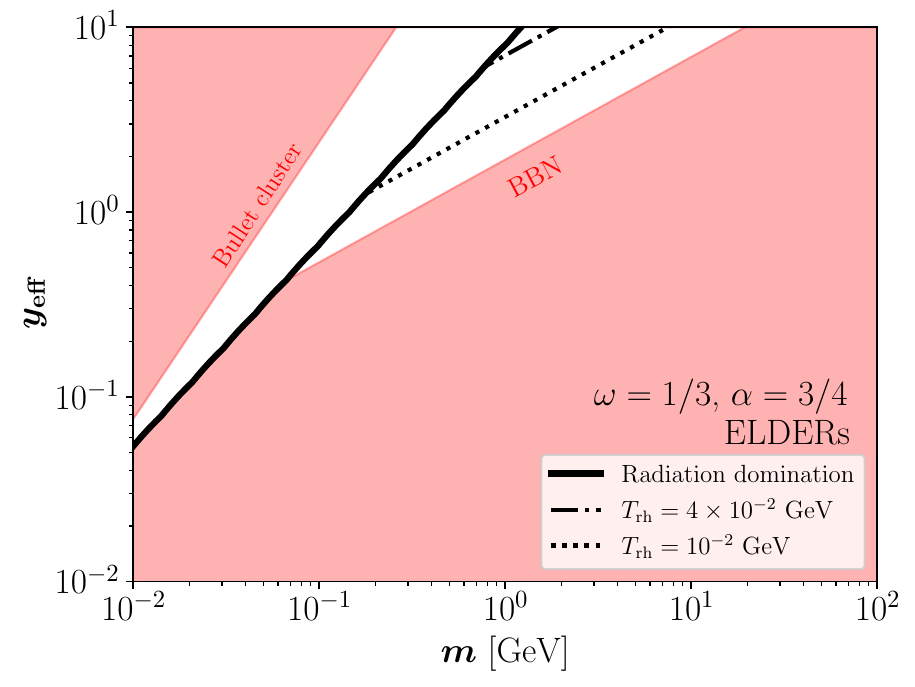}
    \caption{ELDERs. Left: Parameter space that fits the observed DM relic abundance, for $\omega = 1/3$, $\alpha = 3/4$ and different reheating temperatures. Right: Minimal coupling $\yef$ compatible with the ELDER solution. The red bands are excluded because of DM overproduction, BBN or data from the Bullet cluster. The blue and green regions correspond to bounds (solid borders) or projections (dotted borders) from DM indirect and direct detection, respectively.}
    \label{fig:elder-nsc}
\end{figure}

The left panel of Fig.~\ref{fig:elder-nsc} shows the required values of $\eef$ to fit the entire observed abundance of DM, for the ELDER mechanism, with different reheating temperatures, and assuming $\omega = 1/3$ and $\alpha = 3/4$. Even during the radiation-dominated era, it is expected that for a single DM mass a range of $\eef$ is viable. However, the mass dependence is logarithmic and, de facto, the range collapses to a single point; cf. Fig.~\ref{fig:RD2}. Furthermore, and as expected from the top right panel of Fig.~\ref{fig:wimpsimp-nsc}, a low-temperature reheating allows exploration of larger values for $\eef$. Direct detection bounds rule out a large part of the parameter space, and future experiments can probe the allowed regions almost entirely. The right panel of Fig.~\ref{fig:elder-nsc} shows the minimal value of the self-coupling $\yef$ compatible with the ELDER solution; smaller values would give rise to a WIMP. In radiation domination, the perturbative bound of $\yef$ limits the DM mass to be below the GeV ballpark. For MeV DM masses, the Bullet cluster limits high values of the coupling. However, small couplings become viable for low-temperature reheating scenarios, increasing the DM mass window to $m \simeq 20$~GeV.

Finally, the lower right panel of Fig.~\ref{fig:maxwimp} shows the contours of the maximal ELDER mass that can be produced. The curves are basically the same as in the SIMP case (cf. the lower left panel of Fig.~\ref{fig:maxwimp}), with the big difference that, to have a DM genesis in the low-temperature reheating era, one has to guarantee that $\frac{3(1+\omega)}{2\alpha} < 3$; cf. the left panel of Fig.~\ref{fig:xkmin}. For higher values, the ELDER solution can naturally take place, but DM is produced in the radiation-dominated era, which explains the size of the blue area. As in the case of SIMPs, the maximum ELDER mass is $m \simeq 3 \times 10^{13}$~GeV and occurs near $\omega \simeq 0$ and $\alpha \simeq 0.7$.

\section{Conclusions} \label{sec:concl}
Despite extensive experimental efforts during the last few decades, the nature of dark matter (DM) continues to be a mystery. Specifically, the hypothesis that DM consists of weakly interacting massive particles (WIMPs) that achieve thermal equilibrium with the standard model (SM) degrees of freedom has garnered substantial theoretical and experimental interest. However, no definitive evidence has emerged to support the existence of WIMP DM. This lack of conclusive results has led to the exploration of other mechanisms. On the thermal side, alternative production mechanisms exist; DM could be a strongly interacting massive particle (SIMP), an ELastically DEcoupling Relic (ELDER) or a cannibal. In these new paradigms, self-interactions within the DM sector play a crucial role in determining the abundance of DM. The key differences lie in the timing of chemical and kinetic freeze-out. For SIMPs, chemical freeze-out occurs while kinetic equilibrium is maintained, similar to that of WIMPs. In contrast, for ELDERs and cannibals, chemical freeze-out occurs after kinetic equilibrium is broken, with the latter occurring while the DM is non-relativistic or relativistic, respectively.

It is crucial to understand that the current abundance of DM is influenced not just by particle-physics dynamics but also by the cosmological history of the Universe. Given that the early Universe's evolution remains largely uncertain, the conventional assumption is a Universe primarily governed by SM radiation from the end of cosmological inflation up to matter-radiation equality, with a sudden cosmic reheating happening at a very high SM temperature. Here, this picture is not taken for granted. In fact, we allowed noninstantaneous reheating and analyzed its impact on the thermal genesis of the DM.

The focus of this study has been on understanding the viability and implications of these thermal DM candidates in low-reheating scenarios, with careful consideration of their kinetic and/or chemical decoupling. We have studied general reheating scenarios by parameterizing the equation of state of the inflaton during reheating and the dependence of the SM temperature with the scale factor of the Universe. For the standard case of thermal DM production after reheating, we obtained in a model-independent way the region of parameter space that fits the total observed DM abundance for each type of DM candidate. We have then shown how these regions are modified when DM production occurs during reheating and identified the regions in tension with cosmological and laboratory constraints. For the WIMP, SIMP and ELDER mechanisms and for a given cosmology, two solutions that fit the entire observed DM abundance can occur: one during reheating and the other during the radiation-dominated era, or even both during reheating, as shown in Fig.~\ref{fig:RH}.

We subsequently studied the implications for a simple particle-physics model, assuming contact operators for DM self-interactions and DM interactions with SM states, focusing on DM production both after and during reheating. In both cases, cannibal solutions are not viable as a result of strong temperature suppression when the DM is relativistic. Moreover, in this specific realization, WIMP, SIMP, and ELDER solutions become further constrained from BBN, and the possibility of having hot DM. At the same time, low reheating temperatures render viable larger mass values for the DM mass, up to the $10^{14}$~GeV ballpark, well beyond the usual limit of $130$~TeV for WIMPs and a few MeVs for SIMPs and ELDERs; see Fig.~\ref{fig:maxwimp}. Additionally, broader ranges for coupling values are now feasible, expanding the parameter space compared to the DM genesis during radiation domination. We have also shown how current direct and indirect limits on DM constrain significant parts of the allowed parameter space even in low-temperature reheating scenarios. Finally, we showed that upcoming direct detection experiments will be able to probe a substantial part of the remaining parameter space currently allowed.

\section*{Acknowledgments}
The authors thank Leszek Roszkowski for valuable discussions, and Panos Oikonomou for his work in the preliminary stages of this project. NB received funding from the Spanish FEDER / MCIU-AEI under the grant FPA2017-84543-P.

\appendix
\section{Numerical Validation} \label{sec:numerics}
\begin{figure}[t!]
    \def\sepf{0.48}
    \centering
    \includegraphics[scale=\sepf]{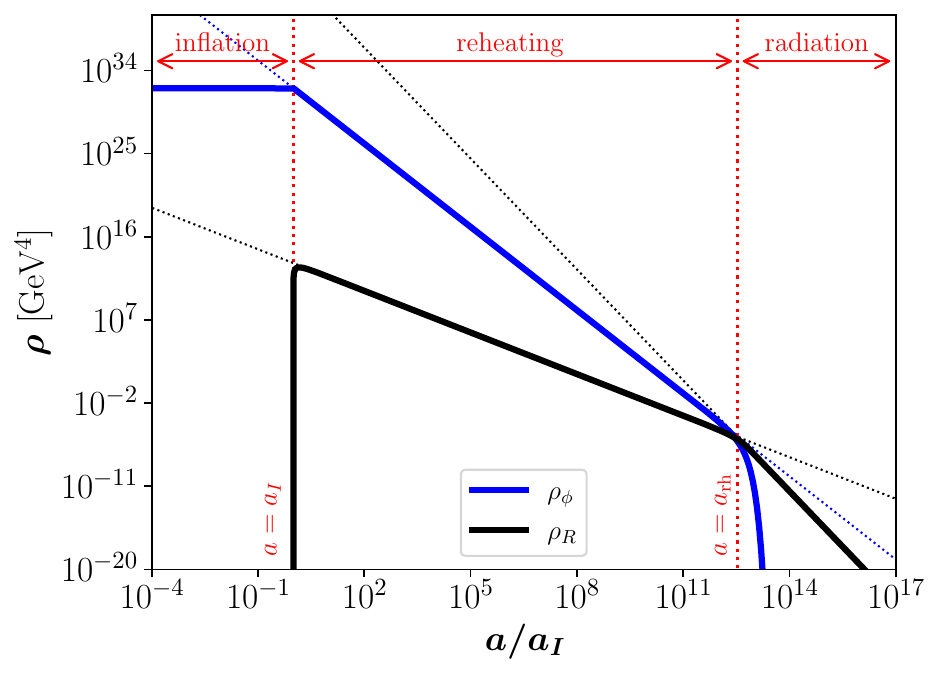}
    \includegraphics[scale=\sepf]{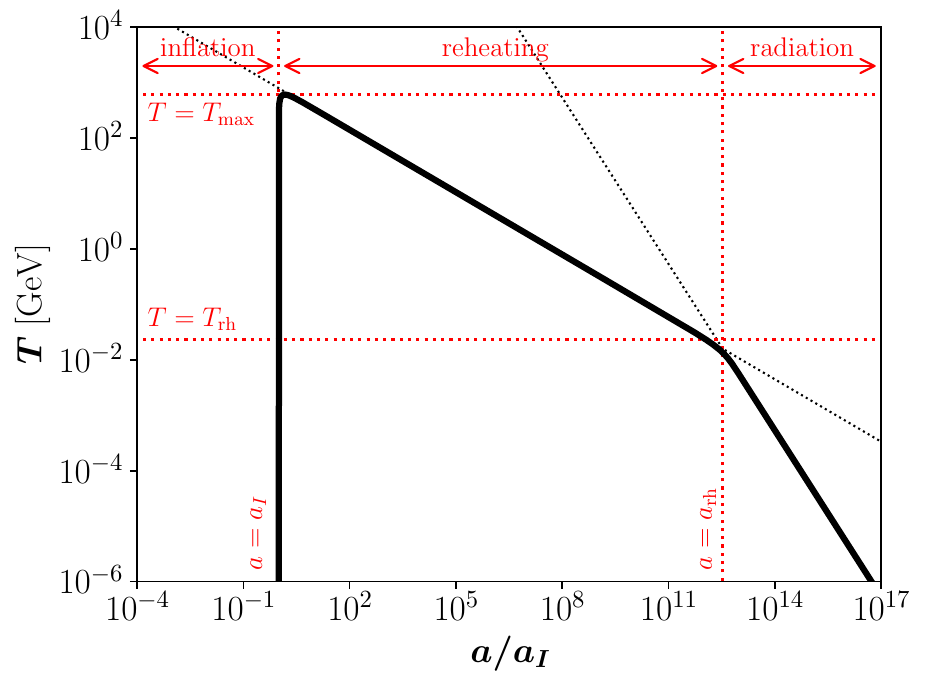}
    \caption{Left: Evolution of inflaton ($\rp$ in blue) and SM radiation ($\rR$ in black) energy densities as a function of the scale factor $a$. Right: Evolution of SM temperature $T$. In both plots, $H_I = 3\times 10^{-3}$~GeV and $\Gamma = 8\times 10^{-22}$~GeV were used. The thin black and blue dotted lines correspond to the analytical solution.}
    \label{fig:bkg}
\end{figure}

In this appendix, we compare our analytical estimations with full numerical results. As an example, we consider the reheating scenario in which the inflaton $\phi$ oscillates in a quadratic potential while perturbatively decaying into SM particles with a total decay width $\Gamma$. The evolution of the inflaton and SM radiation energy densities can be tracked with the system of Boltzmann equations
\begin{align}
     \frac{d\rp}{dt} + 3\, H\, \rp &= -\Gamma\, \rp\,, \label{eq:rho1}\\
     \frac{d\rR}{dt} + 4\, H\, \rR &= +\Gamma\, \rp\,, \label{eq:rho2}
\end{align}
where $H$ is given in Eq.~\eqref{eq:H}. This case corresponds to $\omega = 0$ and $\alpha = 3/8$, as discussed in Section~\ref{sec:LTR}. The left panel of Fig.~\ref{fig:bkg} shows the evolution of $\rp$ (blue) and $\rR$ (black) as a function of the scale factor $a$, while the right panel shows the corresponding evolution of the SM temperature $T$, assuming $H_I = 3\times 10^{-3}$~GeV and $\Gamma = 8\times 10^{-22}$~GeV. For comparison, we also show the analytical estimations from Section~\ref{sec:LTR} with thin black and blue dotted lines, in very good agreement with the fully numerical solution.

\begin{figure}[t!]
    \def\sepf{0.49}
    \centering
    \includegraphics[scale=\sepf]{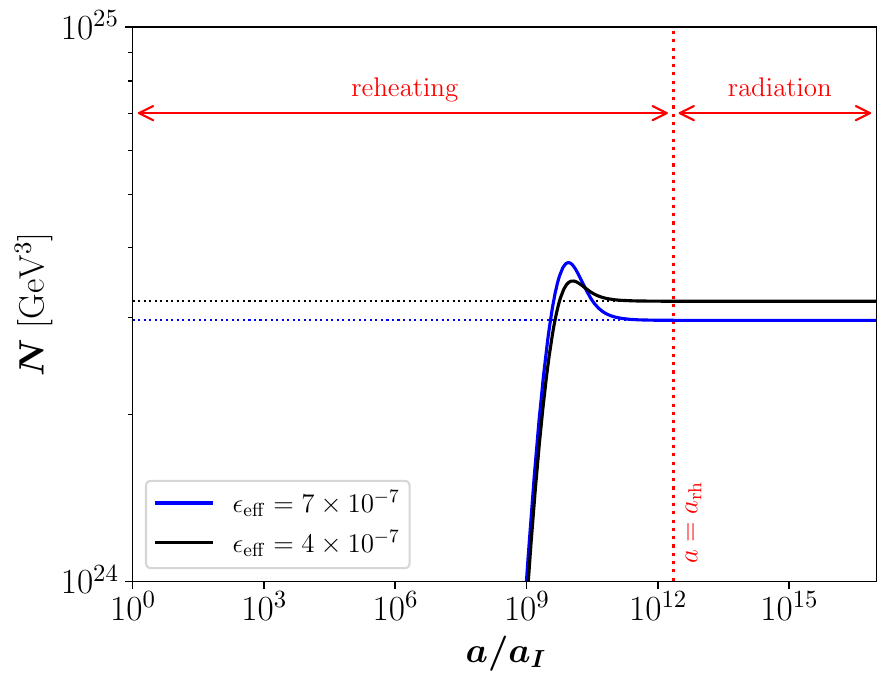}
    \includegraphics[scale=\sepf]{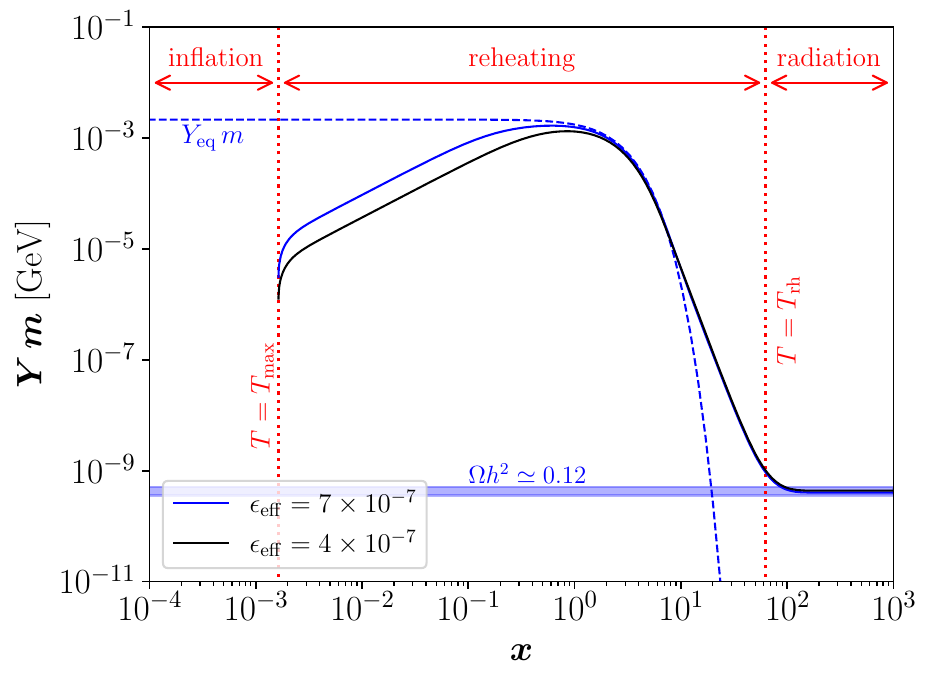}
    \caption{Evolution of the comoving number density $N$ (left) and the DM yield $Y$ (right), for $H_I = 3\times 10^{-3}$~GeV, $\Gamma = 8\times 10^{-22}$~GeV, $m = 1$~GeV, $\xrh = 43$ and $\eef = 4\times 10^{-7}$ (black) or $\eef = 7\times 10^{-7}$ (blue). The two benchmark points correspond to WIMP solutions and fit the entire observed DM abundance.}
    \label{fig:DMnum}
\end{figure} 
The evolution of the DM number density can be followed with the Boltzmann equation
\begin{equation} \label{eq:DM0}
         \frac{dn}{dt} + 3\, H\, n = - \svann\left[n^2 - n_\text{eq}^2\right],
\end{equation}
that has to be solved in the background defined by Eqs.~\eqref{eq:rho1} and~\eqref{eq:rho2}. As during reheating the SM entropy is not conserved, due to the decays of the inflaton into SM states, it is convenient to rewrite Eq.~\eqref{eq:DM0} as a function of the comoving number density $N \equiv n\, a^3$, which implies that
\begin{equation}
         \frac{dN}{da} = - \frac{\svann}{H\, a^4} \left[N^2 - N_\text{eq}^2\right].
\end{equation}
Figure~\ref{fig:DMnum} shows the fully numerical solution of $N$ (left) and the DM yield $Y$ (right), assuming $H_I = 3\times 10^{-3}$~GeV, $\Gamma = 8\times 10^{-22}$~GeV, $m = 1$~GeV, $\xrh = 43$ and two couplings: $\eef = 4\times 10^{-7}$ (black) or $\eef = 7\times 10^{-7}$ (blue). In the right panel, the DM yield at equilibrium is depicted with a blue dashed line, while the observed DM abundance is represented by a horizontal blue band. Several comments are in order: $i)$ Even if it is for a short period, DM reaches chemical equilibrium with the SM. That can be seen in the left panel, where $Y$ tracks the equilibrium yield, but also in the left panel, where the decrease in $N$ corresponds to the period in which chemical equilibrium is attained. As chemical equilibrium with the SM is reached, both solutions correspond to WIMPs. $ii)$ The two selected benchmark points fit the entire observed abundance of DM, as can be seen in the right panel. This is a clear example of the behavior already observed in Section~\ref{sec:during}, where, for a given mass, the abundance of WIMP DM can be fitted with {\it two} different values of the coupling. $iii)$ In general, this two-solution scenario is not easily realized, as it typically occurs on the verge of chemical equilibrium. If one solution occurs prior to chemical equilibrium, it would correspond to a (nonthermal) FIMP instead of a WIMP. In that case, for a given mass one would have two solutions: one WIMP and one FIMP~\cite{Silva-Malpartida:2023yks}. $iv)$ We emphasize that this two-solution scenario can only be realized in the presence of a low-temperature reheating or, equivalently, in a nonstandard cosmological setup~\cite{Silva-Malpartida:2024emu}. If chemical freeze-out occurs in a radiation-dominated era, the usual single solution takes place. And $v)$, the WIMP solution is typically realized with order-one couplings, if the freeze-out occurs in the radiation-dominated epoch. However, much smaller values such as $\eef \sim \mathcal{O}(10^{-7})$ are needed if DM decouples during reheating, as in the present case.

\begin{figure}[t!]
    \def\sepf{0.5}
    \centering
    \includegraphics[scale=\sepf]{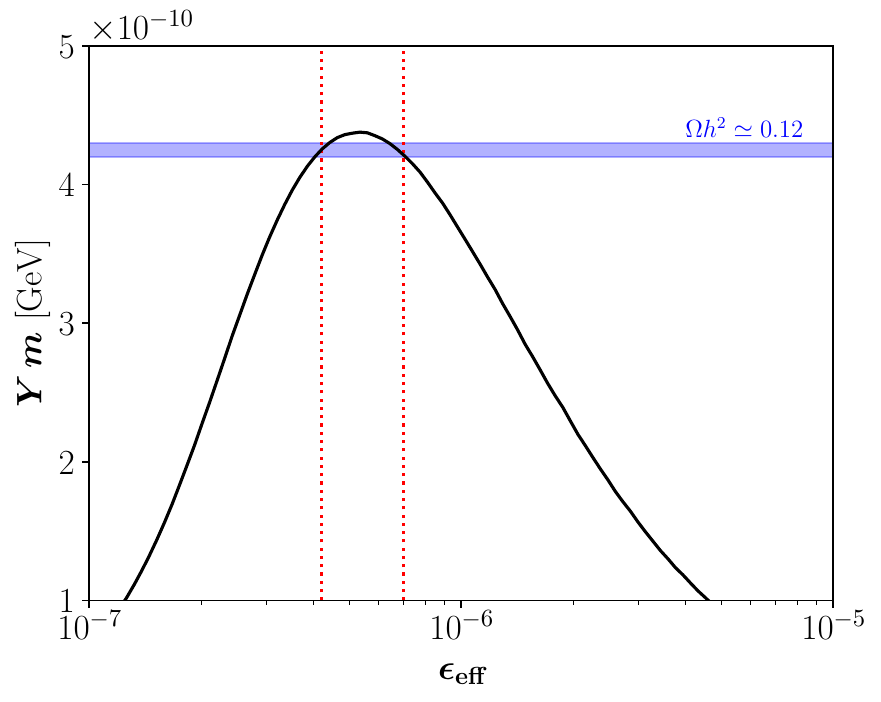}
    \caption{Evolution of DM yield $Y$ as a function of the coupling $\eef$, for $H_I = 3\times 10^{-3}$~GeV, $\Gamma = 8\times 10^{-22}$~GeV, $m = 1$~GeV and $\xrh = 43$. The horizontal blue band corresponds to the observed abundance of DM that can be matched for two values of the coupling.}
    \label{fig:DMnumeps}
\end{figure}

For completeness, the evolution of the DM yield $Y$ as a function of the coupling $\eef$ is shown in Fig.~\ref{fig:DMnumeps}, for $H_I = 3\times 10^{-3}$~GeV, $\Gamma = 8\times 10^{-22}$~GeV, $m = 1$~GeV and $\xrh = 43$. The horizontal blue band corresponds to the observed abundance of DM that can be matched for the two values of the coupling mentioned above: $\eef \simeq 4\times 10^{-7}$ and $\eef \simeq 7\times 10^{-7}$. Between these two values, DM is overproduced, overclosing the Universe. We emphasize that for couplings close but smaller than $\eef \simeq 4\times 10^{-7}$, chemical equilibrium is not granted, and therefore DM is no longer a thermal relic.

\section{\boldmath Bound on $\Tmax$} \label{sec:Tmax}
During reheating, the energy densities of the inflaton and SM radiation can be written as
\begin{align}
    \rp(a) &\simeq 3\, M_P^2\, H_I^2 \left(\frac{a_I}{a}\right)^{3(1+\omega)},\\
    \rR(a) &\simeq \frac{\pi^2}{30}\, \gs\, \Trh^4 \left(\frac{\arh}{a}\right)^{4 \alpha},
\end{align}
where $a_I$ corresponds to the scale factor at the end of inflation and beginning of reheating, and $H_I \equiv H(a_I)$ is the inflationary scale. The BICEP/Keck bound on the tensor-to-scalar ratio implies that $H_I\leq 2.0\times 10^{-5}~M_P$~\cite{BICEP:2021xfz}. Taking into account that at the end of reheating $\rp(\arh) = \rR(\arh)$, the maximal energy density reached by the SM thermal bath near $a=a_I$ can be estimated and therefore the maximal temperature $\Tmax$ is
\begin{equation} \label{Tmax}
    \Tmax \simeq \Trh \left[\frac{90}{\pi^2\, \gs}\, \frac{H_I^2\, M_P^2}{\Trh^4}\right]^\frac{\alpha}{3 (1+\omega)}.
\end{equation}
This limit on the maximum temperature reached by the SM bath is particularly relevant, as a thermally produced DM particle has to satisfy $m < \Tmax$.

\bibliographystyle{JHEP}
\bibliography{biblio}

\providecommand{\href}[2]{#2}\begingroup\raggedright\begin{thebibliography}{100}

\bibitem{Planck:2018vyg}
{\scshape Planck} collaboration, \emph{{Planck 2018 results. VI. Cosmological
  parameters}},
  \href{https://doi.org/10.1051/0004-6361/201833910}{\emph{Astron. Astrophys.}
  {\bfseries 641} (2020) A6}
  [\href{https://arxiv.org/abs/1807.06209}{{\ttfamily 1807.06209}}].

\bibitem{Cirelli:2024ssz}
M.~Cirelli, A.~Strumia and J.~Zupan, \emph{{Dark Matter}},
  \href{https://arxiv.org/abs/2406.01705}{{\ttfamily 2406.01705}}.

\bibitem{Steigman:2012nb}
G.~Steigman, B.~Dasgupta and J.F.~Beacom, \emph{{Precise Relic WIMP Abundance
  and its Impact on Searches for Dark Matter Annihilation}},
  \href{https://doi.org/10.1103/PhysRevD.86.023506}{\emph{Phys. Rev. D}
  {\bfseries 86} (2012) 023506}
  [\href{https://arxiv.org/abs/1204.3622}{{\ttfamily 1204.3622}}].

\bibitem{Lee:1977ua}
B.W.~Lee and S.~Weinberg, \emph{{Cosmological Lower Bound on Heavy Neutrino
  Masses}}, \href{https://doi.org/10.1103/PhysRevLett.39.165}{\emph{Phys. Rev.
  Lett.} {\bfseries 39} (1977) 165}.

\bibitem{Griest:1990kh}
K.~Griest and D.~Seckel, \emph{{Three exceptions in the calculation of relic
  abundances}}, \href{https://doi.org/10.1103/PhysRevD.43.3191}{\emph{Phys.
  Rev. D} {\bfseries 43} (1991) 3191}.

\bibitem{Hambye:2008bq}
T.~Hambye, \emph{{Hidden vector dark matter}},
  \href{https://doi.org/10.1088/1126-6708/2009/01/028}{\emph{JHEP} {\bfseries
  01} (2009) 028} [\href{https://arxiv.org/abs/0811.0172}{{\ttfamily
  0811.0172}}].

\bibitem{Hambye:2009fg}
T.~Hambye and M.H.G.~Tytgat, \emph{{Confined hidden vector dark matter}},
  \href{https://doi.org/10.1016/j.physletb.2009.11.050}{\emph{Phys. Lett. B}
  {\bfseries 683} (2010) 39} [\href{https://arxiv.org/abs/0907.1007}{{\ttfamily
  0907.1007}}].

\bibitem{DEramo:2010keq}
F.~D'Eramo and J.~Thaler, \emph{{Semi-annihilation of Dark Matter}},
  \href{https://doi.org/10.1007/JHEP06(2010)109}{\emph{JHEP} {\bfseries 06}
  (2010) 109} [\href{https://arxiv.org/abs/1003.5912}{{\ttfamily 1003.5912}}].

\bibitem{Belanger:2012zr}
G.~B\'elanger, K.~Kannike, A.~Pukhov and M.~Raidal, \emph{{$\mathbb{Z}_3$
  Scalar Singlet Dark Matter}},
  \href{https://doi.org/10.1088/1475-7516/2013/01/022}{\emph{JCAP} {\bfseries
  01} (2013) 022} [\href{https://arxiv.org/abs/1211.1014}{{\ttfamily
  1211.1014}}].

\bibitem{Belanger:2014bga}
G.~B\'elanger, K.~Kannike, A.~Pukhov and M.~Raidal, \emph{{Minimal
  semi-annihilating $\mathbb{Z}_N$ scalar dark matter}},
  \href{https://doi.org/10.1088/1475-7516/2014/06/021}{\emph{JCAP} {\bfseries
  06} (2014) 021} [\href{https://arxiv.org/abs/1403.4960}{{\ttfamily
  1403.4960}}].

\bibitem{Arcadi:2017kky}
G.~Arcadi, M.~Dutra, P.~Ghosh, M.~Lindner, Y.~Mambrini, M.~Pierre et~al.,
  \emph{{The waning of the WIMP? A review of models, searches, and
  constraints}},
  \href{https://doi.org/10.1140/epjc/s10052-018-5662-y}{\emph{Eur. Phys. J. C}
  {\bfseries 78} (2018) 203}
  [\href{https://arxiv.org/abs/1703.07364}{{\ttfamily 1703.07364}}].

\bibitem{Roszkowski:2017nbc}
L.~Roszkowski, E.M.~Sessolo and S.~Trojanowski, \emph{{WIMP dark matter
  candidates and searches - current status and future prospects}},
  \href{https://doi.org/10.1088/1361-6633/aab913}{\emph{Rept. Prog. Phys.}
  {\bfseries 81} (2018) 066201}
  [\href{https://arxiv.org/abs/1707.06277}{{\ttfamily 1707.06277}}].

\bibitem{Arcadi:2024ukq}
G.~Arcadi, D.~Cabo-Almeida, M.~Dutra, P.~Ghosh, M.~Lindner, Y.~Mambrini et~al.,
  \emph{{The Waning of the WIMP: Endgame?}},
  \href{https://arxiv.org/abs/2403.15860}{{\ttfamily 2403.15860}}.

\bibitem{Hochberg:2014dra}
Y.~Hochberg, E.~Kuflik, T.~Volansky and J.G.~Wacker, \emph{{Mechanism for
  Thermal Relic Dark Matter of Strongly Interacting Massive Particles}},
  \href{https://doi.org/10.1103/PhysRevLett.113.171301}{\emph{Phys. Rev. Lett.}
  {\bfseries 113} (2014) 171301}
  [\href{https://arxiv.org/abs/1402.5143}{{\ttfamily 1402.5143}}].

\bibitem{Choi:2015bya}
S.-M.~Choi and H.M.~Lee, \emph{{SIMP dark matter with gauged $\mathbb{Z}_3$
  symmetry}}, \href{https://doi.org/10.1007/JHEP09(2015)063}{\emph{JHEP}
  {\bfseries 09} (2015) 063}
  [\href{https://arxiv.org/abs/1505.00960}{{\ttfamily 1505.00960}}].

\bibitem{Bernal:2015bla}
N.~Bernal, C.~Garc\'ia-Cely and R.~Rosenfeld, \emph{{WIMP and SIMP Dark Matter
  from the Spontaneous Breaking of a Global Group}},
  \href{https://doi.org/10.1088/1475-7516/2015/04/012}{\emph{JCAP} {\bfseries
  04} (2015) 012} [\href{https://arxiv.org/abs/1501.01973}{{\ttfamily
  1501.01973}}].

\bibitem{Bernal:2015lbl}
N.~Bernal, C.~Garc\'ia-Cely and R.~Rosenfeld, \emph{{$\mathbb Z_3$ WIMP and
  SIMP Dark Matter from a Global U(1) Breaking}},
  \href{https://doi.org/10.1016/j.nuclphysbps.2015.11.001}{\emph{Nucl. Part.
  Phys. Proc.} {\bfseries 267-269} (2015) 353}.

\bibitem{Ko:2014nha}
P.~Ko and Y.~Tang, \emph{{Self-interacting scalar dark matter with local
  $\mathbb{Z}_3$ symmetry}},
  \href{https://doi.org/10.1088/1475-7516/2014/05/047}{\emph{JCAP} {\bfseries
  05} (2014) 047} [\href{https://arxiv.org/abs/1402.6449}{{\ttfamily
  1402.6449}}].

\bibitem{Choi:2017mkk}
S.-M.~Choi, H.M.~Lee and M.-S.~Seo, \emph{{Cosmic abundances of SIMP dark
  matter}}, \href{https://doi.org/10.1007/JHEP04(2017)154}{\emph{JHEP}
  {\bfseries 04} (2017) 154}
  [\href{https://arxiv.org/abs/1702.07860}{{\ttfamily 1702.07860}}].

\bibitem{Chu:2017msm}
X.~Chu and C.~Garc\'ia-Cely, \emph{{Self-interacting Spin-2 Dark Matter}},
  \href{https://doi.org/10.1103/PhysRevD.96.103519}{\emph{Phys. Rev. D}
  {\bfseries 96} (2017) 103519}
  [\href{https://arxiv.org/abs/1708.06764}{{\ttfamily 1708.06764}}].

\bibitem{Bernal:2015ova}
N.~Bernal, X.~Chu, C.~Garc\'ia-Cely, T.~Hambye and B.~Zaldivar,
  \emph{{Production Regimes for Self-Interacting Dark Matter}},
  \href{https://doi.org/10.1088/1475-7516/2016/03/018}{\emph{JCAP} {\bfseries
  03} (2016) 018} [\href{https://arxiv.org/abs/1510.08063}{{\ttfamily
  1510.08063}}].

\bibitem{Yamanaka:2014pva}
N.~Yamanaka, S.~Fujibayashi, S.~Gongyo and H.~Iida, \emph{{Dark matter in the
  hidden gauge theory}},  \href{https://arxiv.org/abs/1411.2172}{{\ttfamily
  1411.2172}}.

\bibitem{Hochberg:2014kqa}
Y.~Hochberg, E.~Kuflik, H.~Murayama, T.~Volansky and J.G.~Wacker, \emph{{Model
  for Thermal Relic Dark Matter of Strongly Interacting Massive Particles}},
  \href{https://doi.org/10.1103/PhysRevLett.115.021301}{\emph{Phys. Rev. Lett.}
  {\bfseries 115} (2015) 021301}
  [\href{https://arxiv.org/abs/1411.3727}{{\ttfamily 1411.3727}}].

\bibitem{Lee:2015gsa}
H.M.~Lee and M.-S.~Seo, \emph{{Communication with SIMP dark mesons via
  Z'-portal}},
  \href{https://doi.org/10.1016/j.physletb.2015.07.013}{\emph{Phys. Lett. B}
  {\bfseries 748} (2015) 316}
  [\href{https://arxiv.org/abs/1504.00745}{{\ttfamily 1504.00745}}].

\bibitem{Hansen:2015yaa}
M.~Hansen, K.~Lang\ae{}ble and F.~Sannino, \emph{{SIMP model at NNLO in chiral
  perturbation theory}},
  \href{https://doi.org/10.1103/PhysRevD.92.075036}{\emph{Phys. Rev. D}
  {\bfseries 92} (2015) 075036}
  [\href{https://arxiv.org/abs/1507.01590}{{\ttfamily 1507.01590}}].

\bibitem{Bernal:2015xba}
N.~Bernal and X.~Chu, \emph{{$\mathbb {Z}_2$ SIMP Dark Matter}},
  \href{https://doi.org/10.1088/1475-7516/2016/01/006}{\emph{JCAP} {\bfseries
  01} (2016) 006} [\href{https://arxiv.org/abs/1510.08527}{{\ttfamily
  1510.08527}}].

\bibitem{Heikinheimo:2016yds}
M.~Heikinheimo, T.~Tenkanen, K.~Tuominen and V.~Vaskonen, \emph{{Observational
  Constraints on Decoupled Hidden Sectors}},
  \href{https://doi.org/10.1103/PhysRevD.94.063506}{\emph{Phys. Rev. D}
  {\bfseries 94} (2016) 063506}
  [\href{https://arxiv.org/abs/1604.02401}{{\ttfamily 1604.02401}}].

\bibitem{Bernal:2017mqb}
N.~Bernal, X.~Chu and J.~Pradler, \emph{{Simply split strongly interacting
  massive particles}},
  \href{https://doi.org/10.1103/PhysRevD.95.115023}{\emph{Phys. Rev. D}
  {\bfseries 95} (2017) 115023}
  [\href{https://arxiv.org/abs/1702.04906}{{\ttfamily 1702.04906}}].

\bibitem{Heikinheimo:2017ofk}
M.~Heikinheimo, T.~Tenkanen and K.~Tuominen, \emph{{WIMP miracle of the second
  kind}}, \href{https://doi.org/10.1103/PhysRevD.96.023001}{\emph{Phys. Rev. D}
  {\bfseries 96} (2017) 023001}
  [\href{https://arxiv.org/abs/1704.05359}{{\ttfamily 1704.05359}}].

\bibitem{Bernal:2018hjm}
N.~Bernal, A.~Chatterjee and A.~Paul, \emph{{Non-thermal production of Dark
  Matter after Inflation}},
  \href{https://doi.org/10.1088/1475-7516/2018/12/020}{\emph{JCAP} {\bfseries
  12} (2018) 020} [\href{https://arxiv.org/abs/1809.02338}{{\ttfamily
  1809.02338}}].

\bibitem{deLaix:1995vi}
A.A.~de~Laix, R.J.~Scherrer and R.K.~Schaefer, \emph{{Constraints of
  selfinteracting dark matter}},
  \href{https://doi.org/10.1086/176322}{\emph{Astrophys. J.} {\bfseries 452}
  (1995) 495} [\href{https://arxiv.org/abs/astro-ph/9502087}{{\ttfamily
  astro-ph/9502087}}].

\bibitem{Kuflik:2015isi}
E.~Kuflik, M.~Perelstein, N.R.-L.~Lorier and Y.-D.~Tsai, \emph{{Elastically
  Decoupling Dark Matter}},
  \href{https://doi.org/10.1103/PhysRevLett.116.221302}{\emph{Phys. Rev. Lett.}
  {\bfseries 116} (2016) 221302}
  [\href{https://arxiv.org/abs/1512.04545}{{\ttfamily 1512.04545}}].

\bibitem{Kuflik:2017iqs}
E.~Kuflik, M.~Perelstein, N.R.-L.~Lorier and Y.-D.~Tsai, \emph{{Phenomenology
  of ELDER Dark Matter}},
  \href{https://doi.org/10.1007/JHEP08(2017)078}{\emph{JHEP} {\bfseries 08}
  (2017) 078} [\href{https://arxiv.org/abs/1706.05381}{{\ttfamily
  1706.05381}}].

\bibitem{Carlson:1992fn}
E.D.~Carlson, M.E.~Machacek and L.J.~Hall, \emph{{Self-interacting dark
  matter}}, \href{https://doi.org/10.1086/171833}{\emph{Astrophys. J.}
  {\bfseries 398} (1992) 43}.

\bibitem{Pappadopulo:2016pkp}
D.~Pappadopulo, J.T.~Ruderman and G.~Trevisan, \emph{{Dark matter freeze-out in
  a nonrelativistic sector}},
  \href{https://doi.org/10.1103/PhysRevD.94.035005}{\emph{Phys. Rev. D}
  {\bfseries 94} (2016) 035005}
  [\href{https://arxiv.org/abs/1602.04219}{{\ttfamily 1602.04219}}].

\bibitem{Farina:2016llk}
M.~Farina, D.~Pappadopulo, J.T.~Ruderman and G.~Trevisan, \emph{{Phases of
  Cannibal Dark Matter}},
  \href{https://doi.org/10.1007/JHEP12(2016)039}{\emph{JHEP} {\bfseries 12}
  (2016) 039} [\href{https://arxiv.org/abs/1607.03108}{{\ttfamily
  1607.03108}}].

\bibitem{McDonald:2001vt}
J.~McDonald, \emph{{Thermally generated gauge singlet scalars as
  selfinteracting dark matter}},
  \href{https://doi.org/10.1103/PhysRevLett.88.091304}{\emph{Phys. Rev. Lett.}
  {\bfseries 88} (2002) 091304}
  [\href{https://arxiv.org/abs/hep-ph/0106249}{{\ttfamily hep-ph/0106249}}].

\bibitem{Choi:2005vq}
K.-Y.~Choi and L.~Roszkowski, \emph{{E-WIMPs}},
  \href{https://doi.org/10.1063/1.2149672}{\emph{AIP Conf. Proc.} {\bfseries
  805} (2005) 30} [\href{https://arxiv.org/abs/hep-ph/0511003}{{\ttfamily
  hep-ph/0511003}}].

\bibitem{Kusenko:2006rh}
A.~Kusenko, \emph{{Sterile neutrinos, dark matter, and the pulsar velocities in
  models with a Higgs singlet}},
  \href{https://doi.org/10.1103/PhysRevLett.97.241301}{\emph{Phys. Rev. Lett.}
  {\bfseries 97} (2006) 241301}
  [\href{https://arxiv.org/abs/hep-ph/0609081}{{\ttfamily hep-ph/0609081}}].

\bibitem{Petraki:2007gq}
K.~Petraki and A.~Kusenko, \emph{{Dark-matter sterile neutrinos in models with
  a gauge singlet in the Higgs sector}},
  \href{https://doi.org/10.1103/PhysRevD.77.065014}{\emph{Phys. Rev. D}
  {\bfseries 77} (2008) 065014}
  [\href{https://arxiv.org/abs/0711.4646}{{\ttfamily 0711.4646}}].

\bibitem{Hall:2009bx}
L.J.~Hall, K.~Jedamzik, J.~March-Russell and S.M.~West, \emph{{Freeze-In
  Production of FIMP Dark Matter}},
  \href{https://doi.org/10.1007/JHEP03(2010)080}{\emph{JHEP} {\bfseries 03}
  (2010) 080} [\href{https://arxiv.org/abs/0911.1120}{{\ttfamily 0911.1120}}].

\bibitem{Elahi:2014fsa}
F.~Elahi, C.~Kolda and J.~Unwin, \emph{{UltraViolet Freeze-in}},
  \href{https://doi.org/10.1007/JHEP03(2015)048}{\emph{JHEP} {\bfseries 03}
  (2015) 048} [\href{https://arxiv.org/abs/1410.6157}{{\ttfamily 1410.6157}}].

\bibitem{Bernal:2017kxu}
N.~Bernal, M.~Heikinheimo, T.~Tenkanen, K.~Tuominen and V.~Vaskonen, \emph{{The
  Dawn of FIMP Dark Matter: A Review of Models and Constraints}},
  \href{https://doi.org/10.1142/S0217751X1730023X}{\emph{Int. J. Mod. Phys. A}
  {\bfseries 32} (2017) 1730023}
  [\href{https://arxiv.org/abs/1706.07442}{{\ttfamily 1706.07442}}].

\bibitem{Allahverdi:2020bys}
R.~Allahverdi et~al., \emph{{The First Three Seconds: a Review of Possible
  Expansion Histories of the Early Universe}},
  \href{https://doi.org/10.21105/astro.2006.16182}{\emph{Open J.Astrophys.}
  {\bfseries 4} (2021) } [\href{https://arxiv.org/abs/2006.16182}{{\ttfamily
  2006.16182}}].

\bibitem{Sarkar:1995dd}
S.~Sarkar, \emph{{Big bang nucleosynthesis and physics beyond the standard
  model}}, \href{https://doi.org/10.1088/0034-4885/59/12/001}{\emph{Rept. Prog.
  Phys.} {\bfseries 59} (1996) 1493}
  [\href{https://arxiv.org/abs/hep-ph/9602260}{{\ttfamily hep-ph/9602260}}].

\bibitem{Kawasaki:2000en}
M.~Kawasaki, K.~Kohri and N.~Sugiyama, \emph{{MeV scale reheating temperature
  and thermalization of neutrino background}},
  \href{https://doi.org/10.1103/PhysRevD.62.023506}{\emph{Phys. Rev. D}
  {\bfseries 62} (2000) 023506}
  [\href{https://arxiv.org/abs/astro-ph/0002127}{{\ttfamily
  astro-ph/0002127}}].

\bibitem{Hannestad:2004px}
S.~Hannestad, \emph{{What is the lowest possible reheating temperature?}},
  \href{https://doi.org/10.1103/PhysRevD.70.043506}{\emph{Phys. Rev. D}
  {\bfseries 70} (2004) 043506}
  [\href{https://arxiv.org/abs/astro-ph/0403291}{{\ttfamily
  astro-ph/0403291}}].

\bibitem{DeBernardis:2008zz}
F.~De~Bernardis, L.~Pagano and A.~Melchiorri, \emph{{New constraints on the
  reheating temperature of the universe after WMAP-5}},
  \href{https://doi.org/10.1016/j.astropartphys.2008.09.005}{\emph{Astropart.
  Phys.} {\bfseries 30} (2008) 192}.

\bibitem{deSalas:2015glj}
P.F.~de~Salas, M.~Lattanzi, G.~Mangano, G.~Miele, S.~Pastor and O.~Pisanti,
  \emph{{Bounds on very low reheating scenarios after Planck}},
  \href{https://doi.org/10.1103/PhysRevD.92.123534}{\emph{Phys. Rev. D}
  {\bfseries 92} (2015) 123534}
  [\href{https://arxiv.org/abs/1511.00672}{{\ttfamily 1511.00672}}].

\bibitem{Barrow:1982ei}
J.D.~Barrow, \emph{{Massive Particles as a Probe of the Early Universe}},
  \href{https://doi.org/10.1016/0550-3213(82)90233-4}{\emph{Nucl. Phys. B}
  {\bfseries 208} (1982) 501}.

\bibitem{Kamionkowski:1990ni}
M.~Kamionkowski and M.S.~Turner, \emph{{Thermal Relics: Do we Know their
  Abundances?}}, \href{https://doi.org/10.1103/PhysRevD.42.3310}{\emph{Phys.
  Rev. D} {\bfseries 42} (1990) 3310}.

\bibitem{McDonald:1989jd}
J.~McDonald, \emph{{{WIMP} Densities in Decaying Particle Dominated
  Cosmology}}, \href{https://doi.org/10.1103/PhysRevD.43.1063}{\emph{Phys. Rev.
  D} {\bfseries 43} (1991) 1063}.

\bibitem{Salati:2002md}
P.~Salati, \emph{{Quintessence and the relic density of neutralinos}},
  \href{https://doi.org/10.1016/j.physletb.2003.07.073}{\emph{Phys. Lett. B}
  {\bfseries 571} (2003) 121}
  [\href{https://arxiv.org/abs/astro-ph/0207396}{{\ttfamily
  astro-ph/0207396}}].

\bibitem{Fornengo:2002db}
N.~Fornengo, A.~Riotto and S.~Scopel, \emph{{Supersymmetric dark matter and the
  reheating temperature of the universe}},
  \href{https://doi.org/10.1103/PhysRevD.67.023514}{\emph{Phys. Rev. D}
  {\bfseries 67} (2003) 023514}
  [\href{https://arxiv.org/abs/hep-ph/0208072}{{\ttfamily hep-ph/0208072}}].

\bibitem{Comelli:2003cv}
D.~Comelli, M.~Pietroni and A.~Riotto, \emph{{Dark energy and dark matter}},
  \href{https://doi.org/10.1016/j.physletb.2003.05.006}{\emph{Phys. Lett. B}
  {\bfseries 571} (2003) 115}
  [\href{https://arxiv.org/abs/hep-ph/0302080}{{\ttfamily hep-ph/0302080}}].

\bibitem{Rosati:2003yw}
F.~Rosati, \emph{{Quintessential enhancement of dark matter abundance}},
  \href{https://doi.org/10.1016/j.physletb.2003.07.048}{\emph{Phys. Lett. B}
  {\bfseries 570} (2003) 5}
  [\href{https://arxiv.org/abs/hep-ph/0302159}{{\ttfamily hep-ph/0302159}}].

\bibitem{Pallis:2004yy}
C.~Pallis, \emph{{Massive particle decay and cold dark matter abundance}},
  \href{https://doi.org/10.1016/j.astropartphys.2004.05.006}{\emph{Astropart.
  Phys.} {\bfseries 21} (2004) 689}
  [\href{https://arxiv.org/abs/hep-ph/0402033}{{\ttfamily hep-ph/0402033}}].

\bibitem{Gelmini:2006pw}
G.B.~Gelmini and P.~Gondolo, \emph{{Neutralino with the right cold dark matter
  abundance in (almost) any supersymmetric model}},
  \href{https://doi.org/10.1103/PhysRevD.74.023510}{\emph{Phys. Rev. D}
  {\bfseries 74} (2006) 023510}
  [\href{https://arxiv.org/abs/hep-ph/0602230}{{\ttfamily hep-ph/0602230}}].

\bibitem{Drees:2006vh}
M.~Drees, H.~Iminniyaz and M.~Kakizaki, \emph{{Abundance of cosmological relics
  in low-temperature scenarios}},
  \href{https://doi.org/10.1103/PhysRevD.73.123502}{\emph{Phys. Rev. D}
  {\bfseries 73} (2006) 123502}
  [\href{https://arxiv.org/abs/hep-ph/0603165}{{\ttfamily hep-ph/0603165}}].

\bibitem{Gelmini:2006pq}
G.~Gelmini, P.~Gondolo, A.~Soldatenko and C.E.~Yaguna, \emph{{The Effect of a
  late decaying scalar on the neutralino relic density}},
  \href{https://doi.org/10.1103/PhysRevD.74.083514}{\emph{Phys. Rev. D}
  {\bfseries 74} (2006) 083514}
  [\href{https://arxiv.org/abs/hep-ph/0605016}{{\ttfamily hep-ph/0605016}}].

\bibitem{Arbey:2008kv}
A.~Arbey and F.~Mahmoudi, \emph{{SUSY constraints from relic density: High
  sensitivity to pre-BBN expansion rate}},
  \href{https://doi.org/10.1016/j.physletb.2008.09.032}{\emph{Phys. Lett. B}
  {\bfseries 669} (2008) 46} [\href{https://arxiv.org/abs/0803.0741}{{\ttfamily
  0803.0741}}].

\bibitem{Cohen:2008nb}
T.~Cohen, D.E.~Morrissey and A.~Pierce, \emph{{Changes in Dark Matter
  Properties After Freeze-Out}},
  \href{https://doi.org/10.1103/PhysRevD.78.111701}{\emph{Phys. Rev. D}
  {\bfseries 78} (2008) 111701}
  [\href{https://arxiv.org/abs/0808.3994}{{\ttfamily 0808.3994}}].

\bibitem{Arbey:2009gt}
A.~Arbey and F.~Mahmoudi, \emph{{SUSY Constraints, Relic Density, and Very
  Early Universe}}, \href{https://doi.org/10.1007/JHEP05(2010)051}{\emph{JHEP}
  {\bfseries 05} (2010) 051} [\href{https://arxiv.org/abs/0906.0368}{{\ttfamily
  0906.0368}}].

\bibitem{Roszkowski:2014lga}
L.~Roszkowski, S.~Trojanowski and K.~Turzy\'nski, \emph{{Neutralino and
  gravitino dark matter with low reheating temperature}},
  \href{https://doi.org/10.1007/JHEP11(2014)146}{\emph{JHEP} {\bfseries 11}
  (2014) 146} [\href{https://arxiv.org/abs/1406.0012}{{\ttfamily 1406.0012}}].

\bibitem{Berlin:2016vnh}
A.~Berlin, D.~Hooper and G.~Krnjaic, \emph{{PeV-Scale Dark Matter as a Thermal
  Relic of a Decoupled Sector}},
  \href{https://doi.org/10.1016/j.physletb.2016.06.037}{\emph{Phys. Lett. B}
  {\bfseries 760} (2016) 106}
  [\href{https://arxiv.org/abs/1602.08490}{{\ttfamily 1602.08490}}].

\bibitem{Berlin:2016gtr}
A.~Berlin, D.~Hooper and G.~Krnjaic, \emph{{Thermal Dark Matter From A Highly
  Decoupled Sector}},
  \href{https://doi.org/10.1103/PhysRevD.94.095019}{\emph{Phys. Rev. D}
  {\bfseries 94} (2016) 095019}
  [\href{https://arxiv.org/abs/1609.02555}{{\ttfamily 1609.02555}}].

\bibitem{DEramo:2017gpl}
F.~D'Eramo, N.~Fern\'andez and S.~Profumo, \emph{{When the Universe Expands Too
  Fast: Relentless Dark Matter}},
  \href{https://doi.org/10.1088/1475-7516/2017/05/012}{\emph{JCAP} {\bfseries
  05} (2017) 012} [\href{https://arxiv.org/abs/1703.04793}{{\ttfamily
  1703.04793}}].

\bibitem{Hamdan:2017psw}
S.~Hamdan and J.~Unwin, \emph{{Dark Matter Freeze-out During Matter
  Domination}}, \href{https://doi.org/10.1142/S021773231850181X}{\emph{Mod.
  Phys. Lett. A} {\bfseries 33} (2018) 1850181}
  [\href{https://arxiv.org/abs/1710.03758}{{\ttfamily 1710.03758}}].

\bibitem{Visinelli:2017qga}
L.~Visinelli, \emph{{(Non-)thermal production of WIMPs during kination}},
  \href{https://doi.org/10.3390/sym10110546}{\emph{Symmetry} {\bfseries 10}
  (2018) 546} [\href{https://arxiv.org/abs/1710.11006}{{\ttfamily
  1710.11006}}].

\bibitem{Drees:2017iod}
M.~Drees and F.~Hajkarim, \emph{{Dark Matter Production in an Early Matter
  Dominated Era}},
  \href{https://doi.org/10.1088/1475-7516/2018/02/057}{\emph{JCAP} {\bfseries
  02} (2018) 057} [\href{https://arxiv.org/abs/1711.05007}{{\ttfamily
  1711.05007}}].

\bibitem{Hardy:2018bph}
E.~Hardy, \emph{{Higgs portal dark matter in non-standard cosmological
  histories}}, \href{https://doi.org/10.1007/JHEP06(2018)043}{\emph{JHEP}
  {\bfseries 06} (2018) 043}
  [\href{https://arxiv.org/abs/1804.06783}{{\ttfamily 1804.06783}}].

\bibitem{Bernal:2018kcw}
N.~Bernal, C.~Cosme, T.~Tenkanen and V.~Vaskonen, \emph{{Scalar singlet dark
  matter in non-standard cosmologies}},
  \href{https://doi.org/10.1140/epjc/s10052-019-6550-9}{\emph{Eur. Phys. J. C}
  {\bfseries 79} (2019) 30} [\href{https://arxiv.org/abs/1806.11122}{{\ttfamily
  1806.11122}}].

\bibitem{Drees:2018dsj}
M.~Drees and F.~Hajkarim, \emph{{Neutralino Dark Matter in Scenarios with Early
  Matter Domination}},
  \href{https://doi.org/10.1007/JHEP12(2018)042}{\emph{JHEP} {\bfseries 12}
  (2018) 042} [\href{https://arxiv.org/abs/1808.05706}{{\ttfamily
  1808.05706}}].

\bibitem{Betancur:2018xtj}
A.~Betancur and {\'O}.~Zapata, \emph{{Phenomenology of doublet-triplet
  fermionic dark matter in nonstandard cosmology and multicomponent dark
  sectors}}, \href{https://doi.org/10.1103/PhysRevD.98.095003}{\emph{Phys. Rev.
  D} {\bfseries 98} (2018) 095003}
  [\href{https://arxiv.org/abs/1809.04990}{{\ttfamily 1809.04990}}].

\bibitem{Maldonado:2019qmp}
C.~Maldonado and J.~Unwin, \emph{{Establishing the Dark Matter Relic Density in
  an Era of Particle Decays}},
  \href{https://doi.org/10.1088/1475-7516/2019/06/037}{\emph{JCAP} {\bfseries
  06} (2019) 037} [\href{https://arxiv.org/abs/1902.10746}{{\ttfamily
  1902.10746}}].

\bibitem{Poulin:2019omz}
A.~Poulin, \emph{{Dark matter freeze-out in modified cosmological scenarios}},
  \href{https://doi.org/10.1103/PhysRevD.100.043022}{\emph{Phys. Rev. D}
  {\bfseries 100} (2019) 043022}
  [\href{https://arxiv.org/abs/1905.03126}{{\ttfamily 1905.03126}}].

\bibitem{Arias:2019uol}
P.~Arias, N.~Bernal, A.~Herrera and C.~Maldonado, \emph{{Reconstructing
  Non-standard Cosmologies with Dark Matter}},
  \href{https://doi.org/10.1088/1475-7516/2019/10/047}{\emph{JCAP} {\bfseries
  10} (2019) 047} [\href{https://arxiv.org/abs/1906.04183}{{\ttfamily
  1906.04183}}].

\bibitem{Han:2019vxi}
C.~Han, \emph{{Higgsino Dark Matter in a Non-Standard History of the
  Universe}}, \href{https://doi.org/10.1016/j.physletb.2019.134997}{\emph{Phys.
  Lett. B} {\bfseries 798} (2019) 134997}
  [\href{https://arxiv.org/abs/1907.09235}{{\ttfamily 1907.09235}}].

\bibitem{Chanda:2019xyl}
P.~Chanda, S.~Hamdan and J.~Unwin, \emph{{Reviving $Z$ and Higgs Mediated Dark
  Matter Models in Matter Dominated Freeze-out}},
  \href{https://doi.org/10.1088/1475-7516/2020/01/034}{\emph{JCAP} {\bfseries
  01} (2020) 034} [\href{https://arxiv.org/abs/1911.02616}{{\ttfamily
  1911.02616}}].

\bibitem{Arcadi:2020aot}
G.~Arcadi, S.~Profumo, F.S.~Queiroz and C.~Siqueira, \emph{{Right-handed
  Neutrino Dark Matter, Neutrino Masses, and non-Standard Cosmology in a
  2HDM}}, \href{https://doi.org/10.1088/1475-7516/2020/12/030}{\emph{JCAP}
  {\bfseries 12} (2020) 030}
  [\href{https://arxiv.org/abs/2007.07920}{{\ttfamily 2007.07920}}].

\bibitem{Bhatia:2020itt}
D.~Bhatia and S.~Mukhopadhyay, \emph{{Unitarity limits on thermal dark matter
  in (non-)standard cosmologies}},
  \href{https://doi.org/10.1007/JHEP03(2021)133}{\emph{JHEP} {\bfseries 03}
  (2021) 133} [\href{https://arxiv.org/abs/2010.09762}{{\ttfamily
  2010.09762}}].

\bibitem{Barman:2021ifu}
B.~Barman, P.~Ghosh, F.S.~Queiroz and A.K.~Saha, \emph{{Scalar multiplet dark
  matter in a fast expanding Universe: Resurrection of the desert region}},
  \href{https://doi.org/10.1103/PhysRevD.104.015040}{\emph{Phys. Rev. D}
  {\bfseries 104} (2021) 015040}
  [\href{https://arxiv.org/abs/2101.10175}{{\ttfamily 2101.10175}}].

\bibitem{Cheek:2021cfe}
A.~Cheek, L.~Heurtier, Y.F.~P\'erez-Gonz\'alez and J.~Turner, \emph{{Primordial
  black hole evaporation and dark matter production. II. Interplay with the
  freeze-in or freeze-out mechanism}},
  \href{https://doi.org/10.1103/PhysRevD.105.015023}{\emph{Phys. Rev. D}
  {\bfseries 105} (2022) 015023}
  [\href{https://arxiv.org/abs/2107.00016}{{\ttfamily 2107.00016}}].

\bibitem{Arcadi:2021doo}
G.~Arcadi, J.P.~Neto, F.S.~Queiroz and C.~Siqueira, \emph{{Roads for
  right-handed neutrino dark matter: Fast expansion, standard freeze-out, and
  early matter domination}},
  \href{https://doi.org/10.1103/PhysRevD.105.035016}{\emph{Phys. Rev. D}
  {\bfseries 105} (2022) 035016}
  [\href{https://arxiv.org/abs/2108.11398}{{\ttfamily 2108.11398}}].

\bibitem{Bernal:2020bjf}
N.~Bernal and {\'O}.~Zapata, \emph{{Dark Matter in the Time of Primordial Black
  Holes}}, \href{https://doi.org/10.1088/1475-7516/2021/03/015}{\emph{JCAP}
  {\bfseries 03} (2021) 015}
  [\href{https://arxiv.org/abs/2011.12306}{{\ttfamily 2011.12306}}].

\bibitem{Bernal:2022wck}
N.~Bernal and Y.~Xu, \emph{{WIMPs during reheating}},
  \href{https://doi.org/10.1088/1475-7516/2022/12/017}{\emph{JCAP} {\bfseries
  12} (2022) 017} [\href{https://arxiv.org/abs/2209.07546}{{\ttfamily
  2209.07546}}].

\bibitem{Haque:2023yra}
M.R.~Haque, D.~Maity and R.~Mondal, \emph{{WIMPs, FIMPs, and Inflaton
  phenomenology via reheating, CMB and \ensuremath{\Delta}N$_{eff}$}},
  \href{https://doi.org/10.1007/JHEP09(2023)012}{\emph{JHEP} {\bfseries 09}
  (2023) 012} [\href{https://arxiv.org/abs/2301.01641}{{\ttfamily
  2301.01641}}].

\bibitem{Silva-Malpartida:2023yks}
J.~Silva-Malpartida, N.~Bernal, J.~Jones-P\'erez and R.A.~Lineros, \emph{{From
  WIMPs to FIMPs with low~reheating~temperatures}},
  \href{https://doi.org/10.1088/1475-7516/2023/09/015}{\emph{JCAP} {\bfseries
  09} (2023) 015} [\href{https://arxiv.org/abs/2306.14943}{{\ttfamily
  2306.14943}}].

\bibitem{Bernal:2023ura}
N.~Bernal, P.~Konar and S.~Show, \emph{{Unitarity bound on dark matter in
  low-temperature reheating scenarios}},
  \href{https://doi.org/10.1103/PhysRevD.109.035018}{\emph{Phys. Rev. D}
  {\bfseries 109} (2024) 035018}
  [\href{https://arxiv.org/abs/2311.01587}{{\ttfamily 2311.01587}}].

\bibitem{Arcadi:2024jzv}
G.~Arcadi, \emph{{Thermal and non-thermal DM production in non-Standard
  Cosmologies: a mini review}},
  \href{https://arxiv.org/abs/2406.11042}{{\ttfamily 2406.11042}}.

\bibitem{Bernal:2018ins}
N.~Bernal, C.~Cosme and T.~Tenkanen, \emph{{Phenomenology of Self-Interacting
  Dark Matter in a Matter-Dominated Universe}},
  \href{https://doi.org/10.1140/epjc/s10052-019-6608-8}{\emph{Eur. Phys. J. C}
  {\bfseries 79} (2019) 99} [\href{https://arxiv.org/abs/1803.08064}{{\ttfamily
  1803.08064}}].

\bibitem{Bernal:2020kse}
N.~Bernal and {\'O}.~Zapata, \emph{{Self-interacting Dark Matter from
  Primordial Black Holes}},
  \href{https://doi.org/10.1088/1475-7516/2021/03/007}{\emph{JCAP} {\bfseries
  03} (2021) 007} [\href{https://arxiv.org/abs/2010.09725}{{\ttfamily
  2010.09725}}].

\bibitem{Spokoiny:1993kt}
B.~Spokoiny, \emph{{Deflationary universe scenario}},
  \href{https://doi.org/10.1016/0370-2693(93)90155-B}{\emph{Phys. Lett. B}
  {\bfseries 315} (1993) 40}
  [\href{https://arxiv.org/abs/gr-qc/9306008}{{\ttfamily gr-qc/9306008}}].

\bibitem{Ferreira:1997hj}
P.G.~Ferreira and M.~Joyce, \emph{{Cosmology with a primordial scaling field}},
  \href{https://doi.org/10.1103/PhysRevD.58.023503}{\emph{Phys. Rev. D}
  {\bfseries 58} (1998) 023503}
  [\href{https://arxiv.org/abs/astro-ph/9711102}{{\ttfamily
  astro-ph/9711102}}].

\bibitem{Khoury:2001wf}
J.~Khoury, B.A.~Ovrut, P.J.~Steinhardt and N.~Turok, \emph{{The Ekpyrotic
  universe: Colliding branes and the origin of the hot big bang}},
  \href{https://doi.org/10.1103/PhysRevD.64.123522}{\emph{Phys. Rev. D}
  {\bfseries 64} (2001) 123522}
  [\href{https://arxiv.org/abs/hep-th/0103239}{{\ttfamily hep-th/0103239}}].

\bibitem{Khoury:2003rt}
J.~Khoury, P.J.~Steinhardt and N.~Turok, \emph{{Designing cyclic universe
  models}}, \href{https://doi.org/10.1103/PhysRevLett.92.031302}{\emph{Phys.
  Rev. Lett.} {\bfseries 92} (2004) 031302}
  [\href{https://arxiv.org/abs/hep-th/0307132}{{\ttfamily hep-th/0307132}}].

\bibitem{Gasperini:2002bn}
M.~Gasperini and G.~Veneziano, \emph{{The Pre-big bang scenario in string
  cosmology}}, \href{https://doi.org/10.1016/S0370-1573(02)00389-7}{\emph{Phys.
  Rept.} {\bfseries 373} (2003) 1}
  [\href{https://arxiv.org/abs/hep-th/0207130}{{\ttfamily hep-th/0207130}}].

\bibitem{Erickson:2003zm}
J.K.~Erickson, D.H.~Wesley, P.J.~Steinhardt and N.~Turok, \emph{{Kasner and
  mixmaster behavior in universes with equation of state $w \geq 1$}},
  \href{https://doi.org/10.1103/PhysRevD.69.063514}{\emph{Phys. Rev. D}
  {\bfseries 69} (2004) 063514}
  [\href{https://arxiv.org/abs/hep-th/0312009}{{\ttfamily hep-th/0312009}}].

\bibitem{Barrow:2010rx}
J.D.~Barrow and K.~Yamamoto, \emph{{Anisotropic Pressures at Ultra-stiff
  Singularities and the Stability of Cyclic Universes}},
  \href{https://doi.org/10.1103/PhysRevD.82.063516}{\emph{Phys. Rev. D}
  {\bfseries 82} (2010) 063516}
  [\href{https://arxiv.org/abs/1004.4767}{{\ttfamily 1004.4767}}].

\bibitem{Ijjas:2019pyf}
A.~Ijjas and P.J.~Steinhardt, \emph{{A new kind of cyclic universe}},
  \href{https://doi.org/10.1016/j.physletb.2019.06.056}{\emph{Phys. Lett. B}
  {\bfseries 795} (2019) 666}
  [\href{https://arxiv.org/abs/1904.08022}{{\ttfamily 1904.08022}}].

\bibitem{Scherrer:2022nnz}
R.J.~Scherrer, \emph{{How slowly can the early Universe expand?}},
  \href{https://doi.org/10.1103/PhysRevD.106.103516}{\emph{Phys. Rev. D}
  {\bfseries 106} (2022) 103516}
  [\href{https://arxiv.org/abs/2209.03421}{{\ttfamily 2209.03421}}].

\bibitem{Turner:1983he}
M.S.~Turner, \emph{{Coherent Scalar Field Oscillations in an Expanding
  Universe}}, \href{https://doi.org/10.1103/PhysRevD.28.1243}{\emph{Phys. Rev.
  D} {\bfseries 28} (1983) 1243}.

\bibitem{Lozanov:2016hid}
K.D.~Lozanov and M.A.~Amin, \emph{{Equation of State and Duration to Radiation
  Domination after Inflation}},
  \href{https://doi.org/10.1103/PhysRevLett.119.061301}{\emph{Phys. Rev. Lett.}
  {\bfseries 119} (2017) 061301}
  [\href{https://arxiv.org/abs/1608.01213}{{\ttfamily 1608.01213}}].

\bibitem{Lozanov:2017hjm}
K.D.~Lozanov and M.A.~Amin, \emph{{Self-resonance after inflation: oscillons,
  transients and radiation domination}},
  \href{https://doi.org/10.1103/PhysRevD.97.023533}{\emph{Phys. Rev. D}
  {\bfseries 97} (2018) 023533}
  [\href{https://arxiv.org/abs/1710.06851}{{\ttfamily 1710.06851}}].

\bibitem{Bodeker:2006ij}
D.~Bodeker, \emph{{Moduli decay in the hot early Universe}},
  \href{https://doi.org/10.1088/1475-7516/2006/06/027}{\emph{JCAP} {\bfseries
  06} (2006) 027} [\href{https://arxiv.org/abs/hep-ph/0605030}{{\ttfamily
  hep-ph/0605030}}].

\bibitem{Mukaida:2012qn}
K.~Mukaida and K.~Nakayama, \emph{{Dynamics of oscillating scalar field in
  thermal environment}},
  \href{https://doi.org/10.1088/1475-7516/2013/01/017}{\emph{JCAP} {\bfseries
  01} (2013) 017} [\href{https://arxiv.org/abs/1208.3399}{{\ttfamily
  1208.3399}}].

\bibitem{Daido:2017wwb}
R.~Daido, F.~Takahashi and W.~Yin, \emph{{The ALP miracle: unified inflaton and
  dark matter}},
  \href{https://doi.org/10.1088/1475-7516/2017/05/044}{\emph{JCAP} {\bfseries
  05} (2017) 044} [\href{https://arxiv.org/abs/1702.03284}{{\ttfamily
  1702.03284}}].

\bibitem{Co:2020xaf}
R.T.~Co, E.~Gonz\'alez and K.~Harigaya, \emph{{Increasing Temperature toward
  the Completion of Reheating}},
  \href{https://doi.org/10.1088/1475-7516/2020/11/038}{\emph{JCAP} {\bfseries
  11} (2020) 038} [\href{https://arxiv.org/abs/2007.04328}{{\ttfamily
  2007.04328}}].

\bibitem{Garcia:2020wiy}
M.A.G.~Garc\'ia, K.~Kaneta, Y.~Mambrini and K.A.~Olive, \emph{{Inflaton
  Oscillations and Post-Inflationary Reheating}},
  \href{https://doi.org/10.1088/1475-7516/2021/04/012}{\emph{JCAP} {\bfseries
  04} (2021) 012} [\href{https://arxiv.org/abs/2012.10756}{{\ttfamily
  2012.10756}}].

\bibitem{Ahmed:2021fvt}
A.~Ahmed, B.~Grzadkowski and A.~Socha, \emph{{Implications of time-dependent
  inflaton decay on reheating and dark matter production}},
  \href{https://doi.org/10.1016/j.physletb.2022.137201}{\emph{Phys. Lett. B}
  {\bfseries 831} (2022) 137201}
  [\href{https://arxiv.org/abs/2111.06065}{{\ttfamily 2111.06065}}].

\bibitem{Barman:2022tzk}
B.~Barman, N.~Bernal, Y.~Xu and {\'O}.~Zapata, \emph{{Ultraviolet freeze-in
  with a time-dependent inflaton decay}},
  \href{https://doi.org/10.1088/1475-7516/2022/07/019}{\emph{JCAP} {\bfseries
  07} (2022) 019} [\href{https://arxiv.org/abs/2202.12906}{{\ttfamily
  2202.12906}}].

\bibitem{Banerjee:2022fiw}
A.~Banerjee and D.~Chowdhury, \emph{{Fingerprints of freeze-in dark matter in
  an early matter-dominated era}},
  \href{https://doi.org/10.21468/SciPostPhys.13.2.022}{\emph{SciPost Phys.}
  {\bfseries 13} (2022) 022}
  [\href{https://arxiv.org/abs/2204.03670}{{\ttfamily 2204.03670}}].

\bibitem{Arias:2022qjt}
P.~Arias, N.~Bernal, J.K.~Osi\'nski and L.~Roszkowski, \emph{{Dark matter
  axions in the early universe with a period of increasing temperature}},
  \href{https://doi.org/10.1088/1475-7516/2023/05/028}{\emph{JCAP} {\bfseries
  05} (2023) 028} [\href{https://arxiv.org/abs/2207.07677}{{\ttfamily
  2207.07677}}].

\bibitem{Chowdhury:2023jft}
D.~Chowdhury and A.~Hait, \emph{{Thermalization in the presence of a
  time-dependent dissipation and its impact on dark matter production}},
  \href{https://doi.org/10.1007/JHEP09(2023)085}{\emph{JHEP} {\bfseries 09}
  (2023) 085} [\href{https://arxiv.org/abs/2302.06654}{{\ttfamily
  2302.06654}}].

\bibitem{Co:2017pyf}
R.T.~Co and K.~Harigaya, \emph{{Gravitino Production Suppressed by Dynamics of
  Sgoldstino}}, \href{https://doi.org/10.1007/JHEP10(2017)207}{\emph{JHEP}
  {\bfseries 10} (2017) 207}
  [\href{https://arxiv.org/abs/1707.08965}{{\ttfamily 1707.08965}}].

\bibitem{Cosme:2024ndc}
C.~Cosme, F.~Costa and O.~Lebedev, \emph{{Temperature evolution in the Early
  Universe and freeze-in at stronger coupling}},
  \href{https://doi.org/10.1088/1475-7516/2024/06/031}{\emph{JCAP} {\bfseries
  06} (2024) 031} [\href{https://arxiv.org/abs/2402.04743}{{\ttfamily
  2402.04743}}].

\bibitem{Shtanov:1994ce}
Y.~Shtanov, J.H.~Traschen and R.H.~Brandenberger, \emph{{Universe reheating
  after inflation}},
  \href{https://doi.org/10.1103/PhysRevD.51.5438}{\emph{Phys. Rev. D}
  {\bfseries 51} (1995) 5438}
  [\href{https://arxiv.org/abs/hep-ph/9407247}{{\ttfamily hep-ph/9407247}}].

\bibitem{Ichikawa:2008ne}
K.~Ichikawa, T.~Suyama, T.~Takahashi and M.~Yamaguchi, \emph{{Primordial
  Curvature Fluctuation and Its Non-Gaussianity in Models with Modulated
  Reheating}}, \href{https://doi.org/10.1103/PhysRevD.78.063545}{\emph{Phys.
  Rev. D} {\bfseries 78} (2008) 063545}
  [\href{https://arxiv.org/abs/0807.3988}{{\ttfamily 0807.3988}}].

\bibitem{Bernal:2023wus}
N.~Bernal, S.~Cl\'ery, Y.~Mambrini and Y.~Xu, \emph{{Probing reheating with
  graviton bremsstrahlung}},
  \href{https://doi.org/10.1088/1475-7516/2024/01/065}{\emph{JCAP} {\bfseries
  01} (2024) 065} [\href{https://arxiv.org/abs/2311.12694}{{\ttfamily
  2311.12694}}].

\bibitem{Barman:2024mqo}
B.~Barman, N.~Bernal and Y.~Xu, \emph{{Resonant reheating}},
  \href{https://doi.org/10.1088/1475-7516/2024/08/014}{\emph{JCAP} {\bfseries
  08} (2024) 014} [\href{https://arxiv.org/abs/2404.16090}{{\ttfamily
  2404.16090}}].

\bibitem{Giudice:2000ex}
G.F.~Giudice, E.W.~Kolb and A.~Riotto, \emph{{Largest temperature of the
  radiation era and its cosmological implications}},
  \href{https://doi.org/10.1103/PhysRevD.64.023508}{\emph{Phys. Rev. D}
  {\bfseries 64} (2001) 023508}
  [\href{https://arxiv.org/abs/hep-ph/0005123}{{\ttfamily hep-ph/0005123}}].

\bibitem{Drees:2015exa}
M.~Drees, F.~Hajkarim and E.R.~Schmitz, \emph{{The Effects of QCD Equation of
  State on the Relic Density of WIMP Dark Matter}},
  \href{https://doi.org/10.1088/1475-7516/2015/06/025}{\emph{JCAP} {\bfseries
  06} (2015) 025} [\href{https://arxiv.org/abs/1503.03513}{{\ttfamily
  1503.03513}}].

\bibitem{ParticleDataGroup:2022pth}
{\scshape Particle Data Group} collaboration, \emph{{Review of Particle
  Physics}}, \href{https://doi.org/10.1093/ptep/ptac097}{\emph{PTEP} {\bfseries
  2022} (2022) 083C01}.

\bibitem{Viel:2013fqw}
M.~Viel, G.D.~Becker, J.S.~Bolton and M.G.~Haehnelt, \emph{{Warm dark matter as
  a solution to the small scale crisis: New constraints from high redshift
  Lyman-\ensuremath{\alpha} forest data}},
  \href{https://doi.org/10.1103/PhysRevD.88.043502}{\emph{Phys. Rev. D}
  {\bfseries 88} (2013) 043502}
  [\href{https://arxiv.org/abs/1306.2314}{{\ttfamily 1306.2314}}].

\bibitem{Irsic:2017ixq}
V.~Ir\v{s}i\v{c} et~al., \emph{{New Constraints on the free-streaming of warm
  dark matter from intermediate and small scale Lyman-$\alpha$ forest data}},
  \href{https://doi.org/10.1103/PhysRevD.96.023522}{\emph{Phys. Rev. D}
  {\bfseries 96} (2017) 023522}
  [\href{https://arxiv.org/abs/1702.01764}{{\ttfamily 1702.01764}}].

\bibitem{Sabti:2019mhn}
N.~Sabti, J.~Alvey, M.~Escudero, M.~Fairbairn and D.~Blas, \emph{{Refined
  Bounds on MeV-scale Thermal Dark Sectors from BBN and the CMB}},
  \href{https://doi.org/10.1088/1475-7516/2020/01/004}{\emph{JCAP} {\bfseries
  01} (2020) 004} [\href{https://arxiv.org/abs/1910.01649}{{\ttfamily
  1910.01649}}].

\bibitem{Sabti:2021reh}
N.~Sabti, J.~Alvey, M.~Escudero, M.~Fairbairn and D.~Blas, \emph{{Addendum:
  Refined bounds on MeV-scale thermal dark sectors from BBN and the CMB}},
  \href{https://doi.org/10.1088/1475-7516/2021/08/A01}{\emph{JCAP} {\bfseries
  08} (2021) A01} [\href{https://arxiv.org/abs/2107.11232}{{\ttfamily
  2107.11232}}].

\bibitem{Clowe:2003tk}
D.~Clowe, A.~Gonz\'alez and M.~Markevitch, \emph{{Weak lensing mass
  reconstruction of the interacting cluster 1E0657-558: Direct evidence for the
  existence of dark matter}},
  \href{https://doi.org/10.1086/381970}{\emph{Astrophys. J.} {\bfseries 604}
  (2004) 596} [\href{https://arxiv.org/abs/astro-ph/0312273}{{\ttfamily
  astro-ph/0312273}}].

\bibitem{Markevitch:2003at}
M.~Markevitch, A.H.~Gonz\'alez, D.~Clowe, A.~Vikhlinin, L.~David, W.~Forman
  et~al., \emph{{Direct constraints on the dark matter self-interaction
  cross-section from the merging galaxy cluster 1E0657-56}},
  \href{https://doi.org/10.1086/383178}{\emph{Astrophys. J.} {\bfseries 606}
  (2004) 819} [\href{https://arxiv.org/abs/astro-ph/0309303}{{\ttfamily
  astro-ph/0309303}}].

\bibitem{Randall:2008ppe}
S.W.~Randall, M.~Markevitch, D.~Clowe, A.H.~Gonz\'alez and M.~Brada{\v c},
  \emph{{Constraints on the Self-Interaction Cross-Section of Dark Matter from
  Numerical Simulations of the Merging Galaxy Cluster 1E 0657-56}},
  \href{https://doi.org/10.1086/587859}{\emph{Astrophys. J.} {\bfseries 679}
  (2008) 1173} [\href{https://arxiv.org/abs/0704.0261}{{\ttfamily 0704.0261}}].

\bibitem{Harvey:2015hha}
D.~Harvey, R.~Massey, T.~Kitching, A.~Taylor and E.~Tittley, \emph{{The
  non-gravitational interactions of dark matter in colliding galaxy clusters}},
  \href{https://doi.org/10.1126/science.1261381}{\emph{Science} {\bfseries 347}
  (2015) 1462} [\href{https://arxiv.org/abs/1503.07675}{{\ttfamily
  1503.07675}}].

\bibitem{Bondarenko:2017rfu}
K.~Bondarenko, A.~Boyarsky, T.~Bringmann and A.~Sokolenko, \emph{{Constraining
  self-interacting dark matter with scaling laws of observed halo surface
  densities}}, \href{https://doi.org/10.1088/1475-7516/2018/04/049}{\emph{JCAP}
  {\bfseries 04} (2018) 049}
  [\href{https://arxiv.org/abs/1712.06602}{{\ttfamily 1712.06602}}].

\bibitem{Harvey:2018uwf}
D.~Harvey, A.~Robertson, R.~Massey and I.G.~McCarthy, \emph{{Observable tests
  of self-interacting dark matter in galaxy clusters: BCG wobbles in a constant
  density core}}, \href{https://doi.org/10.1093/mnras/stz1816}{\emph{Mon. Not.
  Roy. Astron. Soc.} {\bfseries 488} (2019) 1572}
  [\href{https://arxiv.org/abs/1812.06981}{{\ttfamily 1812.06981}}].

\bibitem{Griest:1989wd}
K.~Griest and M.~Kamionkowski, \emph{{Unitarity Limits on the Mass and Radius
  of Dark Matter Particles}},
  \href{https://doi.org/10.1103/PhysRevLett.64.615}{\emph{Phys. Rev. Lett.}
  {\bfseries 64} (1990) 615}.

\bibitem{Coy:2024itg}
R.~Coy, J.~Kimus and M.H.G.~Tytgat, \emph{{Light from darkness: history of a
  hot dark sector}},  \href{https://arxiv.org/abs/2405.10792}{{\ttfamily
  2405.10792}}.

\bibitem{Leane:2018kjk}
R.K.~Leane, T.R.~Slatyer, J.F.~Beacom and K.C.Y.~Ng, \emph{{GeV-scale thermal
  WIMPs: Not even slightly ruled out}},
  \href{https://doi.org/10.1103/PhysRevD.98.023016}{\emph{Phys. Rev. D}
  {\bfseries 98} (2018) 023016}
  [\href{https://arxiv.org/abs/1805.10305}{{\ttfamily 1805.10305}}].

\bibitem{Hess:2021cdp}
{\scshape Hess, HAWC, VERITAS, MAGIC, H.E.S.S., Fermi-LAT} collaboration,
  \emph{{Combined dark matter searches towards dwarf spheroidal galaxies with
  Fermi-LAT, HAWC, H.E.S.S., MAGIC, and VERITAS}},
  \href{https://doi.org/10.22323/1.395.0528}{\emph{PoS} {\bfseries ICRC2021}
  (2021) 528} [\href{https://arxiv.org/abs/2108.13646}{{\ttfamily
  2108.13646}}].

\bibitem{Bergstrom:2013jra}
L.~Bergstrom, T.~Bringmann, I.~Cholis, D.~Hooper and C.~Weniger, \emph{{New
  Limits on Dark Matter Annihilation from AMS Cosmic Ray Positron Data}},
  \href{https://doi.org/10.1103/PhysRevLett.111.171101}{\emph{Phys. Rev. Lett.}
  {\bfseries 111} (2013) 171101}
  [\href{https://arxiv.org/abs/1306.3983}{{\ttfamily 1306.3983}}].

\bibitem{Calore:2022stf}
F.~Calore, M.~Cirelli, L.~Derome, Y.~Genolini, D.~Maurin, P.~Salati et~al.,
  \emph{{AMS-02 antiprotons and dark matter: Trimmed hints and robust bounds}},
  \href{https://doi.org/10.21468/SciPostPhys.12.5.163}{\emph{SciPost Phys.}
  {\bfseries 12} (2022) 163}
  [\href{https://arxiv.org/abs/2202.03076}{{\ttfamily 2202.03076}}].

\bibitem{Zemp:2008gw}
M.~Zemp, J.~Diemand, M.~Kuhlen, P.~Madau, B.~Moore, D.~Potter et~al.,
  \emph{{The Graininess of Dark Matter Haloes}},
  \href{https://doi.org/10.1111/j.1365-2966.2008.14361.x}{\emph{Mon. Not. Roy.
  Astron. Soc.} {\bfseries 394} (2009) 641}
  [\href{https://arxiv.org/abs/0812.2033}{{\ttfamily 0812.2033}}].

\bibitem{Pato:2010yq}
M.~Pato, O.~Agertz, G.~Bertone, B.~Moore and R.~Teyssier, \emph{{Systematic
  uncertainties in the determination of the local dark matter density}},
  \href{https://doi.org/10.1103/PhysRevD.82.023531}{\emph{Phys. Rev. D}
  {\bfseries 82} (2010) 023531}
  [\href{https://arxiv.org/abs/1006.1322}{{\ttfamily 1006.1322}}].

\bibitem{Bernal:2014mmt}
N.~Bernal, J.E.~Forero-Romero, R.~Garani and S.~Palomares-Ruiz,
  \emph{{Systematic uncertainties from halo asphericity in dark matter
  searches}}, \href{https://doi.org/10.1088/1475-7516/2014/09/004}{\emph{JCAP}
  {\bfseries 09} (2014) 004} [\href{https://arxiv.org/abs/1405.6240}{{\ttfamily
  1405.6240}}].

\bibitem{Boudaud:2014dta}
M.~Boudaud et~al., \emph{{A new look at the cosmic ray positron fraction}},
  \href{https://doi.org/10.1051/0004-6361/201425197}{\emph{Astron. Astrophys.}
  {\bfseries 575} (2015) A67}
  [\href{https://arxiv.org/abs/1410.3799}{{\ttfamily 1410.3799}}].

\bibitem{Bernal:2015oyn}
N.~Bernal, J.E.~Forero-Romero, R.~Garani and S.~Palomares-Ruiz,
  \emph{{Systematic uncertainties from halo asphericity in dark matter
  searches}},
  \href{https://doi.org/10.1016/j.nuclphysbps.2015.10.129}{\emph{Nucl. Part.
  Phys. Proc.} {\bfseries 267-269} (2015) 345}.

\bibitem{Bernal:2016guq}
N.~Bernal, L.~Necib and T.R.~Slatyer, \emph{{Spherical Cows in Dark Matter
  Indirect Detection}},
  \href{https://doi.org/10.1088/1475-7516/2016/12/030}{\emph{JCAP} {\bfseries
  12} (2016) 030} [\href{https://arxiv.org/abs/1606.00433}{{\ttfamily
  1606.00433}}].

\bibitem{Benito:2016kyp}
M.~Benito, N.~Bernal, N.~Bozorgnia, F.~Calore and F.~Iocco, \emph{{Particle
  Dark Matter Constraints: the Effect of Galactic Uncertainties}},
  \href{https://doi.org/10.1088/1475-7516/2017/02/007}{\emph{JCAP} {\bfseries
  02} (2017) 007} [\href{https://arxiv.org/abs/1612.02010}{{\ttfamily
  1612.02010}}].

\bibitem{CTA:2020qlo}
{\scshape CTA} collaboration, \emph{{Sensitivity of the Cherenkov Telescope
  Array to a dark matter signal from the Galactic centre}},
  \href{https://doi.org/10.1088/1475-7516/2021/01/057}{\emph{JCAP} {\bfseries
  01} (2021) 057} [\href{https://arxiv.org/abs/2007.16129}{{\ttfamily
  2007.16129}}].

\bibitem{Essig:2017kqs}
R.~Essig, T.~Volansky and T.-T.~Yu, \emph{{New Constraints and Prospects for
  sub-GeV Dark Matter Scattering off Electrons in Xenon}},
  \href{https://doi.org/10.1103/PhysRevD.96.043017}{\emph{Phys. Rev. D}
  {\bfseries 96} (2017) 043017}
  [\href{https://arxiv.org/abs/1703.00910}{{\ttfamily 1703.00910}}].

\bibitem{DarkSide:2022knj}
{\scshape DarkSide} collaboration, \emph{{Search for Dark Matter Particle
  Interactions with Electron Final States with DarkSide-50}},
  \href{https://doi.org/10.1103/PhysRevLett.130.101002}{\emph{Phys. Rev. Lett.}
  {\bfseries 130} (2023) 101002}
  [\href{https://arxiv.org/abs/2207.11968}{{\ttfamily 2207.11968}}].

\bibitem{DarkSide-50:2022qzh}
{\scshape DarkSide-50} collaboration, \emph{{Search for low-mass dark matter
  WIMPs with 12~ton-day exposure of DarkSide-50}},
  \href{https://doi.org/10.1103/PhysRevD.107.063001}{\emph{Phys. Rev. D}
  {\bfseries 107} (2023) 063001}
  [\href{https://arxiv.org/abs/2207.11966}{{\ttfamily 2207.11966}}].

\bibitem{XENON:2017vdw}
{\scshape XENON} collaboration, \emph{{First Dark Matter Search Results from
  the XENON1T Experiment}},
  \href{https://doi.org/10.1103/PhysRevLett.119.181301}{\emph{Phys. Rev. Lett.}
  {\bfseries 119} (2017) 181301}
  [\href{https://arxiv.org/abs/1705.06655}{{\ttfamily 1705.06655}}].

\bibitem{LZ:2022lsv}
{\scshape LZ} collaboration, \emph{{First Dark Matter Search Results from the
  LUX-ZEPLIN (LZ) Experiment}},
  \href{https://doi.org/10.1103/PhysRevLett.131.041002}{\emph{Phys. Rev. Lett.}
  {\bfseries 131} (2023) 041002}
  [\href{https://arxiv.org/abs/2207.03764}{{\ttfamily 2207.03764}}].

\bibitem{Essig:2015cda}
R.~Essig, M.~Fernandez-Serra, J.~Mardon, A.~Soto, T.~Volansky and T.-T.~Yu,
  \emph{{Direct Detection of sub-GeV Dark Matter with Semiconductor Targets}},
  \href{https://doi.org/10.1007/JHEP05(2016)046}{\emph{JHEP} {\bfseries 05}
  (2016) 046} [\href{https://arxiv.org/abs/1509.01598}{{\ttfamily
  1509.01598}}].

\bibitem{DARWIN:2016hyl}
{\scshape DARWIN} collaboration, \emph{{DARWIN: towards the ultimate dark
  matter detector}},
  \href{https://doi.org/10.1088/1475-7516/2016/11/017}{\emph{JCAP} {\bfseries
  11} (2016) 017} [\href{https://arxiv.org/abs/1606.07001}{{\ttfamily
  1606.07001}}].

\bibitem{Billard:2021uyg}
J.~Billard et~al., \emph{{Direct detection of dark matter\textemdash{}APPEC
  committee report*}},
  \href{https://doi.org/10.1088/1361-6633/ac5754}{\emph{Rept. Prog. Phys.}
  {\bfseries 85} (2022) 056201}
  [\href{https://arxiv.org/abs/2104.07634}{{\ttfamily 2104.07634}}].

\bibitem{Silva-Malpartida:2024emu}
J.~Silva-Malpartida, N.~Bernal, J.~Jones-P\'erez and R.A.~Lineros, \emph{{From
  WIMPs to FIMPs: Impact of Early Matter Domination}},
  \href{https://arxiv.org/abs/2408.08950}{{\ttfamily 2408.08950}}.

\bibitem{BICEP:2021xfz}
{\scshape BICEP, Keck} collaboration, \emph{{Improved Constraints on Primordial
  Gravitational Waves using Planck, WMAP, and BICEP/Keck Observations through
  the 2018 Observing Season}},
  \href{https://doi.org/10.1103/PhysRevLett.127.151301}{\emph{Phys. Rev. Lett.}
  {\bfseries 127} (2021) 151301}
  [\href{https://arxiv.org/abs/2110.00483}{{\ttfamily 2110.00483}}].

\end{thebibliography}\endgroup
\end{document}